\documentclass[longauth]{aa}  

\usepackage{graphicx}
\usepackage{txfonts}
\usepackage[T1]{fontenc}
\usepackage{ae,aecompl}

\usepackage{amsmath}
\usepackage{amssymb}
\usepackage{multirow,bigdelim}
\usepackage{mathtools}
\usepackage{siunitx}
\usepackage[colorlinks=true,linkcolor=blue, citecolor=blue]{hyperref}%
\usepackage{subfig}
\usepackage{bm}
\usepackage{soul}
\usepackage{comment}
\usepackage{color}

\numberwithin{equation}{section}

\usepackage{etoolbox}

\newcommand{\mpch}{\,h^{-1}{\rm{Mpc}}}

\newcommand{\meanz}{\left\langle{z}\right\rangle}
\newcommand{\sqdeg}{\,{\rm{deg}}^{2}}
\newcommand{\psqarcmin}{\,{\rm{arcmin}}^{-2}}

\newcommand{\nzcov}{\Sigma_{n(z)}}
\newcommand{\deltazi}{\delta{z}_i}
\newcommand{\hcomb}{\mathcal{N}_{\rm comb}+\deltazi}
\newcommand{\fcomb}{\mathcal{N}_{\rm comb}}
\newcommand{\GG}{\gamma\gamma}
\newcommand{\pp}{\rm 3^{ph}\times2pt}
\newcommand{\thr}{\rm 3\times2pt}
\newcommand{\six}{\rm 6\times2pt}

\newcommand{\nztrue}{n^{{\rm true}}_{z}}
\newcommand{\nzshift}{n^{{\rm shift}}_{z}}
\newcommand{\nzunb}{n^{{\rm unb.}}_{z}}
\newcommand{\nzinc}{n^{{\rm inc}}_{z}}
\newcommand{\nzcoh}{n^{{\rm coh}}_{z}}

\newcommand{\Dinc}{\Delta_{\rm inc}}

\newcommand{\Dfull}{\Delta_{\rm full}}
\newcommand{\Om}{\Omega_{\rm m}}

\newcommand{\seom}{S_8\mbox{--}\Omega_{\rm m}}
\newcommand{\siom}{\sigma_8\mbox{--}\Omega_{\rm m}}
\newcommand{\lnas}{\ln\left(10^{10}A_{\rm s}\right)}

\newcommand{\bgphot}{b^{\rm phot}_{\rm g}}

\begin{document}

\title{6x2pt: Forecasting gains from joint weak lensing and galaxy clustering analyses with spectroscopic-photometric galaxy cross-correlations
}
\titlerunning{Joint weak lensing and clustering analyses with sample cross-correlations}

\author{
Harry Johnston\inst{1}\thanks{harrysj100@gmail.com}, 
Nora Elisa Chisari\inst{1,2}\thanks{n.e.chisari@uu.nl}, 
Shahab Joudaki\inst{3,4,5,6}\thanks{shahab.joudaki@ciemat.es}, 
Robert Reischke\inst{7},
Benjamin St\"{o}lzner\inst{7},
Arthur Loureiro\inst{8,9},
Constance Mahony\inst{7,10},
Sandra Unruh\inst{13},
Angus H. Wright\inst{7},
Marika Asgari\inst{11},
Maciej Bilicki\inst{12},
Pierre Burger\inst{5,6,13},
Andrej Dvornik\inst{7},
Christos Georgiou\inst{1},
Benjamin Giblin\inst{14},
Catherine Heymans\inst{7,14},
Hendrik Hildebrandt\inst{7},
Benjamin Joachimi\inst{15},
Konrad Kuijken\inst{2},
Shun-Sheng Li\inst{2, 16},
Laila Linke\inst{17},
Lucas Porth\inst{11},
HuanYuan Shan\inst{18,19,20},
Tilman Tr\"{o}ster\inst{21},
Jan Luca van den Busch\inst{7},
Maximilian von Wietersheim-Kramsta\inst{22,23},
Ziang Yan\inst{7},
Yun-Hao Zhang\inst{2,14}.
}

\authorrunning{H. Johnston et al. }

\institute{
Institute for Theoretical Physics, Utrecht University, Princetonplein 5, 3584CC Utrecht, The Netherlands\and
Leiden Observatory, Leiden University, Niels Bohrweg 2, NL-2333 CA Leiden, The Netherlands\and
Centro de Investigaciones Energ\'{e}ticas, Medioambientales y Tecnol\'{o}gicas (CIEMAT), Av. Complutense 40, E-28040 Madrid, Spain\and
Institute of Cosmology \& Gravitation, Dennis Sciama Building, University of Portsmouth,
Portsmouth, PO1 3FX, UK\and
Waterloo Centre for Astrophysics, University of Waterloo, 200 University Ave W, Waterloo, ON N2L 3G1, Canada\and
Department of Physics and Astronomy, University of Waterloo, 200 University Ave W, Waterloo, ON N2L 3G1, Canada\and
Ruhr University Bochum, Faculty of Physics and Astronomy, Astronomical Institute (AIRUB), German Centre for Cosmological Lensing, 44780 Bochum, Germany \and
The Oskar Klein Centre, Department of Physics, Stockholm University, AlbaNova University Centre, SE-106 91 Stockholm, Sweden \and
Astrophysics Group and Imperial Centre for Inference and Cosmology (ICIC), Blackett Laboratory, Imperial College London, Prince Consort Road, London SW7 2AZ, UK\and
Donostia International Physics Center, Manuel Lardizabal Ibilbidea, 4, 20018 Donostia, Gipuzkoa, Spain\and
School of Mathematics, Statistics and Physics, Newcastle University, Herschel Building, NE1 7RU, Newcastle-upon-Tyne, UK \and
Center for Theoretical Physics, Polish Academy of Sciences, al. Lotników 32/46, 02-668 Warsaw, Poland \and 
Argelander-Institut für Astronomie, Auf dem Hügel 71, 53121 Bonn, Germany \and
Institute for Astronomy, University of Edinburgh, Blackford Hill, Edinburgh, EH9 3HJ, UK\and
Department of Physics and Astronomy, University College London, Gower Street, London WC1E 6BT, UK\and
Aix-Marseille Université, CNRS, CNES, LAM, Marseille, France\and
Universität Innsbruck, Institut für Astro- und Teilchenphysik, Technikerstr. 25/8, 6020 Innsbruck, Austria\and
Shanghai Astronomical Observatory (SHAO), Nandan Road 80, Shanghai 200030, China \and
Key Laboratory of Radio Astronomy and Technology, Chinese Academy of Sciences, A20 Datun Road, Chaoyang District, Beijing, 100101, P. R. China\and 
University of Chinese Academy of Sciences, Beijing 100049, China\and
Institute for Particle Physics and Astrophysics, ETH Z\"{u}rich, Wolfgang-Pauli-Strasse 27, 8093 Z\"{u}rich, Switzerland\and
Department of Physics, Institute for Computational  Cosmology, Durham University, South Road, Durham DH1 3LE, UK \and
Department of Physics, Centre for Extragalactic Astronomy, Durham University, South Road, Durham DH1 3LE, UK 
}
% These dates will be filled out by the publisher
\date{Accepted XXX. Received YYY; in original form ZZZ}

\label{kids:firstpage}
\makeatletter
\renewcommand*\aa@pageof{, page \thepage{} of \pageref*{LastPage}}
\makeatother

\abstract{
Accurate knowledge of galaxy redshift distributions is crucial in the inference of cosmological parameters from large-scale structure data. We explore the potential for enhanced self-calibration of photometric galaxy redshift distributions, $n(z)$, through the joint analysis of up to six two-point functions. Our $\rm 3\times2pt$ configuration is comprised of photometric shear, spectroscopic galaxy clustering, and spectroscopic-photometric galaxy-galaxy lensing (GGL). We expand this to include spectroscopic-photometric cross-clustering; photometric GGL; and photometric auto-clustering, using the photometric shear sample as an additional density tracer. We perform simulated likelihood forecasts of the cosmological and nuisance parameter constraints for Stage-III- and Stage-IV-like surveys. For the Stage-III-like survey, we employ realistic redshift distributions with perturbations across the full shape of the $n(z)$, and distinguish between `coherent' shifting of the bulk distribution in one direction, versus more internal scattering and full-shape $n(z)$ errors. For perfectly known $n(z)$, a $\rm 6\times2pt$ analysis gains $\sim40\%$ in Figure of Merit (FoM) in the $S_8\equiv\sigma_8\sqrt{\Om/0.3}$ and $\Omega_{\rm m}$ plane relative to the $\rm 3\times2pt$ analysis. If untreated, coherent and incoherent redshift errors lead to inaccurate inferences of $S_8$ and $\Omega_{\rm m}$, respectively, and contaminate inferences of the amplitude of intrinsic galaxy alignments. Employing bin-wise scalar shifts $\delta{z}_i$ in the tomographic mean redshifts reduces cosmological parameter biases, with a $\rm 6\times2pt$ analysis constraining the $\delta{z}_i$ parameters with $2-4$ times the precision of a photometric $\rm 3^{ph}\times2pt$ analysis. For the Stage-IV-like survey, a $\rm 6\times2pt$ analysis doubles the FoM($\sigma_8\mbox{--}\Omega_{\rm m}$) compared to the $\rm 3\times2pt$ or $\rm 3^{ph}\times2pt$ analyses, and is only 8\% less constraining than if the $n(z)$ were perfectly known. A Gaussian mixture model for the $n(z)$ is able to reduce mean-redshift errors whilst preserving the $n(z)$ shape, and thereby yields the most accurate and precise cosmological constraints for any given $N\rm\times2pt$ configuration in the presence of $n(z)$ biases.
}

\maketitle

% \clearpage

%Comment out for A&A submission
\tableofcontents

\section{Introduction}
\label{sec:introduction}

Modern analyses of the large-scale structure (LSS) of the Universe frequently combine different cosmological probes to maximally leverage the available information, and break degeneracies between key parameters of the concordance model. One of the most powerful probes in use today is the weak gravitational lensing of light from distant galaxies \citep{Bartelmann01}. 
The coherent ellipticity distortions induced by weak lensing modify galaxy isophotes at the per cent level. We therefore require many millions of galaxy shapes in order to measure shear correlations at a sufficient signal-to-noise ratio to constrain the underlying cosmology \citep{cfhtlens,Hamana:2019etx, Asgari21, Secco2022,Amon2022,Longley22}. The next stage of experiments will expand on this endeavour by observing billions of galaxies.

Constraints derived from the lensing effect alone are subject to a strong degeneracy \citep{Jain97} between the amount of matter in the Universe today, $\Omega_{\rm{m}}$, and its degree of clustering, parameterised by $\sigma_8$ (the root mean square of the density contrast in spheres of radius 8$\mpch$). 
Matter inhomogeneities source lensing convergence and shear fields that correlate with the cosmic density field. By sampling these fields at the positions of galaxies, one can probe density fluctuations in the aggregated dark and luminous matter distribution. The `$\rm 3\times2pt$' method, which jointly analyses the auto- and cross-correlations of lensing and density fields, aids with the breaking of the $\sigma_8-\Om$ degeneracy, and has begun to yield cosmological parameter constraints that are competitive with those obtained via analyses of the cosmic microwave background (CMB) temperature and polarisation anisotropies \citep{Joudaki2017,vanUitert18,DES:2017myr,heymans21,DES3x2pt}. It has also been suggested that the cross-correlation of lensing and galaxy fields breaks the degeneracy between galaxy bias and the growth rate, which allows for stronger constraints on cosmic expansion scenarios \citep{Bernstein11,Gaztanaga12,Cai12}. However, whether these gains are present for any other fundamental parameters or for calibrating lensing systematics has been unclear so far \citep{FontRibera14, dePutter14}.

The increased precision of these constraints has revealed a mild tension between the late- and early-Universe determinations of the matter clustering parameter combination $S_{8}\equiv\sigma_8\sqrt{\Omega_{\rm{m}}/0.3}$, where the CMB predicts a present-day value that is $2-3\sigma$ larger than that observed by lensing experiments \citep{Joudaki:2016mvz,Hildebrandt2016,Joudaki:2016kym, Leauthaud17,Hikage19, Hamana:2019etx, Joudaki:2019pmv, Asgari:2019fkq, Troster2021, Secco2022, Amon2022, HSCy3a, HSCy3b, HSC3x2pta, HSC3x2ptb, HSC3x2ptc}. Resolving this tension, whether by the better understanding of experimental or theoretical limitations and assumptions \citep{Longley23,KiDS1000DES}, or with new physics \citep[see][and references therein]{diValentino21,Joudaki:2020shz, Abdalla22}, is a primary goal of LSS cosmology. To this end, upcoming surveys are already making significant efforts towards implementing the $\rm 3\times2pt$ methodology within their analysis pipelines \citep{Chisari2018,Euclid:2019clj, Tutusaus20,Sanchez21,Prat23}. 

In preparation for the next generation of experiments, the control of systematic errors in weak lensing analyses is more important than ever. The {\it Euclid} satellite \citep{Euclid}, the {\it Vera C. Rubin} Observatory \citep{LSST}, the {\it Nancy Grace Roman} Space Telescope \citep{Spergel2015}, the Canada-France Imaging Survey (CFIS-UNIONS;~\citealt{UNIONS}), and the Chinese Space Station Telescope (CSST;~\citealt{CSST}) will have to outperform their predecessors with regards to the effects and mitigation of photometric redshift uncertainty, intrinsic alignments of galaxies, baryonic feedback, shear misestimation, magnification bias, covariance misestimation, and various other sources of error. Of particular concern in recent years is the calibration of lensing source sample redshift distributions, $n(z)$ \citep{TheLSSTDarkEnergyScienceCollaboration2018,Joudaki2019, SchmidtDESC,Desprez20}. 

External calibration of photometric redshift distributions is an active field of research, and can be broadly categorised as using spectroscopic or high-quality photometric reference samples \citep[e.g.][]{Laigle16, paus19, Weaver:2021obz, Busch:2022pcx} to either infer galaxy colour-redshift space relations \citep{Hildebrandt12,Leistedt16,Hoyle18,Sanchez19,Hikage19,Wright2020,Hildebrandt20,Hildebrandt2021}, or to measure positional cross-correlations \citep{Newman08,Menard13,dePutter14,Johnson:2016ucr, vandenBusch:2020lur, Gatti22}, or some combination thereof \citep{Alarcon2020,LSSTDarkEnergyScience:2021ozf, Myles21}. For a recent review on photometric redshifts and their challenges, see \citet{Newman2022}. While spectroscopy remains too expensive to provide an accurate redshift for every observed object, the modelling of lensing statistics only requires accurate knowledge of the redshift \emph{distributions} of source galaxy samples \citep{Bernardeau97,Huterer06}, rather than redshift point-estimates. Residual inaccuracies in the inferred redshift distributions are then typically modelled with nuisance parameters that are internally calibrated by the measured statistics, and marginalised over in the cosmological parameter inference.

The now-standard $\rm 3\times2pt$ formalism combines cosmic shear, galaxy-galaxy lensing (GGL), and galaxy clustering correlations, wherein positional samples are chosen to be those with spectroscopic redshift (or high-accuracy photo-$z$) information. 
Upcoming surveys are also considering leveraging the information in the clustering of photometric samples with low-quality photo-$z$ \citep{Nicola20,Tutusaus20,Sanchez21}. The prospective gains would be in area coverage, redshift baseline, and the number of objects, where the latter two increase greatly on approach to the survey magnitude limit. The concern lies in having control over spatially-correlated, systematic density fluctuations induced by survey inhomogeneities \citep{Leistedt13,Leistedt14,Awan16,Leistedt16b}. However, new methodologies are now bringing this control within reach \citep{Alonso19,Rezaie2019,Johnston21,Everett22}.

In either of those $\rm 3\times2pt$ strategies, redshift calibration remains a primary challenge. Collecting representative spectroscopic samples is currently infeasible at the depth of upcoming surveys \citep{Newman15}. Clustering-based redshift calibration techniques are then expected to be prioritised over calibration with deep spectroscopic data \citep{Snowmass}. However, the photometric data themselves can also be used to avoid systematic biases or loss of precision. \citet{Schaan20} demonstrated, for example, that photometric redshift scatter and outliers yield detectable clustering cross-correlations across redshift bins in photometric samples. These can improve the constraints on redshift nuisance parameters by an order of magnitude. 
This work takes a step further by exploring the possibility of enhancing internal redshift re-calibration through the inclusion of {\it spectroscopic-photometric cross-correlations} as additional probes within a joint analysis of two-point statistics. We propose to extend the $\rm 3\times2pt$ formalism to include the cross-clustering of spectroscopic and photometric samples ($\rm 4\times2pt$), as well as the GGL and galaxy clustering measured within all-photometric samples, for a maximal $\rm 6\times2pt$ analysis.

Our analysis takes the form of a full simulated likelihood forecast for a current (Stage-III) weak lensing survey, and a complementary, simpler forecast for the upcoming generation of surveys (Stage-IV). We compute all unique cosmological and systematic contributor (intrinsic alignments, magnification) angular power spectra, $C(\ell)$, within and between two tomographic galaxy samples: a photometric sample tracing both the shear and density fields, with an uncertain redshift distribution to be modelled; and a spectroscopic sample tracing only the density field, and with a known redshift distribution.

We explore synthetic source distributions, $n(z)$, with variations over the full shape of the function, choosing distributions for analysis that are `coherently' shifted in terms of tomographic mean redshifts (all mean-redshift differences have the same sign), or `incoherently' shifted (signs can be mixed), where calibration errors of the former kind are expected to manifest more strongly in $S_8$ \citep{Joudaki:2019pmv}. We attempt to recover these shifted distributions through internal recalibration at and/or before the sampler stage, employing a selection of nuisance models for the task: scalar shifts $\delta{z_{i}}$ to be applied to the means of tomographic bins (e.g.~\citealt{cfhtlens,Asgari21, Amon2022}); flexible recalibration with the Gaussian mixture `comb' model \citep{Kuijken1993,Stolzner2020}, with and without additional scalar shifts; and a `do nothing' model, for characterising the cosmological parameter biases incurred by different modes of redshift distribution calibration failure. We report on the suitability of these redshift nuisance models for Stage-III and Stage-IV weak lensing analyses, and on the gains in accuracy and precision of cosmological inference to be derived from the inclusion of additional two-point correlations in the $\rm 6\times2pt$ analysis. In addition, we show how photometric redshift nuisance parameters can couple to other astrophysical systematics, namely, intrinsic galaxy alignments \citep{Wright2020,Fortuna21,Li2021,Secco2022,Fischbacher2022,Leonard24}.

This paper is structured as follows: Sect.~\ref{sec:theory} describes our modelling of the harmonic space two-point functions that form our analysis data-vector. Sect.~\ref{sec:data} details our synthetic data products, and we present our forecasting methodology in Sect.~\ref{sec:forecast}. Sect.~\ref{sec:results} displays and discusses the results of our forecasts, and we make concluding remarks in Sect.~\ref{sec:conclusions}.

\section{Theoretical modelling of two-point functions}
\label{sec:theory}

The two-point functions considered as part of the data-vector in this work include all unique cross- and auto-angular power spectra between the positions of the spectroscopic sample and the positions and shapes of the photometric sample. For the spectroscopic sample, we restrict the analysis to angular power spectra (as opposed to the redshift-space multipole treatments of e.g.~\citealt{Gil-Marin:2020bct, Bautista:2020ahg, Beutler:2021eqq}) due to limitations in the analytical computation of the covariance between multipoles and angular power spectra, which we defer to future work (however, also see \citealt{Taylor:2022rgy}). 

\subsection{Angular power spectra in the general case}
\label{sec:generic_spectra}

The angular power spectrum of the cross-correlation between galaxy positions ($n$) in two tomographic galaxy samples ($\alpha$, $\beta$) is given by several contributions deriving from gravitational clustering (g) and lensing magnification (m),
\begin{equation}
     C^{\alpha\beta}_{nn}(\ell) = C^{\alpha\beta}_{\rm{gg}}(\ell) + C^{\alpha\beta}_{\rm{mg}}(\ell) + C^{\alpha\beta}_{\rm{gm}}(\ell) + C^{\alpha\beta}_{\rm{mm}}(\ell)\,.
    \label{eq:Cnn}
\end{equation}
The indices $\alpha,\beta$ each run over the set of unique radial kernels employed in the analysis of $N_{\mathrm{p}}$ photometric and $N_{\mathrm{s}}$ spectroscopic redshift samples. 

Throughout this analysis, we make use of the Limber approximation \citep{Limber1953,Kaiser1992,LoVerde:2008re} in the computation of the angular power spectra. Although the Limber approximation is known to be insufficient for the level of accuracy required by future surveys, especially in the context of clustering cross-correlations across tomographic bins \citep{Campagne17}, the computational cost of performing non-Limber computations of angular power spectra is currently prohibitive. In the near future, we expect our pipeline to be extended in this direction as new, fast, validated methods become suitable for embedding into full-likelihood analyses.

For now, we limit our analysis to scales $\ell>100$ \citep{Joachimi21}. We also neglect to include contributions derived from redshift space distortions \citep[RSDs;][]{Kaiser87}, which should be small for these scales and principally affect the tomographic cross-correlations \citep{Loureiro2019}. Lastly, we will restrict the clustering analysis to scales were the bias can be approximated as being linear. See Sect.~\ref{sec:approximations} for details of other approximations made in this work.

Under the Limber approximation in Fourier space, the contribution attributed to pure gravitational clustering is
\begin{equation}
C_{\rm{gg}}^{\alpha\beta}(\ell)=\int_0^{\chi_{\rm H}}\mathrm{d}\chi \frac{b_{\rm g}^{\alpha}n_\alpha(\chi)b_{\rm g}^{\beta}n_\beta(\chi)}{f_K^2(\chi)}P_\delta\left(\frac{\ell+1/2}{f_K(\chi)},\chi\right)\,,
    \label{eq:Cgg}
\end{equation}
where $\chi_{\rm H}$ is the comoving distance to the horizon, $b_{\rm g}^{\alpha}$ and $b_{\rm g}^{\beta}$ are the linear biases of the galaxy samples,  $n_\alpha$ and $n_\beta$ are the normalised redshift distributions of each sample, $f_K(\chi)$ is the comoving angular diameter distance, and $P_\delta$ is the matter power spectrum. 

In addition, lensing magnification induces apparent excesses or deficits of galaxies above the flux limit of a survey due to the conservation of surface brightness of the lensed sources. Furthermore, the observed angular separation of galaxies behind the lenses are increased, diluting their number density.
As a result, galaxy number counts pick up an additional contribution which is cross-correlated with the physical locations of galaxies that act as lenses. The result is three additional terms in the $nn$ angular power spectra: the magnification count auto-correlation, `mm'
\begin{equation}
C_{\rm{mm}}^{\alpha\beta}(\ell)=\int^{\chi_{\rm{H}}}_{0} \mathrm{d}\chi \frac{\bar{q}_{\alpha}(\chi)\bar{q}_{\beta}(\chi)}{f^{2}_{K}(\chi)} P_{\delta}\left(\frac{\ell+1/2}{f_{K}(\chi)}, \chi\right)\,;
\end{equation}
and the magnification count--number count cross-correlations, `mg' and `gm'
\begin{equation}
C_{\rm{mg}}^{\alpha\beta}(\ell)= \int^{\chi_{\rm{H}}}_{0} \mathrm{d}\chi \frac{\bar{q}_{\alpha}(\chi)b^{\beta}_{\mathrm{g}}n_{\beta}(\chi)}{f^{2}_{K}(\chi)} P_{\delta}\left(\frac{\ell+1/2}{f_{K}(\chi)}, \chi\right)\,,
\label{eq:Cmg}
\end{equation}
where the gm term is constructed in exact analogy to the mg term by swapping the indices $\alpha$ and $\beta$. Magnification kernels $\bar{q}_{\alpha}(\chi)$ are derived from the respective lensing efficiency kernels $q_{\alpha}(\chi)$, which are given by
\begin{equation}
    q_x(\chi) = \frac{3\Omega_{\rm m}H_0^2}{2c^2}\frac{f_K(\chi)}{a(\chi)}\int_\chi^{\chi_{\rm H}}{\rm d}\chi'n_x(\chi')\frac{f_K(\chi-\chi')}{f_K(\chi')}\,.
    \label{eq:lensing_efficiency}
\end{equation}
Multiplying $F_{x,\rm m}(\chi)$ within the integrand of Eq. (\ref{eq:lensing_efficiency}),
\begin{equation}
    F_{x,\rm m}(\chi)=5s_x(\chi)-2 = 2(\alpha_x(\chi)-1)\,,
    \label{eq:magnification_kernel_modifier}
\end{equation}
yields the magnification kernel $\bar{q}_x(\chi)$. For tomographic sample $x$, $s_x(\chi)$ is the logarithmic slope of the magnitude distribution \citep[e.g.,][]{Chisari2018}, and $\alpha_x(\chi)$ is that of the luminosity function \citep[e.g.,][]{Joachimi2010} -- not to be confused with the tomographic bin index $\alpha$. For computations of $F_{x,{\rm m}}(\chi)$, we make use of the fitting formula for $\alpha$ given by \cite{Joachimi2010} (their Appendix C), assuming a distinct limiting $r$-band depth $r_{\rm lim}$ for each of our synthetic samples (defined in Sect. \ref{sec:data}). 

The shape ($\gamma$) auto-spectrum is similarly given by several contributions
\begin{equation}
   C^{\alpha\beta}_{\gamma\gamma}(\ell) = C^{\alpha\beta}_{\rm{GG}}(\ell) + C^{\alpha\beta}_{\rm{GI}}(\ell) + C^{\alpha\beta}_{\rm{IG}}(\ell) + C^{\alpha\beta}_{\rm{II}}(\ell)\,,
   \label{eq:Cgamgam}
\end{equation}
where `GG' indicates a pure gravitational lensing contribution, given by
\begin{equation}
    C^{\alpha\beta}_{\rm{GG}}(\ell)=\int^{\chi_{\rm{H}}}_{0} \mathrm{d}\chi \frac{q_{\alpha}(\chi)q_{\beta}(\chi)}{f^{2}_{K}(\chi)} P_{\delta}\left(\frac{\ell+1/2}{f_{K}(\chi)}, \chi\right).
\end{equation}
The `I' terms in Eq. (\ref{eq:Cgamgam}) are well-known to arise from intrinsic (local, tidally-induced, as opposed to lensing-induced) alignments of the galaxies with the underlying matter field \citep{Catelan2001,Hirata2004a}. These terms are known to cause biases in cosmological constraints if unaccounted for \citep{Joachimi2010,Krause2016, cfhtlens}. Moreover, they are expected to absorb residual biases in photometric sample redshift calibration if the nuisance model for alignments is too flexible and/or not specific enough \citep{Wright2020,Fortuna21,Li2021,Fischbacher2022}.

The `GI' contribution represents the lensing of background galaxy shapes by the same matter field which is responsible for the intrinsic alignments of foreground galaxies. This is given by
\begin{equation}
    C^{\alpha\beta}_{\rm{GI}}(\ell)=\int^{\chi_{\rm{H}}}_{0} \mathrm{d}\chi \frac{q_{\alpha}(\chi)n_{\beta}(\chi)}{f^{2}_{K}(\chi)} F_{\mathrm{IA}}(\chi) P_{\delta}\left(\frac{\ell+1/2}{f_{K}(\chi)}, \chi\right) \,,
    \label{eq:CGI}
\end{equation}
and the case of `IG' is analogously constructed by swapping the indices $\alpha$ and $\beta$. $F_{\rm IA}(\chi)$ represents an effective amplitude of the alignment of galaxies with respect to the tidal field as a function of comoving distance. Although this formalism is strictly linear, it is common to use the nonlinear matter power spectrum in the computation of intrinsic alignment correlations \citep{Bridle2007}.

The `II' contribution in Eq. (\ref{eq:Cgamgam}) is given by
\begin{equation}
    C^{\alpha\beta}_{\rm{II}}(\ell)
=\int^{\chi_{\rm{H}}}_{0} \mathrm{d}\chi \frac{n_{\alpha}(\chi)n_{\beta}(\chi)}{f^{2}_{K}(\chi)} F^2_{\mathrm{IA}}(\chi) P_{\delta}\left(\frac{\ell+1/2}{f_{K}(\chi)}, \chi\right)\,,
    \label{eq:CII}
\end{equation}
and represents the auto-correlation spectrum of galaxies aligned by the same underlying tidal field; it is thus expected to contribute more weakly to tomographic shear cross-correlations than the GI term, which can operate over wide separations in redshift.

The cross-correlation of lens positions and source shears forms the galaxy-galaxy lensing (GGL) component of the $\rm 3\times2pt$ analysis. This has several components:
\begin{equation}
    C^{\alpha\beta}_{n\gamma}(\ell) = C^{\alpha\beta}_{\rm{gG}}(\ell) + C^{\alpha\beta}_{\rm{gI}}(\ell) + C^{\alpha\beta}_{\rm{mG}}(\ell) + C^{\alpha\beta}_{\rm{mI}}(\ell) \,,
   \label{eq:Cggl}
\end{equation}
where the `gG' term is the cross-correlation of galaxy positions and the shear field and is given by
\begin{equation}
C_{\rm{gG}}^{\alpha\beta}(\ell)=\int_0^{\chi_{\rm H}}\mathrm{d}\chi \frac{b_{\rm g}^{\alpha}n_\alpha(\chi)q_\beta(\chi)}{f_K^2(\chi)}P_\delta\left(\frac{\ell+1/2}{f_K(\chi)},\chi\right)\,.
\end{equation}
The `gI' term in Eq. (\ref{eq:Cggl}) arises through the cross-correlation of lens positions with source intrinsic alignments, and is expected to be non-zero only when the distributions $n_\alpha,n_\beta$ are overlapping. This is given by
\begin{equation}
C_{\rm{gI}}^{\alpha\beta}(\ell)=\int_0^{\chi_{\rm H}}\mathrm{d}\chi \frac{b_{\rm g}^{\alpha}n_\alpha(\chi)n_\beta(\chi)}{f_K^2(\chi)}F_{\mathrm{IA}}(\chi)P_\delta\left(\frac{\ell+1/2}{f_K(\chi)},\chi\right)\,.
\end{equation}
The lensing magnification-induced number counts contribution in the foreground is also correlated with the background shears, creating the `mG' term which is given by
\begin{equation}
C_{\rm{mG}}^{\alpha\beta}(\ell)=\int_0^{\chi_{\rm H}}\mathrm{d}\chi \frac{\bar{q}_\alpha(\chi)q_\beta(\chi)}{f_K^2(\chi)}P_\delta\left(\frac{\ell+1/2}{f_K(\chi)},\chi\right)\,.
\end{equation}
Finally, the magnification-induced number counts contribution yields an additional, weak cross-correlation with the intrinsic alignments' contribution to the shapes, `mI', given by
\begin{equation}
C_{\rm{mI}}^{\alpha\beta}(\ell)=\int_0^{\chi_{\rm H}}\mathrm{d}\chi \frac{\bar{q}_\alpha(\chi)n_\beta(\chi)}{f_K^2(\chi)}F_{\mathrm{IA}}(\chi)P_\delta\left(\frac{\ell+1/2}{f_K(\chi)},\chi\right)\,.
\end{equation}

\subsection{Spectrum modelling choices}
\label{sec:modelling_choices}

For our fiducial true Universe, we adopt the best-fit flat-$\Lambda$CDM cosmology from \citet{Asgari21} (their Table A.1, column 3; see also our Sect. \ref{sec:forecast} and Table \ref{tab:priors}), constrained by cosmic shear band-power observations from the public $1000\sqdeg$ 4th Data Release of the Kilo Degree Survey \citep[`KiDS-1000'][]{Kuijken2019}. Following \cite{Asgari21}, we model intrinsic alignments via the `non-linear linear alignment' \citep[NLA;][]{Catelan2001,Hirata2004a,Bridle2007,Joachimi2011} model. This specifies the alignment kernel as
\begin{equation}
    F_{\mathrm{IA}}(\chi) = -A_{\mathrm{1}}C_{1}\rho_{\mathrm{crit}}\frac{\Omega_{\mathrm{m}}}{D_{+}(\chi)}\,,
    \label{eq:intrinsic_alignment_spectrum_prefactor}
\end{equation}
where $C_{1}$ is a fixed normalisation constant \citep{Bridle2007} and $D_{+}(\chi)$ is the linear growth function, normalised to $1$ today \citep{Joachimi2011}. $A_{1}$ was constrained to $0.973^{+0.292}_{-0.383}$ by \cite{Asgari21}, though they \citep[and][who studied pseudo-$C_\ell$'s]{Loureiro2021} saw that the best-fit alignment amplitude varied for different cosmic shear statistics.

In dealing with the biased photometric sample redshift distributions described in Sect. \ref{sec:data}, we allow for recalibration of the distributions\footnote{Notice that we work here in redshift space, while $n(\chi)=n(z)\,{\rm d}z/{\rm d}\chi$, which requires a model for the expansion of the Universe.} $n_\alpha(z)$ according to widely-used (e.g.~\citealt{cfhtlens, Hikage19,Asgari21,Amon2022,Secco2022}) bin-wise displacements of the mean redshifts, $\deltazi$. These displacements are applied for tomographic bin $i$ as
\begin{equation}
    n_i(z) \rightarrow n_i(z - \deltazi)\,,
    \label{eq:shift_model}
\end{equation}
and are constrained by the two-point correlation data, and by Gaussian priors, the derivations of which are described in Sect. \ref{sec:data}. We shall refer to this approach as the `shift model', to be contrasted with a lack of nuisance modelling (the `do nothing' model), and with the Gaussian mixture `comb' models described below.

For the linear, deterministic galaxy biases $b^\alpha_{\rm g}$, per-tomographic sample $\alpha$, we assume a single functional form for the true bias, setting fiducial values for a magnitude-limited sample according to \citet{TheLSSTDarkEnergyScienceCollaboration2018}:
\begin{equation}
    b^\alpha_{\rm g} = 0.95 / D_+(\meanz_\alpha)\,,
    \label{eq:galaxy_bias_function}
\end{equation}
which is evaluated at the mean redshift $\meanz$ for each sample redshift distribution $n_\alpha(z)$. This bias model is used up to an $\ell_{\rm max}$ compatible with $k_{\rm max}=0.3\,h\,{\rm Mpc}^{-1}$ (see Section \ref{sec:forecast}). Although we do not expect galaxy bias to remain linear up to this scale \citep{Joachimi21}, we generate the data vector and analyse it with the same bias model, which still allows us to draw comparisons across our different probe combinations and redshift error scenarios, and reduces computational expense.

The matter power spectrum is emulated with {\tt CosmoPower} \citep{Mancini2021}, relying on the Boltzmann code {\tt CAMB} \citep{Lewis_2000,Howlett2012} and a correction for the effects of baryons following {\tt HMCode} \citep{Mead2015,Mead2015ascl}. The amplitude of the matter power spectrum is effectively parameterised by $A_{\rm s}$, the primordial power spectrum amplitude, although we sample over $\ln (10^{10}A_{\rm s})$ as it is more convenient for our implementation of {\tt CosmoPower} (and is a more commonly-used sampling variable;~e.g.~\citealt{Planck:2018vyg} -- for an assessment of the impact of this choice, see \citealt{Joudaki:2019pmv}). In contrast to \citet{Asgari21}, we assume a cosmology with massless neutrinos. This is mainly chosen to reduce the computational cost, but we note that it would be valuable to explore the constraining power of $\rm (4\mbox{--}6)\times2pt$ statistics on neutrino mass \citep[e.g.][]{Mishra18}.

To account for the suppression of the matter power spectrum at small scales due to baryons \citep{VanDaalen2011,Chisari19}, we use the {\tt HMCode2016}\footnote{We do not make use of the latest version, {\tt HMCode2020} \citep{Mead2021}, as it was comparatively less well-tested within {\tt CosmoPower} at the start of our analysis.} halo mass-concentration relation amplitude parameter $A_{\rm bary}$, and follow \cite{Joudaki2017} in setting the halo bloating parameter $\eta_{\rm bary}=0.98-0.12A_{\rm bary}$. For the true Universe, we assume $A_{\rm bary}=2.8$, differently to the best-fit $A_{\rm bary}=3.13$ from \cite{Asgari21} as the latter corresponds to a dark matter-only Universe, and we prefer to include some baryonic contributions in the fiducial data-vector. We note that $A_{\rm bary}$ will in practice only be weakly constrained by Stage-III-like forecasts, for which it is principally a parameter that will allow us to capture uncertainties in the modelling of the non-linear matter power spectrum. For Stage-IV-like configurations, it is expected that baryonic feedback models should start to see meaningful constraints from weak lensing and combined probe analyses such as these -- though a detailed investigation of hydrodynamic halo model constraints is beyond the scope of this work.

All of the observable two-point functions are calculated using the {\tt CCL} library\footnote{CCL version: 2.3.1.dev7+ge9317b4f} \citep{Chisari2018} in `calculator' mode. As previously mentioned, we do not include RSDs in our modelling due to constraints related to non-Limber computations. However, photometric surveys are known to be sensitive to RSDs \citep{Ross11,Tanidis_2019}, which should therefore be included in follow-up work.

\subsection{Angular power spectra for the Gaussian comb}
\label{sec:comb_spectra}

The Gaussian `comb' model decomposes the redshift distribution of a given tomographic bin of the photometric sample into a sum over $N_G$ Gaussian basis functions of fixed-width and uniformly-spaced centres. Mathematically,
\begin{equation}
    n_\alpha(z) = \sum_{i=1}^{N_G} A^\alpha_i n_i(z)\,,
    \label{eq:comb_model}
\end{equation}
for tomographic sample $\alpha$, where amplitudes $A_i$ must sum to unity for each sample $\alpha$, and each Gaussian basis function is given by \citep{Stolzner2020}
\begin{equation}
    n_i(z) = \frac{z}{\nu(z_i,\sigma_{\rm comb})} \exp\left\{\frac{(z-z_i)^2}{2\sigma_{\rm comb}^2}\right\}\,,
    \label{eq:comb_component}
\end{equation}
where $z_i$ and $\sigma_{\rm comb}$ are the centre and width of basis function $n_i(z)$, and the normalisation over the interval $z\in[0,\infty]$ is given by
\begin{equation}
    \nu(z_i,\sigma) = \sqrt{\frac{\pi}{2}}\,z_i\,\sigma\,{\rm erf}\left(-\frac{z_i}{\sqrt{2}\sigma}\right)+\sigma^2\exp\left\{-\frac{z_i^2}{2\sigma^2}\right\}\,,
    \label{eq:comb_normalisation}
\end{equation}
where $\sigma\equiv\sigma_{\rm comb}$. The concatenated vector of $\alpha$ tomographic redshift distributions is then $n_{\rm comb}(z)$, which we can fit to an arbitrary redshift distribution $N(z)$, with associated covariance $\nzcov$ (see Sect. \ref{sec:stage_3_photometric}), by varying amplitudes $a^\mu_m=\ln A^\mu_m$ to minimise the $\chi_{n(z)}^2$ given by
\begin{equation}
    \chi_{n(z)}^2 = \sum_{k,l} {\Big (}n_{{\rm comb},k} - N_k{\Big )} \, \Sigma_{n(z),kl}^{-1} \, {\Big (}n_{{\rm comb},l} - N_l{\Big )}\,,
    \label{eq:initial_comb_chi2}
\end{equation}
where $k,l$ index the elements of the concatenated redshift distribution vectors and the covariance $\nzcov$. We refer to this model distribution as the `initial comb' model $n_{\rm comb,ini}(z)$, which can be used directly in theoretical computations of angular power spectra (e.g. Eq. \ref{eq:Cgg}).

\cite{Taylor2010} and \cite{Stolzner2020} detail the construction of analytical expressions for the two-point function likelihood, with marginalisation over some nuisance parameters (the comb amplitudes) given a prior. They also derive expressions for the displacement in the sub-space of nuisance parameters from the peak of the likelihood, dependent upon derivatives of the log-likelihood with respect to those parameters. Minimising that displacement by iteratively varying the amplitudes (and cosmological/other parameters, when the fiducial set is unknown); recomputing angular power spectra; and evaluating the likelihood derivatives, one obtains the `optimised comb' model $n_{\rm comb,opt}(z)$.

During the initial fitting (Eq. \ref{eq:initial_comb_chi2}), comb amplitudes $A_m^\mu$ are often seen to be consistent with zero, leading to concerns about the suitability of Gaussian priors for the parameters $a_m^\mu$. \cite{Taylor2010} also give expressions for flat priors, but these too are difficult to motivate; the choice of a lower boundary in the range $[-\infty,0]$ is somewhat arbitrary for a 1-d likelihood that asymptotes to a constant as $a_m^\mu\rightarrow{}-\infty$, and yet it is highly consequential for the marginalisation. We therefore defer a full $\six$ application of the comb model \citep[with optimisation and marginalisation as demonstrated by][]{Stolzner2020} to future work, for which an extension to describe spectroscopic-photometric cross-correlations is currently under development.

Meanwhile, we obtain the optimised comb model $n_{\rm comb,opt}(z)$ in this work by varying comb amplitudes $a^\mu_m$ to minimise the fiducial two-point $\chi_{\vec{d}}^2$, given by
\begin{equation}
    \chi_{{d}}^2 = \sum_{i,j} {\Big (}{d}_i-{\mu}_i{\Big )}\,\tens{Z}^{-1}_{ij}\,{\Big (}{d}_j-{\mu}_j{\Big )}\,,
    \label{eq:data_chi2}
\end{equation}
for data- and theory-vectors, $\vec{d}$ and $\vec{\mu}$, respectively, and data-vector covariance $\tens{Z}$ (Sect. \ref{sec:covariance}), each indexed by $i,j$. The comb optimisation procedure is thus:

\begin{enumerate}
    \item Begin with some data-vector $\vec{d}$, sourced from an unknown redshift distribution $n(z)$;
    \item Fit an initial comb model $n_{\rm comb,ini}(z)$ directly to a possibly-biased \emph{estimate} of the distribution, $N(z)$, with its associated covariance $\nzcov$;
    \item Estimate the theory-vector $\vec{\mu}$ using $n_{\rm comb,ini}(z)$;
    \item Minimise Eq. \ref{eq:data_chi2} by adjusting the comb amplitudes $a^\mu_m$, resulting in an optimised comb model $n_{\rm comb,opt}(z)$ which has been flexibly recalibrated against information from the data-vector.
\end{enumerate}

The amplitudes $a^\mu_m$ of the optimised comb model $n_{\rm comb,opt}(z)$ are then fixed during the sampling of cosmological and nuisance model parameters. This iterative procedure allows us to reduce the computational expense of varying comb amplitudes at the same time one samples the likelihood.

We shall henceforth refer to this method as the `comb model', denoted as $\fcomb$. In practice, however, we still seek to marginalise over some uncertainty in the redshift distribution, and do so through combination with the commonly-used shift model ($\deltazi$; described in Sect. \ref{sec:generic_spectra}), whereby scalar shifts are applied via Eq. \ref{eq:shift_model} to the optimised comb model $n_{\rm comb,opt}(z)$ during likelihood sampling, and later marginalised over. We refer to this hybrid as the `comb+shift' model, denoted as $\hcomb$. As we shall see in Sect. \ref{sec:forecast}, the comb models reveal the insufficiency of the shift model for application to photometric density statistics, whilst also offering generally superior recoveries of the true lensing efficiency kernel $q(\chi)$.

\subsection{Approximations}
\label{sec:approximations}

\begin{figure*}
    \centering
    \includegraphics[width=\textwidth]{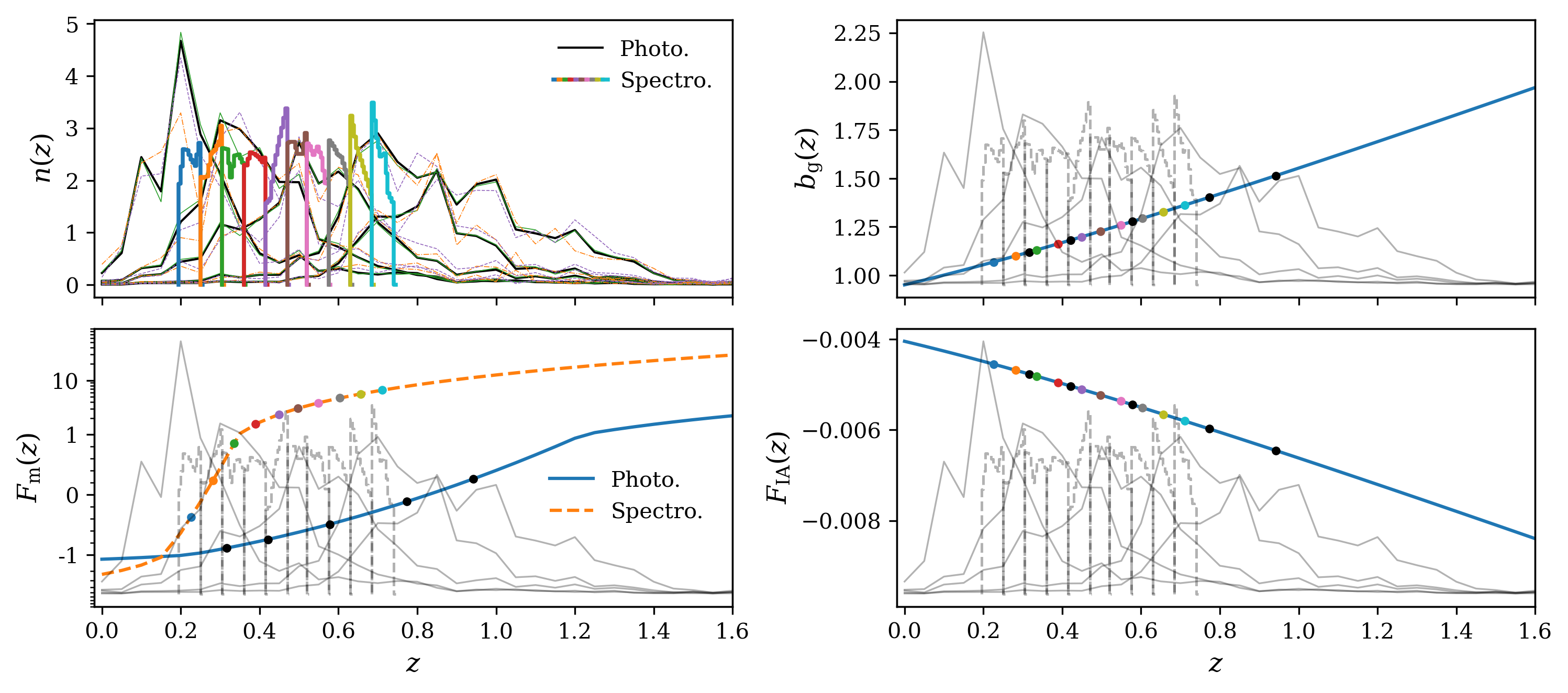}
    \caption{\emph{Top left:} The photometric (black curves and thin faint curves, corresponding to different realisations of the redshift distribution; described in Sect. \ref{sec:stage_3_photometric}) and spectroscopic (coloured curves) tomographic redshift distributions utilised in this work (described in Sect. \ref{sec:data}) for a Stage-III survey. \emph{Top right:} The galaxy bias function (blue curve; Eq. \ref{eq:galaxy_bias_function}) used to define the true galaxy bias, $b_{\rm g}$, for each tomographic redshift sample. \emph{Bottom left:} The magnification bias functions $F_{\rm m}(z)$ (Eq. \ref{eq:magnification_kernel_modifier}) used to calculate lensing contributions to galaxy number count kernels (Sect. \ref{sec:generic_spectra}), shown on a log-linear axis with transitions at $\pm{}1$. The faint-end slope of the luminosity function is estimated via the fitting formulae of \citep{Joachimi2010}, taking the $r$-band limiting magnitude $r_{\rm lim}=24.5$ for the photometric (blue curve) and $r_{\rm lim}=20.0$ for the spectroscopic (orange dashed curve) synthetic galaxy samples. \emph{Bottom right:} The intrinsic alignment power spectrum prefactor function $F_{\rm IA}(z)$ (blue curve; Eq. \ref{eq:intrinsic_alignment_spectrum_prefactor}) used to approximate matter-intrinsic and intrinsic-intrinsic power spectra for the computation of intrinsic alignment contributions to the shear and galaxy-galaxy lensing correlations. The galaxy bias $b_{\rm g}(z)$ and magnification bias $F_{\rm m}(z)$ functions are evaluated for each tomographic sample $i$ at the mean redshift $\meanz_i$ of the sample, denoted by black (photometric) or coloured (spectroscopic) points atop each curve. The intrinsic alignment prefactor $F_{\rm IA}(z)$ also displays these points but is evaluated at the mean of the relevant kernel product $n_i(\chi)n_j(\chi)$, or $n_i(\chi)q_j(\chi)$, for a correlation between samples $i$ and $j$.
    }
    \label{fig:linear_modelling}
\end{figure*}

Over the course of this work, we found that extended-analysis inference simulations were particularly slow to converge, with $\six$ chains potentially taking multiple weeks to reach convergence, even when using fast nested sampling algorithms such as {\tt MultiNest} \citep{Feroz2008,Feroz2009,Feroz2019}. This is primarily due to rapidly decreasing acceptance fractions that become extremely small, $\mathcal{O}\left(10^{-2}\right)$, on the approach to convergence (to be discussed in Sect. \ref{sec:cross_clustering_slow_chains}), and the large number of Limber integrations required to fully characterise the enlarged theory-vectors.

This computational demand led us to the approximations already discussed: one-parameter linear galaxy bias and alignments; emulated matter power spectra with a one-parameter baryonic feedback model; Limber-approximate angular power spectra without RSDs; and massless neutrinos. Many of these choices are insufficient to describe the real Universe with the accuracy required by future surveys, but together they greatly increase the speed of the likelihood evaluation. Given that we apply scale cuts to density probes, which lessen the impact of these choices (Sect.~\ref{sec:forecast}; Eq. \ref{eq:scale_cuts}), and given our intention to investigate the potential gains from $\six$ \emph{relative} to $\rm (1\mbox{--}3)\times2pt$ analyses, we make some further approximations in our application of kernel modifiers for intrinsic alignment and magnification contributions to enhance the computational speed.

To reduce the required number of integrations over $\chi$, we make the following transformations:
\begin{equation}
 \begin{split}
    F_{\mathrm{IA}}(\chi) &\rightarrow \bar{F}_{\mathrm{IA}} = F_{\mathrm{IA}}(\chi(\meanz)) \\
    F_{x,\mathrm{m}}(\chi) &\rightarrow \bar{F}_{x,\mathrm{m}} = 2(\alpha_{x}(\meanz) - 1)\,,
 \end{split}
 \label{eq:F_kernel_approximation}
\end{equation}
where $\bar{F}_{\mathrm{IA}}$ is evaluated at the mean of the kernel product $n(\chi)q(\chi)$, or $n(\chi)n(\chi)$, and $\bar{F}_{x,\mathrm{m}}$ takes the luminosity function slope $\alpha_x(\meanz)$ at the mean redshift of sample $x$ \citep[see Appendix~A in][where the slope is evaluated at the sample median redshift]{Joachimi2010}. Provided that the Limber integration kernels (Sect.~\ref{sec:generic_spectra}) have relatively compact support -- as is the case for narrow spectroscopic bins (Sect. \ref{sec:data}), but less so for broad photometric bins -- these approximations will not yield unrealistic spectra, at least in the context of linear models for IA and magnification.

An illustration of these, and the linear galaxy bias approximations, is given in Fig. \ref{fig:linear_modelling}, where our mock Stage-III photometric and spectroscopic redshift distributions (to be described in Sect. \ref{sec:data}) are reproduced in each panel and overlaid with the functional forms for $b_{\rm g}(z),F_{x, \rm m}(z),$ and $F_{\rm IA}(z)$. Circular points atop each curve mark the mean redshifts of photometric (black) or spectroscopic (colours) redshift samples, at which the galaxy bias and magnification kernels are evaluated ($F_{\rm IA}(z)$ is evaluated at the means of kernel products). 

We find that the $\bar{F}_{\rm IA}$ approximation results in cosmic shear spectrum (Eq. \ref{eq:Cgamgam}) deficits of $\lesssim0.03\sigma$, for the uncertainty $\sigma$ on each respective shear signal (see Sect. \ref{sec:covariance} for details on signal covariance estimation). Interestingly, the $\bar{F}_{x,\mathrm{m}}$ approximation is comparatively more consequential; through scale-dependent modifications to mg/gm (mm is largely unaffected) contributions (Eq. \ref{eq:Cmg}), the total clustering signals (Eq. \ref{eq:Cgg}) are suppressed by $0.05\sigma-0.8\sigma$, dependent upon the tomographic bin pairing, and most severely at high-$\ell$. Our application of scale-cuts thus reduces the proportion of strongly-suppressed points entering the data-vector -- only a slim minority of our analysed clustering $C(\ell)$'s are suppressed by more than $0.3\sigma$, and the majority of clustering signal-to-noise comes from relatively unbiased points. We note that number counts from lensing magnification are commonly modelled according to such averages over $\alpha_x(\chi)$ \citep{Garcia-Fernandez2018,VonWietersheim-Kramsta2021,Mahony2021,Liu2021}. It is likely that future analyses will need to explicitly integrate over the magnification kernel in order to accurately model these contributions \citep[see also the recent work of][for a more rigorous treatment of magnification bias in samples with complex selection functions]{ElvinPoole2022}. We emphasise that the inaccuracies induced by our approximations apply to all forecasts under consideration, such that we make like-with-like comparisons when analysing the results for different two-point probe combinations.

With linear factors extracted from the Limber integrals, we are able to reduce the number of required Limber integrations by about an order of magnitude by reusing the `raw' angular power spectra for each unique kernel product $n_\alpha(\chi)n_\beta(\chi),n_\alpha(\chi)q_\beta(\chi)$, or $q_\alpha(\chi)q_\beta(\chi)$. Each is then re-scaled by relevant factors of $b_{\rm g}$, $\bar{F}_{\mathrm{IA}}$, and/or $\bar{F}_{x,\mathrm{m}}$ correspondingly to produce appropriate spectral contributions for each probe. Thus, we give up some small amount of realism from explicit integrations over various kernels describing the different density, magnification, shear, and alignment contributions, in exchange for large gains in computational speed through the reuse of factorisable Limber integrations. Sampling in parallel with 40-48 cores, the resulting Stage-III chains take hours (sometimes less than one) to converge for $\rm (1\mbox{--}3)\times2pt$, and up to a few days for $\rm 6\times2pt$, which is tractable for our purposes here.

We recommend that future work make use of emulators for Boltzmann computations, and explore the possibility of extending the emulation to the level of $C_\ell$'s or other observables. We particularly recommend this in the context of extended models with additional parameters, e.g. for IA \citep{Blazek19,Vlah20} and galaxy bias \citep{Modi20,Barreira2021,Mahony22}, and of such theoretical developments as non-Limber integration for the utilisation of large-angle correlations \protect\citep[e.g.][]{Campagne17,Fang2020}, each of which is likely to prove especially costly for joint analyses of multiple probes \citep{N5K}.

\subsection{Analytic covariance estimation}
\label{sec:covariance}
We assume a Gaussian covariance throughout this analysis, since Gaussian contributions should dominate the error budget for the scales that we consider (Sect. \ref{sec:forecast}; Eq. \ref{eq:scale_cuts}). We homogenise the analysis choices and approximations across our forecast (Sect. \ref{sec:approximations}) and make like-with-like comparisons when quoting results. Whilst future work should consider connected non-Gaussian and super-sample covariance contributions to the signal covariance, esp. in the $3\mbox{--}\six$ case \citep{Barreira18,Barreira18b}, we assume that these would not significantly affect our conclusions which come from comparing across different probe combinations while always adopting a Gaussian covariance.

The Gaussian covariance matrix $\tens{Z}\equiv{\rm Cov}[\vec{d},\vec{d}']$ is entirely specified by the power spectra. Given Wick's theorem, one finds
\begin{equation}
\label{eq:Gaussian_cov}
    \mathrm{Cov}[C^{ij}(\ell), C^{mn}(\ell^\prime)] = \frac{\delta^\mathrm{K}_{\ell\ell^\prime}}{f_\mathrm{sky}N_\ell}\left(\hat C^{im}(\ell)\hat C^{jn} (\ell^\prime) + \hat C^{in}(\ell)\hat C^{jm} (\ell^\prime)\right)\;,
\end{equation}
where $f_\mathrm{sky}$ is the sky fraction of the survey, $4\pi f_\mathrm{sky} = A_\mathrm{survey}$, $i,j,m,n$ label any tracer in the analysis, and $\hat{C}^{ij}(\ell)$ is the observed angular power spectrum, i.e. including noise. In other words,
\begin{equation}
    \hat C^{ij}(\ell) = 
    \begin{cases}
    C^{ij}(\ell) + \cfrac{\sigma_\epsilon^2}{\bar{n}_{ij}}\delta^\mathrm{K}_{ij}\,, & \quad \text{if $i \land j \in$ source,} \\
    C^{ij}(\ell) + \cfrac{1}{\bar{n}_{ij}}\delta^\mathrm{K}_{ij}\,, &\quad \text{if $i  \land  j \in$ lens,}\\
    C^{ij}(\ell)\,, &\quad\text{else},
    \end{cases}
\end{equation}
where $\delta^{\rm K}_{ij}$ is the Dirac delta function, $\sigma^2_\epsilon$ is the single-component ellipticity dispersion and $\bar{n}_{ij}$ is the average number density of sources (shear sample objects) or lenses (position sample objects) for each tracer (see \citealt{Joachimi21} for more details).
The factor $N_\ell$ in Eq. (\ref{eq:Gaussian_cov}) counts the number of independent modes at multipole $\ell$,
\begin{equation}
    N_\ell = (2\ell + 1)\Delta \ell \;,
\end{equation}
where $\Delta \ell$ is the bandwidth of each multipole bin used in the analysis. It should be noted that the covariance is calculated from the same $n(z)$ as the mock-data (see Sect.~\ref{sec:data} for the sampling of the mock-data), therefore it is derived from the true $n(z)$ and does not change during the sampling. Furthermore, if two statistics are measured over a different sky area we take the maximum between the two areas in the sky fraction and similarly for overlapping surveys \citep[see e.g.][]{vanUitert18}.

\section{Synthetic data products}
\label{sec:data}

\begin{figure*}
    \centering
    \includegraphics[width=0.95\textwidth]{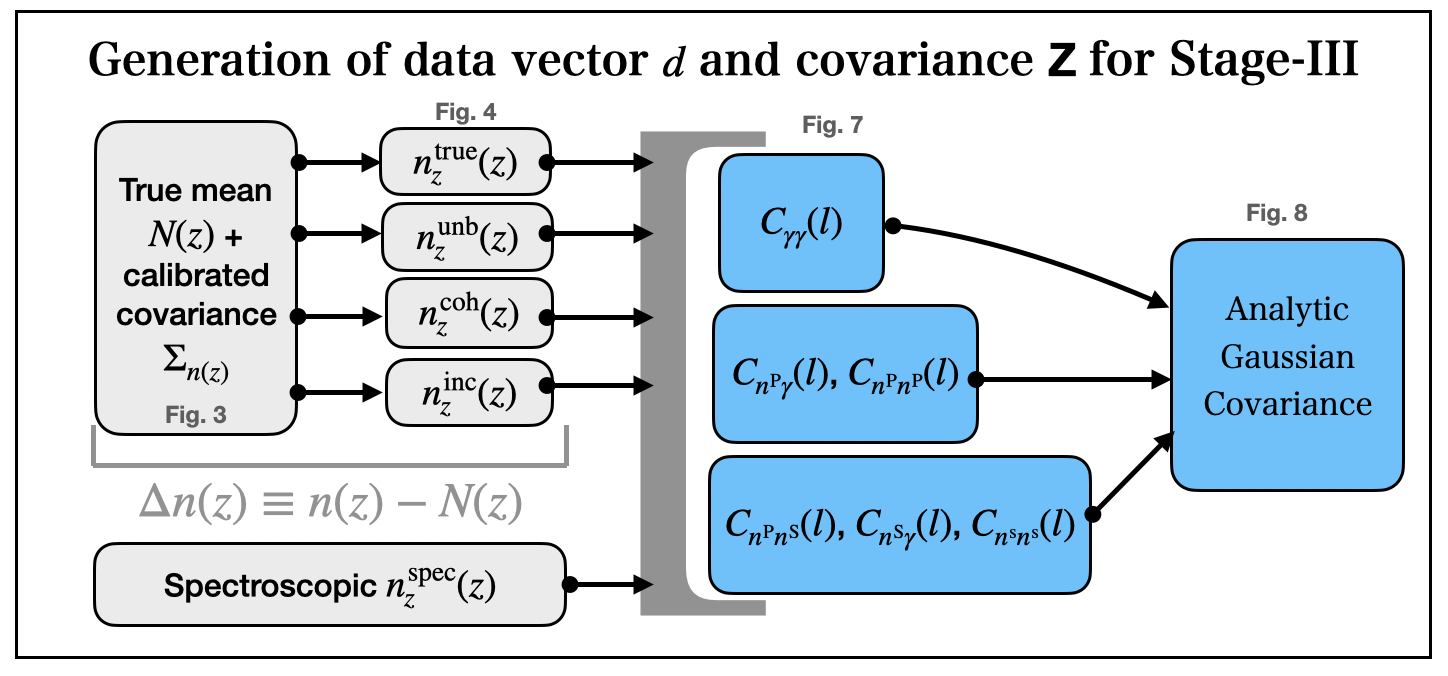}
    \caption{A sketch of how the data-vector and covariance are generated for our Stage-III forecasts.}
    \label{fig:s3_dc}
\end{figure*}

\begin{table*}
    \tiny
    \centering
    \begin{tabular}{lccccccc}
        \hline
        \hline
        Sample & $\meanz$ & $\meanz_{\rm{incoh./shifted}}$ & $\meanz_{\rm{coherent}}$ & $n_{\rm{eff}}$ [$\psqarcmin$] & $\sigma_\epsilon$ & $r_{\rm{lim}}$ & area [$\sqdeg$] \\
\hline
Stage-III Photometric & 0.606 & 0.631 & 0.648 & 6.22 & 0.26 & 24.5 & 777 \\
\hline
bin 1 & 0.316 & 0.385 & 0.375 & 0.62 & 0.27 & - & - \\
bin 2 & 0.421 & 0.449 & 0.460 & 1.18 & 0.26 & - & - \\
bin 3 & 0.578 & 0.600 & 0.633 & 1.85 & 0.27 & - & - \\
bin 4 & 0.774 & 0.779 & 0.808 & 1.26 & 0.25 & - & - \\
bin 5 & 0.942 & 0.939 & 0.965 & 1.31 & 0.27 & - & - \\
\hline
Stage-III Spectroscopic & 0.470 & - & - & 0.0311 & - & 20.0 & 9329 \\
\hline
bin 1 & 0.226 & - & - & 0.0016 & - & - & - \\
bin 2 & 0.282 & - & - & 0.0020 & - & - & - \\
bin 3 & 0.335 & - & - & 0.0026 & - & - & - \\
bin 4 & 0.390 & - & - & 0.0026 & - & - & - \\
bin 5 & 0.449 & - & - & 0.0042 & - & - & - \\
bin 6 & 0.497 & - & - & 0.0053 & - & - & - \\
bin 7 & 0.549 & - & - & 0.0056 & - & - & - \\
bin 8 & 0.603 & - & - & 0.0039 & - & - & - \\
bin 9 & 0.657 & - & - & 0.0022 & - & - & - \\
bin 10 & 0.711 & - & - & 0.0010 & - & - & - \\
\hline
Stage-IV Photometric & 0.759 & 0.756 & - & 10.00 & 0.26 & 25.8 & 12300 \\
\hline
bin 1 & 0.233 & 0.249 & - & 1.93 & 0.26 & - & - \\
bin 2 & 0.445 & 0.433 & - & 2.05 & 0.26 & - & - \\
bin 3 & 0.650 & 0.635 & - & 1.97 & 0.26 & - & - \\
bin 4 & 0.918 & 0.876 & - & 2.03 & 0.26 & - & - \\
bin 5 & 1.547 & 1.590 & - & 2.02 & 0.26 & - & - \\
\hline
Stage-IV Spectroscopic & 0.881 & - & - & 0.6104 & - & 23.4 & 4000 \\
\hline
bin 1 & 0.186 & - & - & 0.1621 & - & - & - \\
bin 2 & 0.412 & - & - & 0.0162 & - & - & - \\
bin 3 & 0.609 & - & - & 0.0123 & - & - & - \\
bin 4 & 0.787 & - & - & 0.1092 & - & - & - \\
bin 5 & 0.986 & - & - & 0.1076 & - & - & - \\
bin 6 & 1.204 & - & - & 0.0739 & - & - & - \\
bin 7 & 1.410 & - & - & 0.0288 & - & - & - \\
bin 8 & 1.595 & - & - & 0.0105 & - & - & - \\
bin 9 & 0.631 & - & - & 0.0415 & - & - & - \\
bin 10 & 0.824 & - & - & 0.0449 & - & - & - \\
bin 11 & 1.046 & - & - & 0.0034 & - & - & - \\ 
\hline
\hline
    \end{tabular}
    \caption{Details of the synthetic Stage-III and Stage-IV galaxy samples defined for use in our large-scale structure analysis forecasts (described in Sect. \ref{sec:data}). Columns give the sample; its mean redshift $\meanz$ according to the starting distribution $N(z)$; according to the incoherently-biased $\nzinc$ or shifted $\nzshift$ distribution; and according to the coherently-biased distribution $\nzcoh$ (see Sect. \ref{sec:stage_3_photometric}); its effective galaxy number density $n_{\rm eff}$ per square arcminute; its intrinsic shear dispersion $\sigma_\epsilon$; its $r$-band limiting magnitude $r_{\rm lim}$; and its area in square degrees. Biased redshift distributions and shear dispersion statistics apply only to photometric samples, and the magnitude limits and areas apply for all tomographic subsets of each Stage-III or Stage-IV sample. Sects. \ref{sec:stage_3_photometric} and \ref{sec:stage_4_photometric} detail our definitions of shifted redshift distributions, which we use to generate data-vectors in this work.
    }
    \label{tab:sample_details}
\end{table*}

We define here several synthetic galaxy samples with which to conduct our angular power spectrum analysis forecasts, summarising their characteristics in Table \ref{tab:sample_details}.

\subsection{Mock Stage-III samples}

\begin{figure}
    \centering
    \includegraphics[width=\columnwidth]{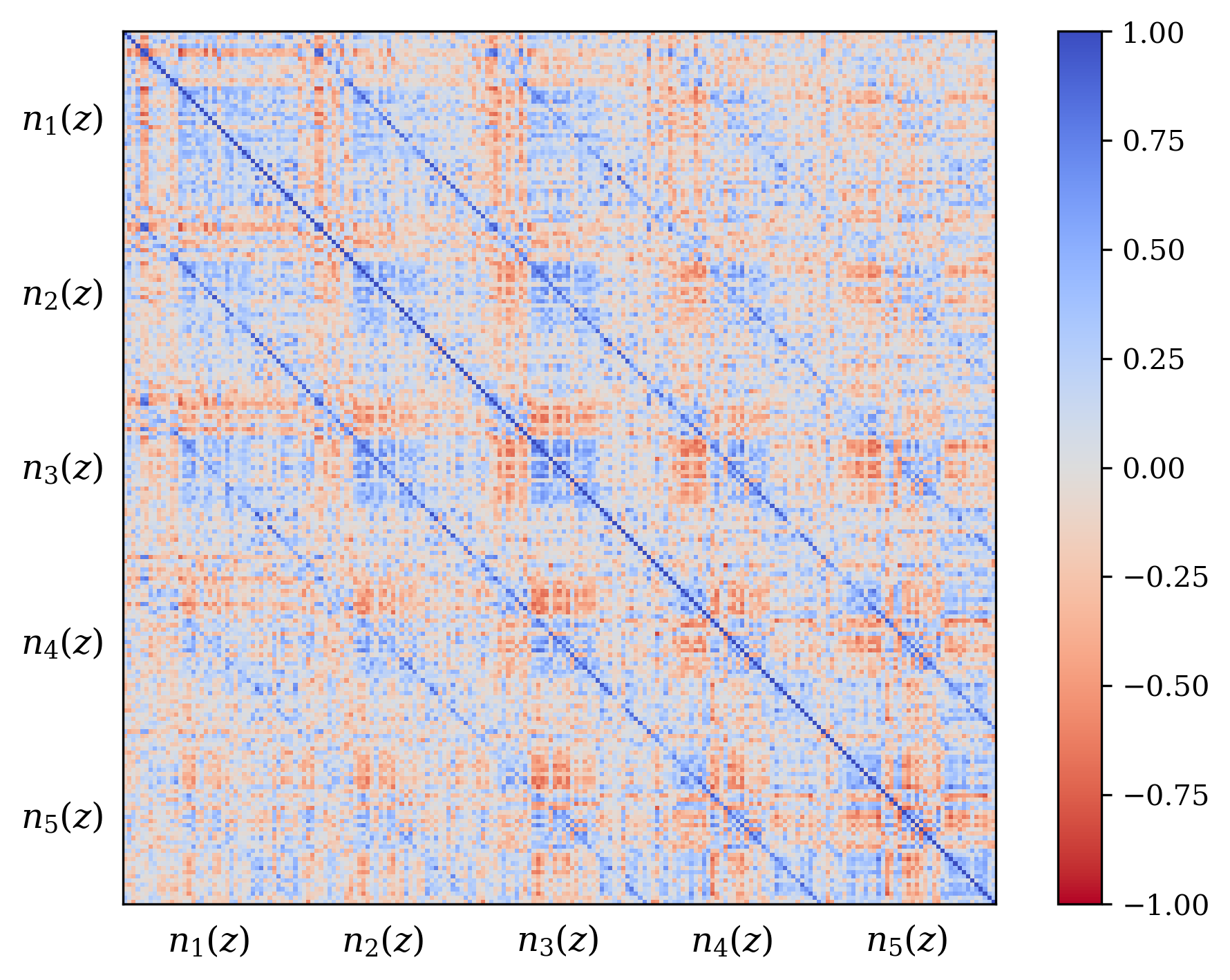}
    \caption{The matrix of correlation coefficients corresponding to the covariance, $\nzcov$, of the five-bin tomographic redshift distribution (see also Figs. \ref{fig:linear_modelling} \& \ref{fig:redshift_distributions}) estimated for the KiDS+VIKING-450 data release \citep{Hildebrandt20, Wright2018} via direct redshift calibration. The covariance is estimated through spatial bootstrapping of spectroscopic calibration samples (see \citealt{Hildebrandt20} for more details), and axis labels illustrate the subsections of the matrix corresponding to each of the five tomographic bins $n_i(z)$. The covariance is used to describe a multivariate Gaussian sampling distribution for realisations of photometric redshift distributions (Sect. \ref{sec:stage_3_photometric}).
    }
    \label{fig:nz_covariance}
\end{figure}

\begin{figure*}
    \centering
    \includegraphics[width=0.9\textwidth]{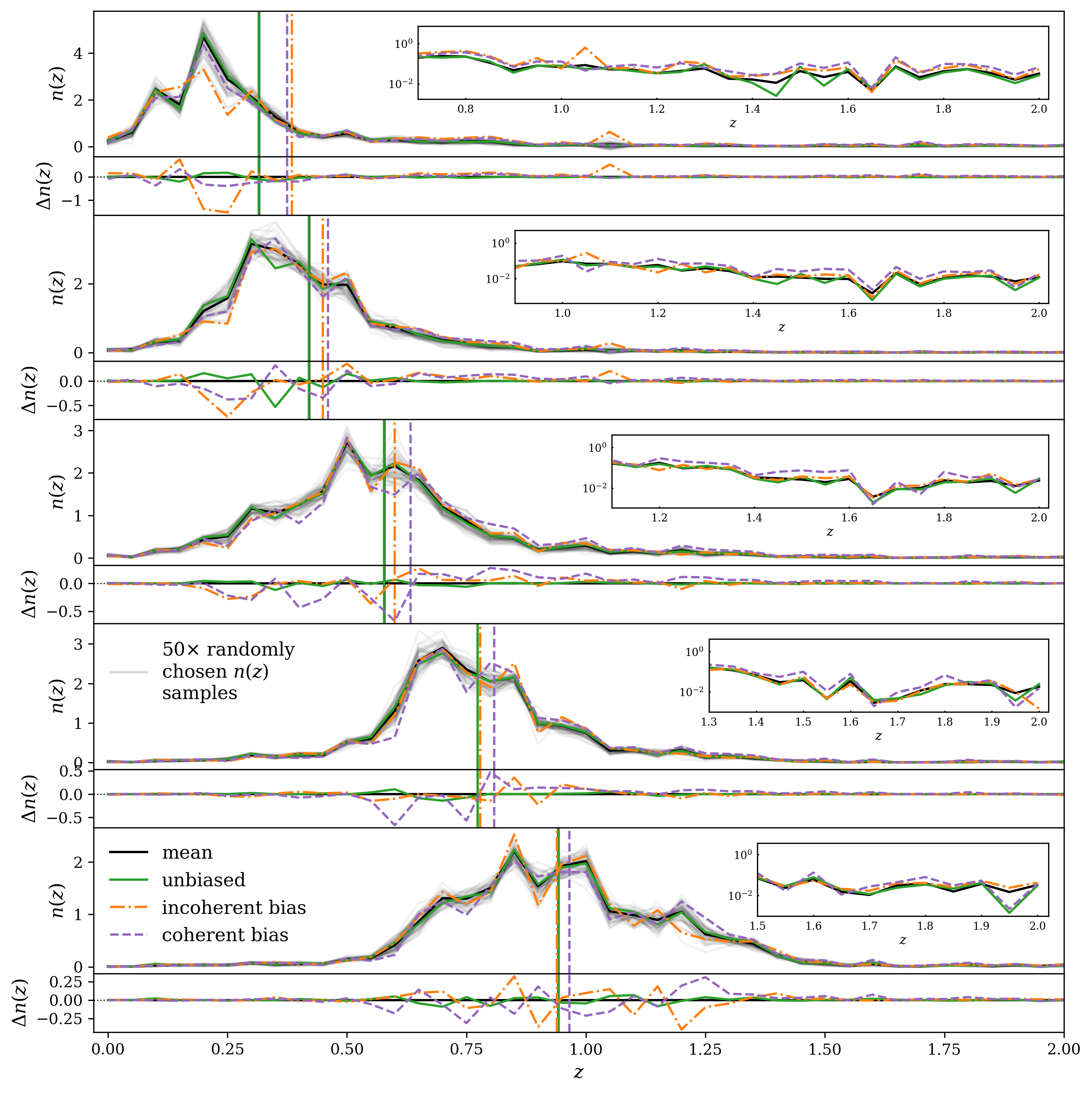}
    \caption{The five-bin Stage-III photometric redshift distributions employed in our forecasts, with alternating panels showing the distribution $n_i(z)$ of bin $i$ and its difference $\Delta{}n_i(z)=n_i(z)-N_i(z)$ with respect to the `mean' distribution $N_i(z)$. Black curves in each panel correspond to the mean distribution $N(z)$, taken as the final, public KiDS+VIKING-450 estimate \citep{Hildebrandt20, Wright2018}. As described in Sect. \ref{sec:stage_3_photometric}, we assume a multivariate Gaussian distribution described by the mean $N(z)$ and covariance $\nzcov$ (Fig. \ref{fig:nz_covariance}) to generate $45\,000$ sample redshift distributions, a random $50$ of which are shown in $n(z)$ panels as faint grey curves. After discarding the 80\% of samples with negative $n(z)$ values, the tomographic mean distance metrics $\Delta_{\rm incoherent}$ and $\Delta_{\rm coherent}$ (Eq. \ref{eq:delta_incoh_coh}) are used to select the most deviant (with respect to the mean $N(z)$) of the remaining samples: the `incoherently' $\nzinc$ and `coherently' $\nzcoh$ shifted distributions, shown in each panel as orange dot-dashed and purple dashed curves, respectively. The least deviant (in terms of $\Delta_{\rm incoherent}$) sample is also selected as the `unbiased' distribution, given by green curves. Inset panels zoom in on the high-$z$ tails of each tomographic distribution, revealing small deviations there with logarithmic $y$-axes.
    }
    \label{fig:redshift_distributions}
\end{figure*}

For our Stage-III forecasts, we base our synthetic samples on those used for the cosmic shear and combined-probe analyses of KiDS-1000 \citep{Asgari21,Giblin2020,Hildebrandt2021,heymans21,Joachimi21,Troster2021}. The strategy for generating the data-vector and covariance for our Stage-III forecast is summarised in Fig. \ref{fig:s3_dc}. 

\subsubsection{KiDS-1000-like photometric sample}
\label{sec:stage_3_photometric}

With the exception of the redshift distributions, we define our Stage-III photometric samples to directly resemble those of the public KiDS-1000 data. We substitute the SOM-calibrated redshift distributions of KiDS-1000 \citep{Hildebrandt2021,Wright2020} for the earlier KiDS+VIKING-450 (KV450) direct redshift calibration of \cite{Hildebrandt20} and \cite{Wright2018}, from which a full covariance of the $n(z)$ could be reliably derived owing to the spatial bootstrapping approach utilised. We use this covariance to generate additional redshift distributions which are biased with respect to the starting distribution, as we shall describe below.

Whilst the KV450 and KiDS-1000 redshift distributions are similar, we do not require the distributions to correspond closely to the latest KiDS $n(z)$ calibration because we are conducting a simulated analysis, and we are free to choose the true and biased $n(z)$ accordingly. We do assume that the $n(z)$ covariance is reasonably realistic, and note that any future forecasting analysis conducted along these lines would do well to utilise a covariance derived from the latest calibration, incorporating state-of-the-art methods as far as possible.

To define un/biased redshift distributions, we apply the following procedure. We begin by assuming the KV450 redshift distribution, $N(z)$, evaluated at $41$ equally-spaced redshifts $z\in[0,2]$, to represent the mean of a multivariate Gaussian distribution, i.e. the calibrated covariance $\nzcov$. The correlation matrix corresponding to $\nzcov$ is shown in Fig. \ref{fig:nz_covariance}, where labels $n_i(z)$ in the figure denote the tomographic bins $i\in\{1,2,3,4,5\}$. We then draw $\sim45\,000$ realisations of the full $n(z)$ ($>200$ times the number of calibrated data points in a single  $n(z)$, across all five bins) from the multivariate normal distribution $\mathcal{N}(0, \nzcov)$ and add these to the mean distribution, given by $N(z)$\footnote{We note here for clarity that the KV450 redshift distribution, denoted $N(z)$, is equal to: (i) the mean of the distribution from which samples $n(z)$ are drawn to simulate photometric redshift biases; (ii) the `estimated', or `starting' distribution for simulated likelihood analyses, to which nuisance models will be applied; and (iii) in limited forecast cases of zero redshift bias, the actual target distribution, denoted $\nztrue$. Meanwhile, $n(z)$ refers to a sample redshift distribution drawn from the covariance, and in later sections to any arbitrary redshift distribution.}
, to yield an ensemble of redshift distributions $\{n(z)\}_X$. Approximately $80\%$ of the realisations in the ensemble have one or more negative $n(z)$ values and are consequently discarded, such that $X$ refers to those remaining.

For each of the remaining realisations, we then compute the 5-vector of mean redshifts $\meanz_i$ of each tomographic bin and define the quantities
\begin{equation}
\begin{split}
    \Delta_{\rm{incoherent}} &= \sqrt{\sum_i{\Big(}\meanz_i - \meanz_{i0}{\Big)}^2} \\
    \Delta_{\rm{coherent}} &= \left|\sum_i {\Big(}\meanz_i - \meanz_{i0}{\Big)}\,\right|\,,
    \end{split}
    \label{eq:delta_incoh_coh}
\end{equation}
where $\meanz_{i0}$ denote the mean redshifts of the starting distribution $N(z)$. Thus $\Delta_{\rm{incoherent}}$ describes the Euclidean distance of a sampled $n(z)$ from the mean of the multivariate normal distribution $N(z)$, after compressing to the five mean redshifts, and can be large regardless of the signs of the deviations $\meanz_i-\meanz_{i0}$ ($\equiv\delta{}z_i$; Sect. \ref{sec:modelling_choices}). Conversely, $\Delta_{\rm{coherent}}$ describes a post-compression distance from $N(z)$ that is large only if the deviations are of the same sign; i.e. if the total $n(z)$ realisation is \emph{coherently} shifted to higher or lower redshifts across all five bins.

Sorting the ensemble $\{n(z)\}_X$ according to these $\Delta$ quantities, we then select three realisations of redshift distributions:
\begin{enumerate}
    \item the \textbf{unbiased}, or in practice least biased, redshift distribution $\nzunb$, to minimise $\Delta_{\rm{incoherent}}$;
    \item the \textbf{incoherently biased} redshift distribution $\nzinc$, to maximise $\Delta_{\rm{incoherent}}$;
    \item the \textbf{coherently biased} redshift distribution $\nzcoh$, to maximise $\Delta_{\rm{coherent}}$.
\end{enumerate}
As a consequence of the fixed covariance, $\nzcov$, the sorting order of $\{n(z)\}_X$ by in/coherent bias in tomographic mean redshifts is similar. We therefore choose $\nzcoh$ first and impose that $\nzinc \neq \nzcoh$, and that the deviations in mean redshifts $\meanz_i-\meanz_0$ should not all have the same sign for $\nzinc$. Under these conditions, the $\Delta_{\rm{incoherent}}$ and $\Delta_{\rm{incoherent}}$ statistics for $\nzcoh$ are $\sim27\%$ and $\sim74\%$ larger, respectively, than those seen for $\nzinc$. However, defining a third distance quantity as a simple Euclidean distance over the full $n(z)$ shape,
\begin{equation}
    \Delta_{\rm{full}} = \sqrt{\sum_{ij}{\Big(}n_i(z_j) - n_{i0}(z_j){\Big)}^2}\,,
    \label{eq:delta_full}
\end{equation}
where $i$ still indexes tomographic bins and $j$ indexes the redshift axis, we see that $\Delta_{\rm full}$ is $\sim33\%$ larger for our chosen $\nzinc$ than for $\nzcoh$. Despite being less deviant in the mean redshifts, the incoherently biased distribution is in totality more deviant from $N(z)$ than is $\nzcoh$. Since we are interested in the differential impacts upon cosmological constraints of (i) coherent shifting of the bulk redshift distribution, and (ii) more stochastic errors within the distribution, these choices suit our purposes and we proceed accordingly. We note that a more detailed follow-up analysis could explore several such choices for in/coherently biased distributions, perhaps using the full shape distance as another metric for the selection, and making use of different $\nzcov$ that yield more heterogeneous ensembles $\{n(z)\}_X$.

The resulting distributions are shown in Fig. \ref{fig:redshift_distributions},
where the mean distribution $N(z)$ is given by black curves; the unbiased distribution by green curves; the incoherent bias by orange dot-dashed curves; and the coherent bias by purple dashed curves. Alternating panels show the distributions, $n(z)$, and the differences, $\Delta{}n(z)=n(z)-N(z)$, for each of the chosen redshift bias scenarios. The $n(z)$ are also shown in faint grey for a random $50$ realisations from the initial $45\,000$ ensemble. Inset panels display the high-$z$ tail for each bin on a logarithmic $y$-axis and reveal excesses relative to the starting $N(z)$, particularly in the case of the coherent bias (purple).

It is expected that a coherent bias in the redshift distribution used to model cosmic shear statistics should manifest more strongly in the final inference of the structure growth parameter $S_8$ than an incoherent redshift bias (see e.g. \citealt{Joudaki2019}, and the recent work of \citealt{Giannini2022}). This is because the same measured weak lensing signal, assumed to originate from a higher redshift, would be consistent with a lower $S_8$ value if all else is held constant. We note too that all distributions, including $\nzunb$, feature full-shape differences with respect to the mean distribution $N(z)$ (e.g. second $\Delta{}n(z)$ panel in Fig. \ref{fig:redshift_distributions}); such localised features are more likely to manifest in the density statistics, particularly the photometric auto-clustering.

Our central proposal is that nuisance models designed to compensate for any such errors in redshift calibration will enjoy more accurate and precise constraints upon the inclusion of additional two-point correlations in a joint analysis \citep[similarly to the proposal of][in the context of intrinsic alignment calibration]{Joachimi2010}, particularly the spectroscopic-photometric clustering cross-correlations. It is therefore important that the photometric shear and density samples share a redshift distribution \citep{Schaan2020}. This implies that any weighting of the shear sample (e.g. as derived from shape measurements) that is incorporated into the redshift calibration must also be applied to the photometric density sample\footnote{This could lead to complications that must be considered if, for example, per-galaxy weights are intended for the mitigation of systematic density fluctuations in measured positional statistics \citep{Rezaie2019,Wagoner2020}.}.

For each of our forecasts, the \emph{true} $n(z)$ distribution that enters the computation of our mock data-vector is given by one of the above-defined distributions: unbiased, incoherently biased, or coherently biased (or the exact $N(z)$, for limited use), as described in Sect. \ref{sec:theory}. 
Meanwhile, the $N(z)$ that enters the theory-vector computation is an externally-calibrated estimate for the photometric sample redshift distribution, which we refer to as the exact $N(z)$, with or without the application of nuisance models (Sect. \ref{sec:modelling_choices})\footnote{Notice that we differentiate between the estimated $N(z)$, the initial guess for the redshift distribution, and $n(z)$, the true distribution informing the data-vector, i.e. the target distribution.}.
We shall henceforth refer to the redshift bias configurations as: `incoherent', where the tomographic mean redshifts are systematically low at low redshift (i.e.~pertaining to the starting distribution relative to the target distribution informing the data-vector), but more accurate as the redshift increases; `coherent', where the mean redshifts are systematically low at all redshifts; and `unbiased', where the mean redshifts are accurate (see Fig. \ref{fig:redshift_distributions}). All three configurations feature full-shape errors in the redshift distribution. We also make limited use of the $N(z)$ distribution, without additional modelling, as the source of both data- and theory-vector, referring to this as the `exact-true' configuration.

We emphasise here that our different $n(z)$ bias cases yield different data-vectors; the in/coherently biased cases are sourced from generally higher redshifts than the unbiased/exact-true cases (Table \ref{tab:sample_details}). Consequently, the addition of nuisance model parameters can counter-intuitively \emph{increase} the overall constraining power, if the parameters act via the $n(z)$ to push signals into regimes of higher signal-to-noise (aided in our case by the application of nonzero-mean Gaussian priors for $\deltazi$ parameters; to be discussed more in Sect. \ref{sec:results}). Although this also complicates direct comparisons with previous work on real data (e.g. \citealt{Joudaki2019}), where the theory-vectors are variable through $n(z)$ estimates and nuisance modelling but the galaxy data are fixed, the qualitative results are in agreement.

We note that our methods for defining biased redshift distributions constitute a mixture of pessimistic and optimistic choices. Whilst we draw many thousands of realisations, and choose serious outliers to describe the true distributions, these are all compatible with the calibrated covariance; thus they do not represent catastrophic failures of redshift calibration, but uncommon realisations. We assume Gaussian priors (Sect. \ref{sec:forecast}; Table \ref{tab:priors}) for use with the shift model (Sect. \ref{sec:modelling_choices}) that are centred on the true shifts $\delta{z}_i=\meanz_i-\meanz_{i0}$, and have widths corresponding to the true variance of $\delta{z}_i$ over the ensemble $\{n(z)\}_X$; thus the calibration is assumed to yield a perfectly accurate prior for $\delta{z}_i$ in each case. Lastly, we have selected these $n(z)$ only considering variations in the tomographic mean redshifts $\delta{z}_i$. It may be that an equivalent consideration of the full $n(z)$ shape, e.g. via metrics like $\Delta_{\rm full}$, would yield distributions of different profiles, carrying distinct consequences for cosmological parameter inference under probe configurations variably sensitive to the full shape of the $n(z)$ and the tomographic mean redshifts. Each of these choices could be revisited in future analyses of the impacts of redshift distribution misestimation.

The shear dispersion and effective number density statistics of the photometric sample are taken to be exactly those estimated for the KiDS-1000 shear sample, given in Table 1 of \cite{Asgari21}. Bin-wise mean redshifts $\meanz$ are computed for $\nzinc$ and $\nzcoh$, while the means of $\nzunb$ are practically the same as for $N(z)$. These are given in Table~\ref{tab:sample_details}, which also records the assumed $r$-band depth, $r_{\rm lim} = 24.5$, and ${\rm area}=777\sqdeg$ for our Stage-III photometric survey setup.

\subsubsection{BOSS-2dFLenS-like spectroscopic sample}
\label{sec:stage_3_spectroscopic}

For synthetic spectroscopic samples, we take the combined redshift distribution of SDSS-III BOSS \citep{Eisenstein2011} and 2dFLenS \citep{Blake2016a}, presented for KiDS-1000 usage by \cite{Joachimi21}. As mentioned in Sect. \ref{sec:theory}, an analytical consideration of the covariance between angular power spectra and 3-dimensional multipole power spectra is under development. In the meantime, we attempt to retain the 3-d density information from the mock spectroscopic sample by finely re-binning the spectroscopic $n(z)$, defining 10 tomographic samples in the range $z\sim0.2-0.75$, each having width $\Delta{}z\sim0.05$ (see \citealt{Loureiro2019} for a similar treatment of BOSS DR12). We recompute the mean redshifts $\meanz_i$, re-scale the number density statistics from \cite{Joachimi21} for each newly defined redshift bin, and record these figures in Table \ref{tab:sample_details} along with the assumed area of $9329\sqdeg$. For spectroscopic-photometric cross-correlations, we assume an overlapping area of $661\sqdeg$ (the sum of BOSS+2dFLenS versus KiDS-1000 overlapping areas; \citealt{Joachimi21}) and retain the number densities and bin-wise redshift distributions of the full spectroscopic sample.

We assume an $r$-band depth of $r_{\rm lim}=20.0$ for the Stage-III spectroscopic samples in order to roughly match the luminosity function slopes, $\alpha_x(z)$, observed by \cite{VonWietersheim-Kramsta2021} for BOSS data. The $\alpha_x(z,r_{\rm lim})$ that result from this magnitude limit via the fitting formulae of \cite{Joachimi2010} are slightly low for the lower-$z$ spectroscopic bins, and high for the higher-$z$ bins, and do not allow for significantly improved agreement through changes to $r_{\rm lim}$. This is likely due to the complex selection function defining BOSS galaxy samples, resulting in luminosity functions that are not well-described by the fitting formula of \cite{Joachimi2010}, which is calibrated against magnitude-limited galaxy data. More principled estimation of $\alpha_x(z)$, perhaps using luminosity functions directly, or using simulations \citep{ElvinPoole2022}, would be desirable for more accurate modelling of magnification number count contributions in future work.

\subsection{Mock Stage-IV samples}

\begin{figure*}
    \centering
    \includegraphics[width=0.95\textwidth]{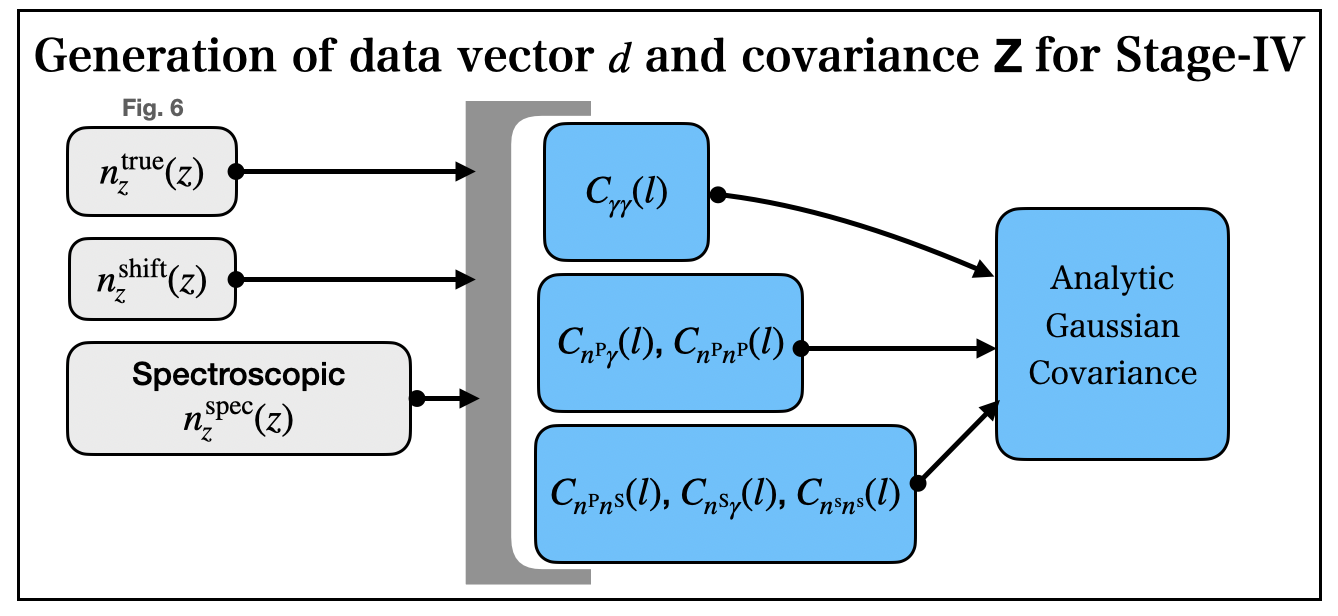}
    \caption{A sketch of how the data-vector and covariance are generated for our Stage-IV forecasts.}
    \label{fig:s4_dc}
\end{figure*}

\begin{figure}
    \centering
    \includegraphics[width=\columnwidth]{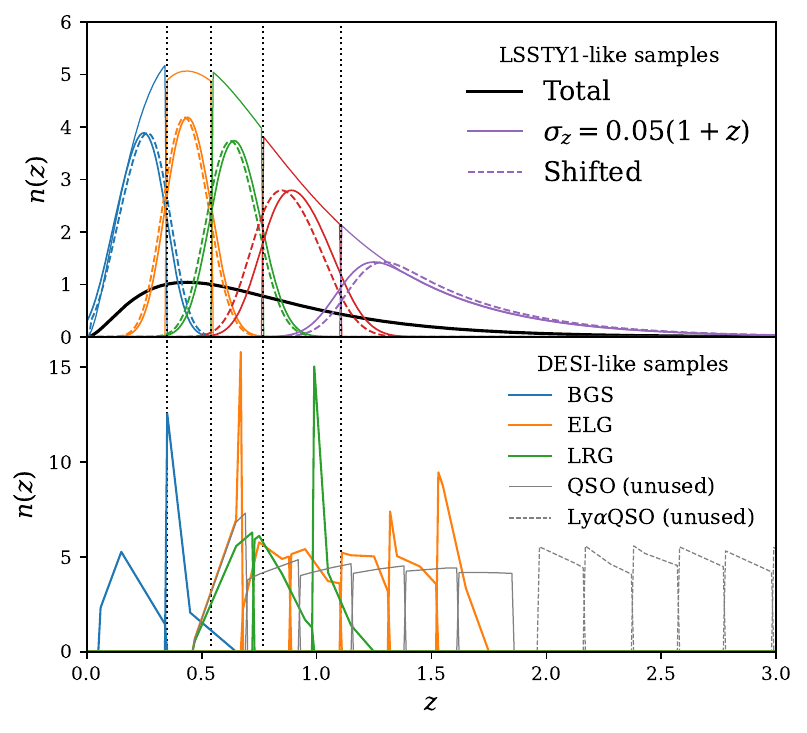}
    \caption{The redshift distributions assumed for our Stage-IV synthetic galaxy samples, with LSST Year 1-like photometric tomography (\emph{top}; Sect. \ref{sec:stage_4_photometric}), supplemented by DESI-like spectroscopic galaxy samples (\emph{bottom}; Sect. \ref{sec:stage_4_spectrosopic}). Vertical dotted lines give the redshift edges where the full photometric redshift distribution (\emph{top panel}; black curve) is cut into tomographic bins (hard-edged histograms). These distributions are convolved with Gaussian kernels of width $\sigma_z=0.05(1+z)$ to produce the true redshift distributions (solid, coloured curves), and then displaced with randomly-drawn shifts $\deltazi$ as $n^{\rm shifted}_{i}(z) = n^{\rm true}_i(z - \deltazi)$ to define the `shifted' distributions (dashed, coloured curves). 
    %Models for nuisance recalibration of the photometric redshift distributions should enjoy more accurate and precise constraints through the inclusion of positional cross-correlations with the spectroscopic galaxy samples (\emph{bottom}), whose redshift distributions are assumed to be exactly known. 
    Apparent small overlaps of similarly-coloured histograms in the bottom panel are plotting artefacts.
    }
    \label{fig:stage_4_distributions}
\end{figure}

We define mock Stage-IV samples based on information from the Science Requirements Document (SRD) of the Rubin Observatory Legacy Survey of Space and Time (LSST) Dark Energy Science Collaboration \citep[DESC;][]{TheLSSTDarkEnergyScienceCollaboration2018}. These are supplemented by mock spectroscopic samples, modelled after the survey specifications of the Dark Energy Spectroscopic Instrument \citep[DESI;][]{DESICollaboration2016}. The strategy for generating the data-vector and covariance for our Stage-IV forecast is summarized in Fig. \ref{fig:s4_dc}. 

\subsubsection{LSST Y1-like photometric sample}
\label{sec:stage_4_photometric}

For our Stage-IV-like photometric dataset, we assume the LSST Year 1 redshift distribution from the SRD, given by
\begin{equation}
    n(z) = z^2 \exp\left\{-\left(\frac{z}{0.13}\right)^{0.78}\right\} ,
    \label{eq:stage_4_nofz}
\end{equation}
which we evaluate over the range $z=0-4$. Also following the SRD, we define the shear sample to have five equi-populated bins over this range and convolve each of these with a Gaussian kernel of evolving width $\sigma_z=0.05(1+z)$ in order to simulate photometric redshift errors.

We go beyond the SRD for our forecasts, additionally defining a biased redshift distribution. To do so, we simply draw random shifts $\deltazi$ from the normal distributions $\mathcal{N}(0,\sigma_{\rm shift})$, where $\sigma_{\rm shift}=[0.01,0.02,0.03,0.04,0.05]$ for the five bins, respectively, and apply these to the starting distributions $n_i(z)$ as $n^{\rm shifted}_{i}(z) = n^{\rm true}_i(z - \deltazi)$. Without any restriction on the sign of the shifts, these more closely resemble the incoherently biased Stage-III redshift distributions from Sect. \ref{sec:stage_3_photometric}. We assess the ability of the shift model to correct these redshift errors (given Gaussian priors centred on the true $\deltazi$, with widths equal to $\sigma_{\rm shift}$), noting that redshift biases constructed in this way are unrealistic and overly generous to the shift model; future forecasts should consider more complex, full-shape $n(z)$ biases, as we have done in our Stage-III setup (Sect. \ref{sec:stage_3_photometric}). We accordingly differentiate between the full-shape, `in/coherently biased' Stage-III redshift distributions, and the `shifted' distributions considered for Stage-IV.

By construction, each of the tomographic bins has a similar number density, which we compute after assuming the full sample to have $10$ galaxies $\psqarcmin$ and re-binning the total $n(z)$. We follow the SRD in assuming an intrinsic shear dispersion of $\sigma_\epsilon=0.26$, an $r$-band depth of $r_{\rm lim}=25.8$, and an area of $12\,300\sqdeg$ for LSST Year 1 -- these statistics, and per-bin mean redshifts $\meanz_i$, are recorded in Table \ref{tab:sample_details}.

We note that photometric data are already intended for usage as density samples in the analysis of LSST \citep{TheLSSTDarkEnergyScienceCollaboration2018}. However, these lens samples are to be defined with uniform spacing in redshift, and with limits such that $z\in[0.2,1.2]$. Our forecasts here presume that the shear sample itself can be used for density statistics \citep[similarly to e.g.][]{jk2012, Schaan2020}, with redshift distribution recalibration bolstered by cross-correlations with a spectroscopic density sample, given the overlap between LSST and DESI.

\subsubsection{DESI-like spectroscopic samples}
\label{sec:stage_4_spectrosopic}

We consider $4000\sqdeg$ of DESI-like spectroscopic observations, which completely overlap with our LSST Year~1-like samples \citep{TheLSSTDarkEnergyScienceCollaboration2018}. For simplicity, we assume this as a conservative area coverage for the DESI Year~1 data, noting that the true coverage is nearly twice as large. We take the forecasted redshift distributions ${\rm d}N/{\rm d}z{\rm d}\Omega$ (where $\Omega$ denotes a solid angle) for the DESI Bright Galaxy Sample (BGS), Emission Line Galaxy (ELG) sample, and Luminous Red Galaxy (LRG) sample\footnote{We neglect quasar (QSO/Ly$\alpha$QSO) samples such as those observed by DESI, so as to lessen the computational demands of these forecasts. Future work could consider these additional, sparser but higher-redshift samples, as well as other spectrosopic observations planned to overlap with Stage-IV surveys. Note also that observational results now present in \url{data.desi.lbl.gov} were not available when this work was in progress.} from \cite{DESICollaboration2016}, Tables 2.4 and 2.6\footnote{In comparison to recent DESI Collaboration results \citep{DESI_target_2, DESI_target_1}, the redshift success rates we have adopted are either lower or equal. Therefore, the number density of targets we have used for this work is conservative.}, and re-bin them to have uniform widths of at least $\Delta{}z=0.2$, resulting in $11$ tomographic bins (BGS:2, ELG:6, LRG:3) that share some internal overlaps in the range $z\sim0.5-1.0$. We note that these spectroscopic bins are $\sim4\times$ wider (in redshift space) than those implemented in our Stage-III setup; this choice was made only to reduce the computational demand of these forecasts, and finer tomography is a primary avenue for improvement in future work. Indeed, the calibration power of cross-correlations estimated here for Stage-IV analyses could be considered as conservative, though this is offset by the simplicity of the implemented redshift errors (Sect. \ref{sec:stage_4_photometric}).

Given the area coverage and newly defined redshift distributions, number densities per bin are simply calculated and recorded in Table \ref{tab:sample_details} alongside mean redshifts $\meanz_i$. Targeting surveys for DESI are estimated to yield an $r$-band depth of at least $r_{\rm lim}=23.4$, which we assume as the limiting magnitude for each of the mock DESI galaxy samples for simplicity. The $\alpha_x(z)$ so-estimated from the fitting formula \citep{Joachimi2010} are again unlikely to describe well the luminosity functions of these highly-selected DESI samples. We are therefore modelling magnification contributions for these (and to a slightly lesser extent the Stage-III) samples according to rough guesses of reasonable values for the slopes of luminosity functions -- since these are minor contributions to clustering correlations, and since we neither vary nuisance parameters to describe magnification contributions \citep[see e.g.][]{ElvinPoole2022}, nor fail to model them entirely \citep[see e.g.][for the impact of faulty modelling]{Mahony2021}, we do not expect these choices to affect our conclusions.

Our Stage-IV sample redshift distributions are depicted in Fig. \ref{fig:stage_4_distributions}, with LSST Year 1-like photometric samples in the top panel, and DESI-like spectroscopic samples in the bottom panel (including the prospective, sparse, high-$z$ quasar -- `QSO' and `Lyman-$\alpha$ QSO' -- samples that we do not consider in this work due to computational constraints).

\section{Forecasting methodology}
\label{sec:forecast}

\begin{figure*}
    \centering
    \includegraphics[width=0.95\textwidth]{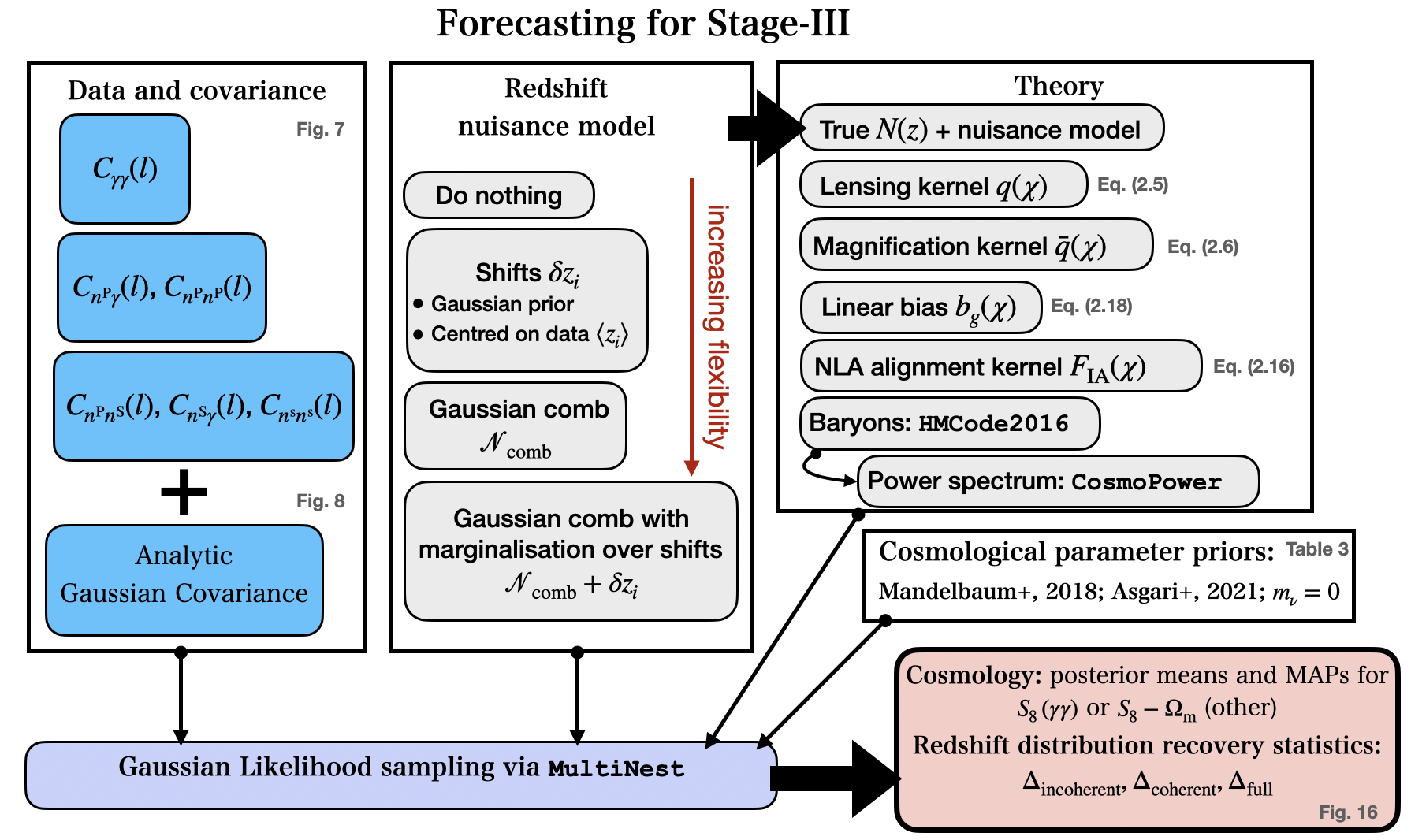}
    \caption{A sketch of the forecasting steps and choices for the Stage-III configuration.
    }
    \label{fig:s3_forecast}
\end{figure*}

\begin{figure*}
    \centering
    \includegraphics[width=0.95\textwidth]{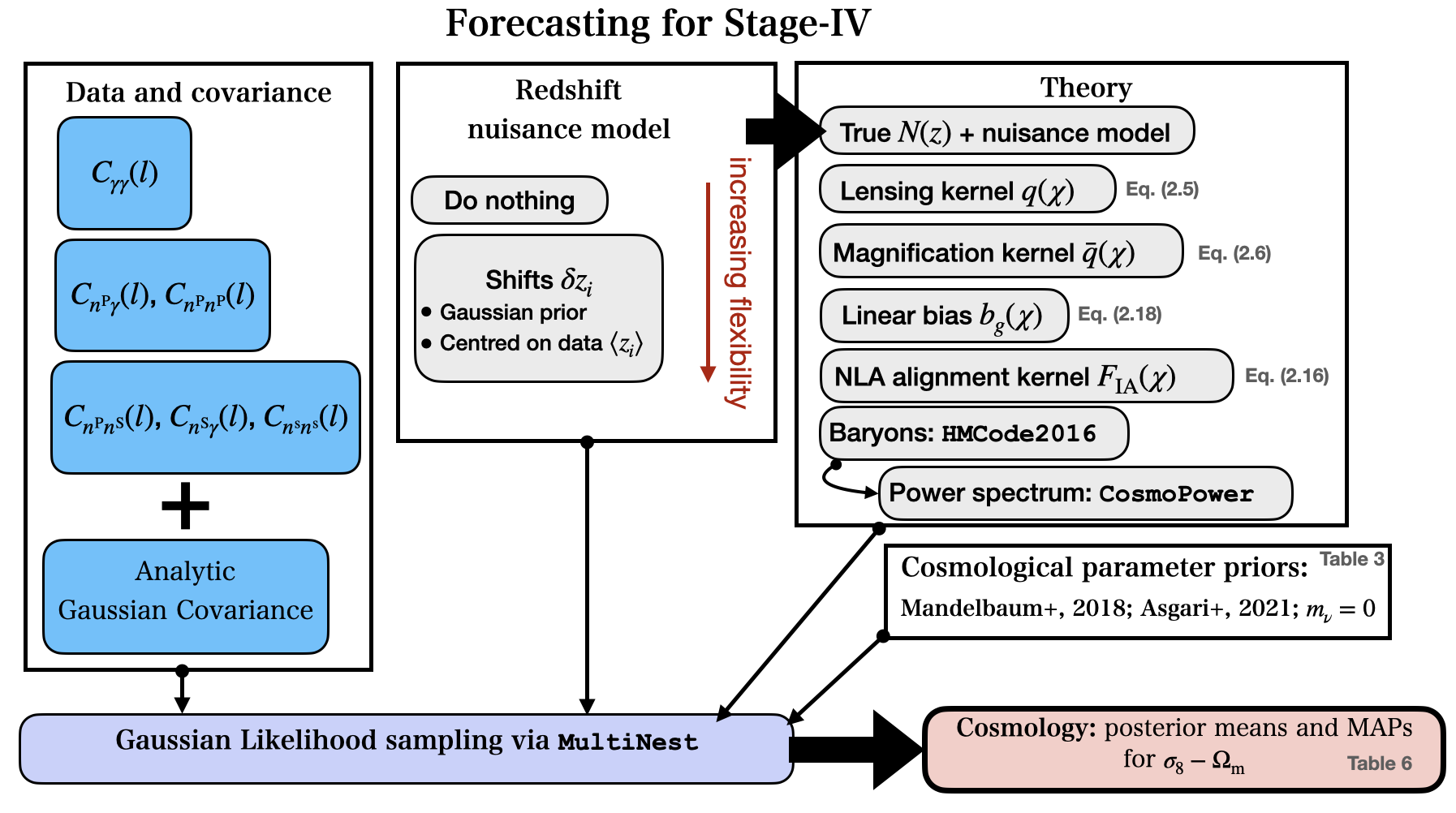}
    \caption{A sketch of the forecasting steps and choices for the Stage-IV configuration. We explored fewer redshift nuisance models here than for the Stage-III configurations, since the 
$n(z)$ biases described in Sect. \ref{sec:stage_4_photometric} are comparatively simple, only featuring shifts in the tomographic mean redshifts that ought to be well-compensated by the $\deltazi$ model. In addition, results from Stage-IV forecasts are always considered on the $\siom$ plane given that cosmic shear alone has sufficient constraining power to alleviate the non-linear degeneracy seen for Stage-III.}
    \label{fig:s4_forecast}
\end{figure*}

\begin{table*}
    \centering
    \begin{tabular}{lccc}
        \hline
        \hline
        Configuration choice & Label & Performed for Stage-III & Performed for Stage-IV \\
        \hline
        {\textbf{Probe set}} & & & \\
        Shear-only & $\gamma\gamma$ & \checkmark & \checkmark \\
        Spectroscopic $\rm 3\times2pt$ & $\thr$ & \checkmark & \checkmark \\
        Photometric $\rm 3\times2pt$ & $\pp$ & \checkmark & \checkmark \\
        Full $\rm 6\times2pt$ & $\six$ & \checkmark & \checkmark\\
        \hline
        {\textbf{Redshift bias}} & & & \\
        Exactly zero & $n^{\rm true}(z)$ & \checkmark & \checkmark \\
        Unbiased & $n^{\rm unb.}(z)$ & \checkmark & \\
        Incoherently biased & $n^{\rm inc}(z)$ & \checkmark & \\
        Coherently biased & $n^{\rm coh}(z)$ & \checkmark & \\
        Shifted & $n^{\rm shifted}(z)$ & & \checkmark \\
        \hline
        {\textbf{Redshift nuisance model}} & & & \\
        Do nothing & $\times$ & \checkmark & \checkmark \\
        Shift model & $\deltazi$ & \checkmark & \checkmark \\
        Comb+shift & $\hcomb$ & \checkmark & \\
        Comb & $\fcomb$ & \checkmark & \\
        \hline
        \hline
    \end{tabular}
    \caption{A summary of the large-scale structure analysis forecast configurations explored in this work. Columns give: the choice of probe combinations, redshift biases (Sects. \ref{sec:stage_3_photometric} \& \ref{sec:stage_4_photometric}), and redshift nuisance models (Sects. \ref{sec:generic_spectra} \& \ref{sec:comb_spectra}); corresponding labels (for figure legends to follow). 
    }
    \label{tab:configurations}
\end{table*}

\begin{table}
    \centering
    \def\arraystretch{1.2}
    \begin{tabular}{ccc}
\hline
\hline
Parameter & Fiducial centre & Forecast prior \\
\hline
$\Omega_{\rm{c}}h^{2}$ & $0.107$ & $[0.051, 0.255]$ \\
$\Omega_{\rm{b}}h^{2}$ & $0.026$ & $[0.019, 0.026]$ \\
$\ln(10^{10}A_{\rm{s}})$ & $3.042$ & $[1.609, 3.912]$ \\
$h$ & $0.64$ & $[0.64, 0.82]$ \\
$n_{\rm{s}}$ & $1.001$ & $[0.84, 1.1]$ \\
\hline
$A_1$ & $0.973$ & $[-6, 6]$ \\
$A_{\rm{bary}}$ & $2.8$ & $[2.0, 3.13]$ \\
\hline
$b^\alpha_{{\rm{g}}}$ & $0.95/D_{+}(\meanz_\alpha)$ & $[0.5, 9]$ \\
$\delta{}z_i$ & $\mu_i$& $\mathcal{N}(\mu_i, \sigma_i)$ \\
\hline
$S_8$ & $0.7718$ & - \\
\hline
\hline
    \end{tabular}
    \caption{The fiducial model parameters assumed in this analysis, where the cosmological model is taken as flat-$\Lambda$CDM, intrinsic galaxy alignment contributions are given by the 1-parameter ($A_1$) non-linear alignment model, non-linear structure growth is described by the 1-parameter ($A_{\rm bary}$) halo model {\tt HMCode} with baryonic contributions \citep{Mead2015,Mead2021}, galaxy biases $b_{\rm g}$ are linear and deterministic, and scalar shifts in the mean redshift $\deltazi$ are employed for recalibration of photometric redshift distributions. Priors are flat except for $\deltazi$, where Gaussian priors are estimated as described in Sect. \ref{sec:stage_3_photometric}. Central values (for non-$\deltazi$ parameters) are taken as the best-fit values from \cite{Asgari21} (their Table A.2, column 3) except in the case of $A_{\rm bary}$, which we reduce slightly in order to introduce some baryonic contributions into the fiducial data-vectors. Whilst \cite{Asgari21} sampled directly over the $S_8$ parameter, it is more convenient for our implementation of {\tt CosmoPower} \citep{Mancini2021} to sample over $A_{\rm s}$. We convert the (derived) best-fit $A_{\rm s}$ from \cite{Asgari21} into a fiducial centre for $\lnas$, and assume a flat prior on $\lnas$ corresponding to $0.5<10^9A_{\rm s}<5.0$ \citep{Secco2022}. Other flat priors are taken from \cite{Asgari21}, or from \cite{TheLSSTDarkEnergyScienceCollaboration2018} in the case of the galaxy bias, where the latter also provides the functional form for the galaxy bias (Eq. \ref{eq:galaxy_bias_function}). The derived, fiducial $S_8$ value is given in the last row, and is slightly larger than that found by \cite{Asgari21} due to our modification of the fiducial $A_{\rm bary}$. 
    Notice that we also adopt the best-fit for $h$ as in \cite{Asgari21}, and while this is at the boundary of our prior, it does not affect our conclusions given that $h$ is unconstrained.
    }
    \label{tab:priors}
\end{table}

\begin{figure*}
    \centering
    \includegraphics[width=\textwidth]{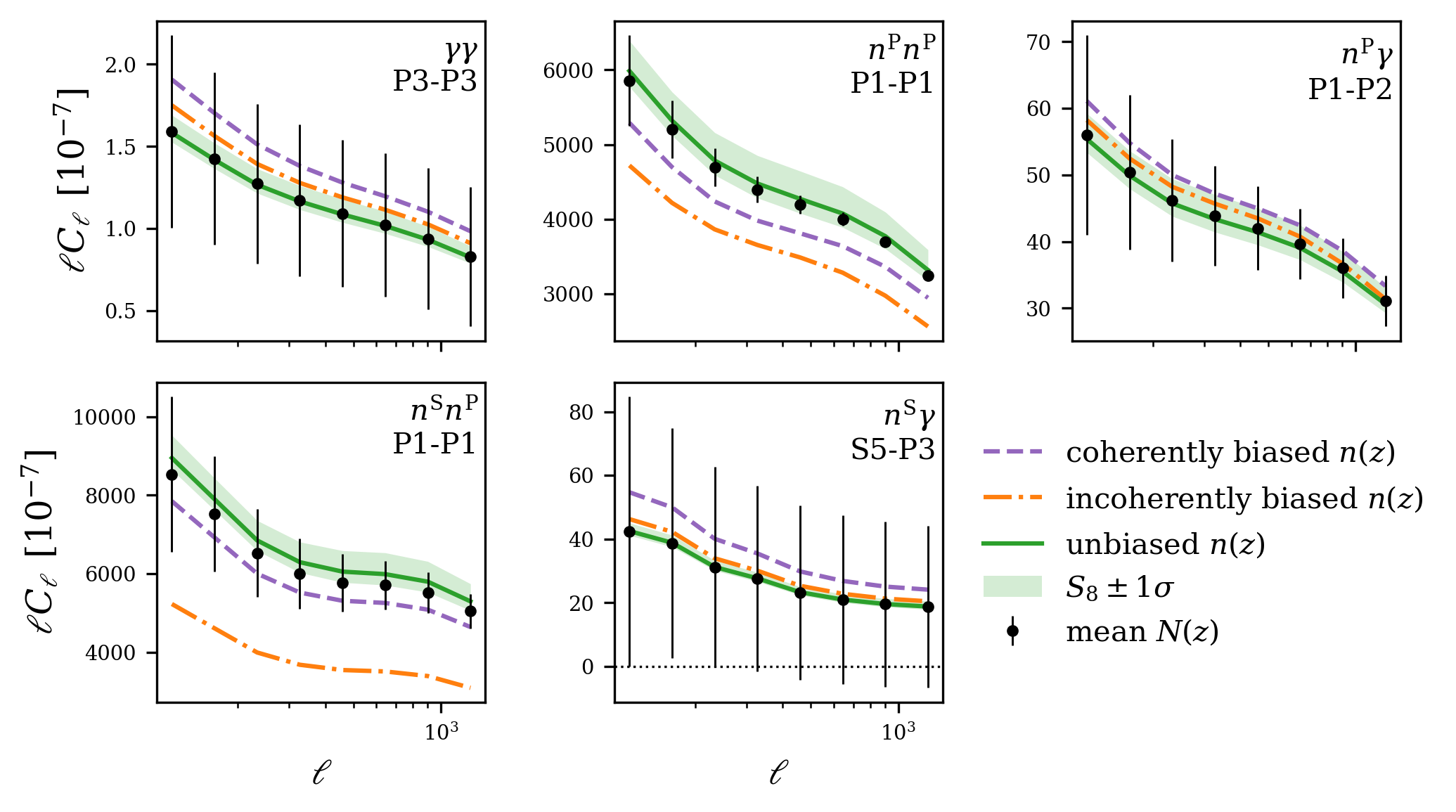}
    \caption{Illustration of tomographic two-point correlations that form a subset of the $\six$ data/theory-vectors employed in this work. Annotations in the top-right of each panel give the spectrum type and tomographic sample pairing, for cosmic shear ($\gamma\gamma$), photometric auto-clustering ($n^{\rm P}n^{\rm P}$), photometric galaxy-galaxy lensing ($n^{\rm P}\gamma$), spectroscopic-photometric cross-galaxy-galaxy lensing ($n^{\rm S}\gamma$), and spectroscopic-photometric cross-clustering ($n^{\rm S}n^{\rm P}$) correlations. No spectroscopic auto-clustering correlations are shown, as they are independent of the photometric redshift distribution. Black points and errors give the correlations (Sect. \ref{sec:generic_spectra}) and root-diagonal of the Gaussian covariance (Sect. \ref{sec:covariance}) computed for the `mean' redshift distribution $N(z)$ (taken as the public estimate from KiDS+VIKING-450; \citealt{Hildebrandt20,Wright2018} -- see Sect. \ref{sec:stage_3_photometric}). Coloured curves then show the same correlations, now computed for different redshift distributions: `unbiased' (green, with shading indicating a $\pm1\sigma$ shift in the $S_8$ parameter, holding all else constant); `incoherently biased' (orange dot-dashed); and `coherently biased' (purple dashed). The differences between these curves reflect the sensitivity -- to the $n(z)$ and to $S_8$ -- of the additional observables that we propose to employ for enhanced internal recalibration of photometric redshift distributions via nuisance models.
    We note that the positions of coloured curves relative to black points change as a function of redshift pairings across each angular power spectrum, and we are only showing a subset here.
    }
    \label{fig:dvec_figure}
\end{figure*}

\begin{figure*}
    \centering
    \includegraphics[width=0.8\textwidth]{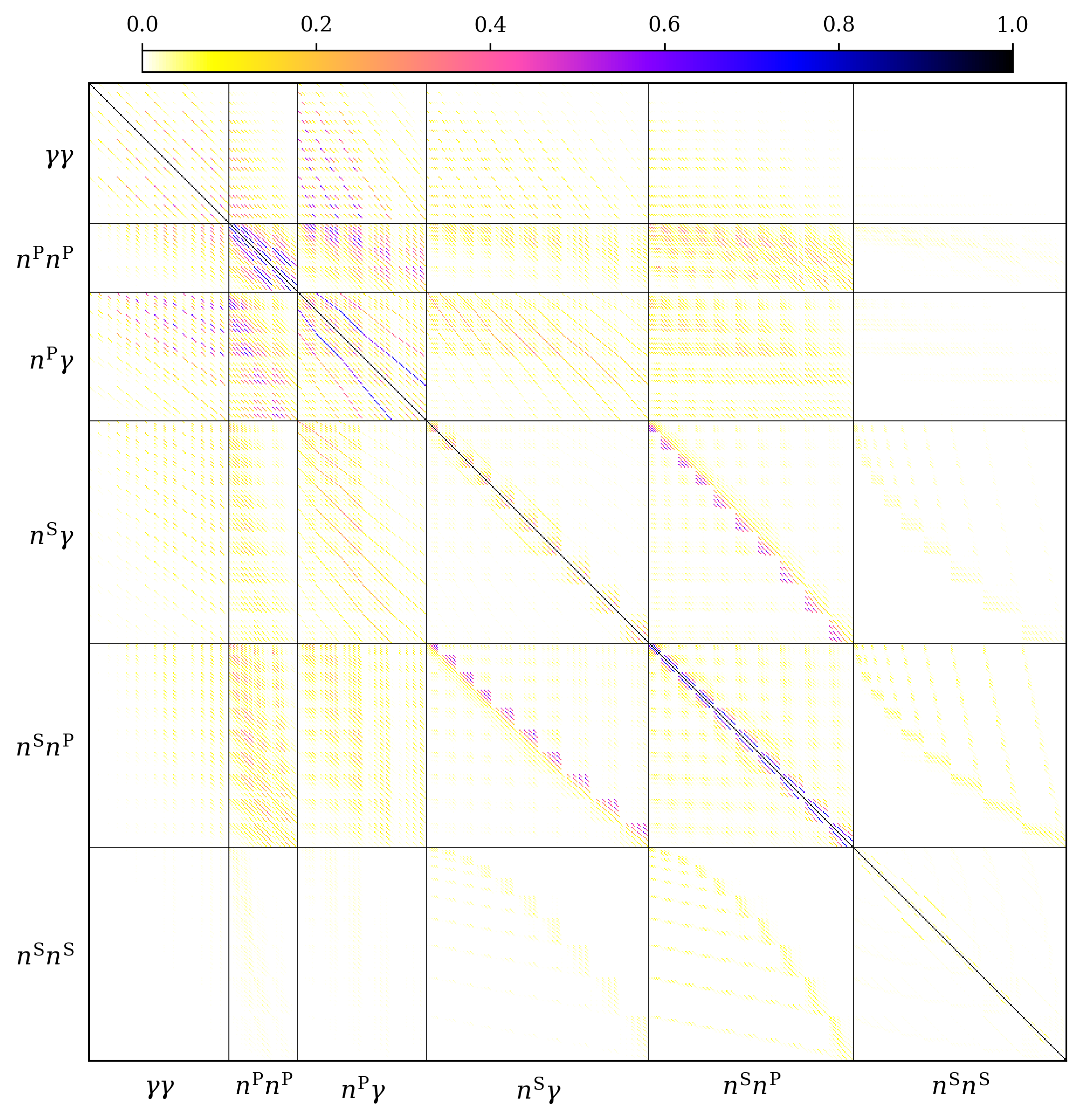}
    \caption{The matrix of correlation coefficients corresponding to an analytical Gaussian covariance, $\tens{Z}$ (Sect. \ref{sec:covariance}), estimated for a (scale-cut; see Sect. \ref{sec:forecast}) Stage-III $\six$ data-vector employed in this work. Axis labels denote sections of the covariance corresponding to cosmic shear ($\gamma\gamma$), photometric auto-clustering ($n^{\rm P}n^{\rm P}$), photometric galaxy-galaxy lensing ($n^{\rm P}\gamma$), spectroscopic-photometric cross-galaxy-galaxy lensing ($n^{\rm S}\gamma$), spectroscopic-photometric cross-clustering ($n^{\rm S}n^{\rm P}$), and spectroscopic auto-clustering ($n^{\rm S}n^{\rm S}$) correlations. The sample statistics used for data-vector covariance estimations are those given in Table \ref{tab:sample_details}.
    }
    \label{fig:data_covariance}
\end{figure*}

\begin{figure}
    \centering
    \includegraphics[width=\columnwidth]{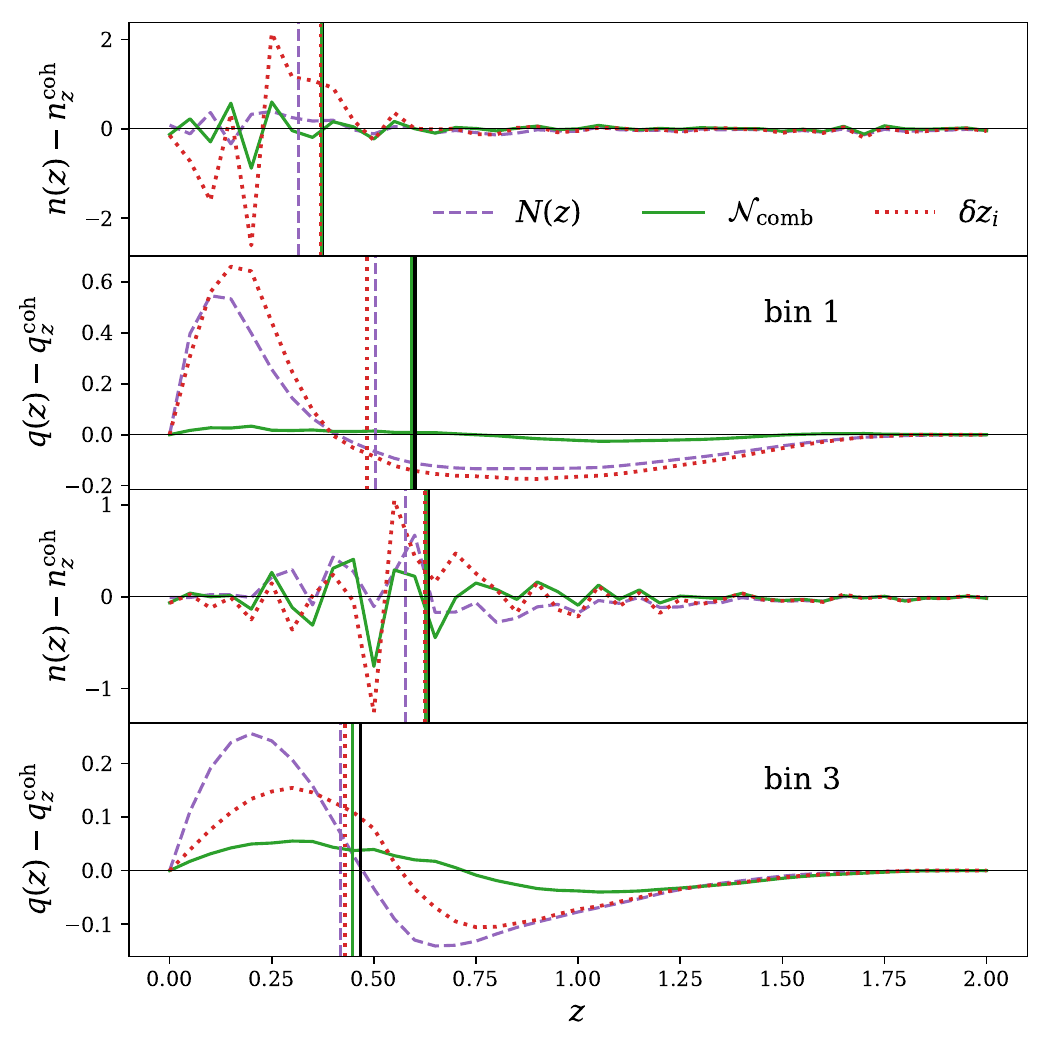}
    \caption{
    An illustration of the application of redshift nuisance models to the first and third bins of the $\nzcoh$ biased Stage-III photometric redshift distribution (Fig. \ref{fig:redshift_distributions}).
    Alternating panels give the target distribution $\nzcoh$, or its corresponding lensing efficiency kernel $q_{z}^{\rm coh}$, subtracted from the equivalent $n(z)$ or $q(z)$ seen under the various recalibration models: the `do nothing' model, denoted as $N(z)$ (dashed purple), the comb model $\fcomb$ (solid green), and the shift model $\deltazi$ (dotted red). The latter two are calibrated here against the $\thr$ data-vector.
    Vertical lines give the mean of each $n(z)$ or $q(z)$.
    The optimised comb model corresponds to a flexible recalibration of the $n(z)$, which minimises the difference (via Eq. \ref{eq:data_chi2}) between the theory- and data-vector (e.g. Fig. \ref{fig:dvec_figure}) by varying the amplitudes of a Gaussian mixture model. One sees that the comb model outperforms the shift model in reducing errors in $q(z)$ whilst minimally increasing errors in $n(z)$.
    }
    \label{fig:comb_vs_shifts}
\end{figure}

We conduct simulated likelihood forecasts for a number of Stage-III and Stage-IV angular power spectrum analysis configurations as detailed in Figs. \ref{fig:s3_forecast} and \ref{fig:s4_forecast}, respectively. Unique configurations are determined by choosing (i) a set of probes, (ii) the bias in the redshift distribution, and (iii) the model chosen for mitigating the uncertainties in the estimation of the redshift distributions. Unless otherwise stated, all modelling and other assumptions are replicated between the Stage-III and Stage-IV forecasts. We do not perform forecasts with nuisance models for the cases where $N(z)$ is both the estimate and the truth, such that the bias is exactly zero. Instead, the unbiased case serves to inform us of how nuisance models behave when the expected corrections are minor. Our selection of forecast configurations is summarised in Table~\ref{tab:configurations}.

Each forecast is performed as follows. First, the mock data-vector $\vec{d}$ is constructed using the fiducial cosmological and nuisance parameters, given in Table \ref{tab:priors}, and the  {\textbf{true}} redshift distribution. In our Stage-III setup, this true redshift distribution is selected to be one of the mean $N(z)$, unbiased $n^{\rm unb.}(z)$, incoherently biased $n^{\rm i-bias}(z)$, or coherently biased $n^{\rm c-bias}(z)$ distributions described in Sect. \ref{sec:stage_3_photometric}. In our Stage-IV setup, the true redshift distribution is either the tomographically-binned distribution given by Eq. \ref{eq:stage_4_nofz}, or the `shifted' distribution described in Sect. \ref{sec:stage_4_photometric}.

Given the cosmological model, the distribution of galaxies in redshift, $n(z)$, are converted into a distribution of galaxies in comoving distance, $n(\chi)=n(z)\,{\rm d}z/{\rm d}\chi$, and lensing efficiencies $q(\chi)$ via Eq. (\ref{eq:lensing_efficiency}). The emulated matter power spectrum, $P_\delta(k,\chi)$, is integrated over auto-/cross-products of $n(\chi)$ and $q(\chi)$ (multiplied by $f^{-2}_K(\chi)$) to produce the full set of `raw' spectra, $C^{\alpha\beta}(\ell)$, required by the configuration. These are initially evaluated at eight logarithmically-spaced angular wavenumbers, $\ell\in[100,1500]$, each rounded to the nearest integer. Linear scaling factors $b_{\rm g},\bar{F}_{\rm m},\bar{F}_{\rm IA}$ (Fig. \ref{fig:linear_modelling}; described in Sects. \ref{sec:generic_spectra} \& \ref{sec:approximations}) are applied to produce the spectral contributions GG, GI, IG, II, gg, mg, gm, mm, gG, gI, mG, and mI (Sect. \ref{sec:generic_spectra}), which are then appropriately summed to produce the `observed' (single) cosmic shear $C_{\gamma\gamma}(\ell)$, (up to two) galaxy-galaxy lensing $C_{n\gamma}(\ell)$, and (up to three) galaxy clustering $C_{nn}(\ell)$ angular power spectra.

Next, to mimic real analyses that avoid describing non-linear structure growth with insufficiently complex models, we apply further scale cuts to the clustering and GGL spectra to exclude small physical scales. For each tomographic density sample $n_\alpha(z)$, the maximum angular wavenumber $\ell_{\rm max}$ is calculated as \citep{TheLSSTDarkEnergyScienceCollaboration2018}
\begin{equation}
    \ell_{\rm max} = k_{\rm max}\,\chi(\meanz)-0.5\,,
    \label{eq:scale_cuts}
\end{equation}
where we take $k_{\rm max}=0.3\,h\,{\rm Mpc}^{-1}$, and $\chi(\meanz)$ is the comoving distance to the mean of the distribution $n_\alpha(z)$. For clustering correlations having two different density kernels, $\ell_{\rm max}$ is taken as the minimum of the two. In the case of cosmic shear, we apply no further scale cuts beyond the initial restricted range in $\ell$, following KiDS-1000 \citep{Joachimi21,Asgari21,heymans21}. For cosmic shear, this implies $\ell_{\rm max}=1500$.

The final data-vectors are thus formed of $\leq6$ unique angular power spectra, $C(\ell)$, defined within and between photometric (shear + density) and spectroscopic (density) samples, including all tomographic auto- and cross-correlations that satisfy the scale cuts described above. Analytic Gaussian covariances are computed for the data-vectors as described in Sect. \ref{sec:covariance}, and the data-vector $\vec{d}$ and inverse data covariance $Z$ are then fixed for the remainder of each forecast.

Some example tomographic power spectra (except the spectroscopic auto-clustering) from a Stage-III $\six$ data-vector are given in Fig. \ref{fig:dvec_figure}, which shows the $C(\ell)$'s computed for the `estimated' $N(z)$ in black, with error-bars corresponding to the root-diagonal of the Gaussian covariance (for which the corresponding correlation matrix is shown in Fig. \ref{fig:data_covariance}). Additional curves in Fig. \ref{fig:dvec_figure} illustrate the thesis of this work: correlations that use the photometric (i.e. the shear) sample as a density tracer (middle-top, top-right, and bottom-left panels) are highly sensitive to changes in the redshift distribution and this information could be used to aid with recalibration of the distribution. This is particularly true of the spectroscopic-photometric cross-clustering, where the spectroscopic $n(z)$ (taken to be exactly known) can act as an anchor for the parameters of any nuisance model -- though only if the data are capable of constraining spectroscopic and photometric galaxy biases simultaneously with $n(z)$ model parameters. The cross-clustering is thus only useful in conjunction with both auto-clustering correlations, each quadratically dependent upon its respective galaxy bias (indeed, an all-clustering $\thr$ analysis could be an interesting avenue for exploration). Upon including cosmic shear correlations, both types of GGL correlation then become useful to constrain galaxy intrinsic alignment model contributions \citep{Joachimi2011}, and we arrive at the full $\six$ analysis.

We turn now to the theory-vectors, $\vec{\mu}$, which are computed in identical fashion to the data-vectors, $\vec{d}$, but for different cosmological and nuisance parameters as part of the inference procedure. For each Stage-III forecast, the initial redshift distribution for computations of $\vec{\mu}$ is the `estimated' $N(z)$ (Sect. \ref{sec:stage_3_photometric}), whilst Stage-IV forecasts start from the distribution given by Eq. (\ref{eq:stage_4_nofz}) (and tomographically-binned; Sect. \ref{sec:stage_4_photometric}).

For Stage-III forecast configurations making use of the $\fcomb$ or $\hcomb$ models (Sect. \ref{sec:comb_spectra})\footnote{We note that whilst the comb model is considered for the Stage-III forecasts, its use would be excessive for our simple Stage-IV redshift bias implementation; hence we do not apply it to the Stage-IV forecasts in this work.}, we perform an additional set of steps prior to Monte Carlo sampling. First, the comb model is fit to the initial distribution $N(z)$ by minimisation of Eq. (\ref{eq:initial_comb_chi2}) via the Gaussian-component amplitudes $a^\mu_m=\ln(A^\mu_m)$ (see Eqs. \ref{eq:comb_model} and \ref{eq:comb_component}). The resulting initial $n_{\rm comb,ini}(z)$ is then used to compute a theory-vector at the fiducial cosmology\footnote{Real analyses will not have access to the fiducial cosmological/nuisance parameter set. As described by \cite{Stolzner2020}, this point must be found via an iterative maximum likelihood search, alternating between the spaces of cosmological and nuisance parameters.} and calculate the goodness of fit,  $\chi^2_{\vec{d}}$, with respect to the data-vector given the inverse of the signal covariance, $\tens{Z}$ (Eq. \ref{eq:data_chi2}). We now vary the comb amplitudes $a_m^{\mu}$ again to minimise $\chi^2_{\vec{d}}$, leveraging information from the `observed' two-point functions (which correspond to the true redshift distributions) to find the optimised comb model $n_{\rm comb,opt}(z)$\footnote{We note that the results of optimisation of the comb model are sensitive to the choice of minimisation algorithm and its hyperparameters. We homogenise these choices across all forecasts, using the Nelder-Mead \citep{NelderMead1965} algorithm with a maximum of $30\,000$ function evaluations, and suggest a more detailed exploration regarding flexible $n(z)$ model parameterisation in future work.}.

The results of this procedure for the forecast configuration ${\thr}:\nzcoh$ can be seen in Fig. \ref{fig:comb_vs_shifts}. This figure shows the errors in the redshift distributions, $n(z)-\nzcoh$, and the lensing efficiencies, $q(z)-q_{z}^{\rm coh}$ (see Eq. \ref{eq:lensing_efficiency}), for bins 1 and 3 of the Stage-III photometric sample after the application of no nuisance model ($N(z)$; dashed purple), of the shift model $\deltazi$ (dotted red), and of the comb model optimisation (solid green). Whilst the correct shift model can provide a reasonably effective correction to the lensing efficiency at higher redshifts, it is clearly seen to add little to the low-redshift bin. It is moreover seen to be generally detrimental to the shape of the $n(z)$, particularly at low redshifts where the $z=0$ boundary can cause a large diversion of power towards the tails upon renormalisation of the distribution. This is to say that a full-shape bias in the redshift distribution is smoothed out at the level of the lensing efficiency, and perhaps amenable to correction with a scalar shift; however, the same shift is likely to exacerbate the bias at the level of the density distribution. Meanwhile, the optimised comb model is seen to outperform the shift model both in recovering the true lensing efficiency, and in dealing minimal damage to the density distribution, irrespective of the redshift.

For comb model configurations, the optimised comb $n_{\rm comb,opt}(z)$ now takes the place of the initial redshift distribution $N(z)$. In the case of the $\hcomb$ model, which applies scalar shifts to $n_{\rm comb,opt}(z)$ during sampling, we apply Gaussian priors to the shifts with the same widths as inferred from the ensemble $\{n(z)\}_X$, but now centred on zero. This choice to re-centre the priors is made because the `correct' shift will have changed after optimisation of the comb model. Whilst we are able to simply recompute the correct shift here (and we do, for plotting purposes), this information is inaccessible to a real analysis where the true redshift distribution is unknown. Moreover, the $\hcomb$ model uses shifts more to marginalise over $n(z)$ uncertainty than as a corrective measure. Hence we limit the freedom of shifts during sampling to preserve a fair comparison with the shift model, and we explore the sensitivity of $\hcomb$ constraints to this re-centring by doubling, trebling, and flattening the shift priors, finding some variability in the resultant \emph{maxima a posteriori} (MAPs) for shift and cosmological parameters (to be discussed in more detail in Sect. \ref{sec:comb_model}). We note that more complex marginalisation schemes \citep[e.g.][]{Stolzner2020,Cordero2022} will be explored in future work. In parallel, other promising marginalization techniques for systematics (of and beyond photometric redshifts) are emerging which can significantly speed up the sampling of the likelihood  \citep{Ruiz23,Hadzhiyska23}.

We are now ready to sample the posterior probability distribution of cosmological and nuisance parameters under our various probe sets, redshift bias, and redshift nuisance model configurations, as we seek to quantify the advantages of using spectroscopic-photometric cross-correlations to aid with redshift recalibration and reduce cosmological parameter biases. We make use of the nested sampling algorithm, {\tt MultiNest} \citep{Feroz2008,Feroz2009,Feroz2019}, with the priors given in Table \ref{tab:priors}, and the following settings:

\begin{itemize}
    \item {\tt Maximum iterations}: 1e5 (S-III), 2e5 (S-IV);
    \item {\tt N live points}: 1e3;
    \item {\tt efficiency}: 0.3;
    \item {\tt constant efficiency}: False;
    \item {\tt tolerance}: 0.01.
\end{itemize}
For the shift and $\hcomb$ models, the redshift distribution is shifted at each sampling step according to Eq. (\ref{eq:shift_model}), prior to computation of $C(\ell)$'s and evaluation of the log-likelihood, $\mathcal{L}=-\chi^2_{\vec{d}}/2$ (Eq. \ref{eq:data_chi2}), whilst it remains fixed for the $\fcomb$ and `do nothing' models (Table \ref{tab:configurations}). All other modelling choices are held constant across all forecasts; the differences we explore here are purely due to the interaction of the probes under analysis, and the biases and modelling of redshift distributions.

Assessments of the recovery of point estimates for cosmological parameters are complicated by projection effects \citep{Troster2021}, and by the variability of estimation by different methods; e.g. global MAPs, marginal posterior means, or marginal posterior peaks \citep{KiDS1000DES}. Here, we quote posterior means, and compare the results of each forecast with the equivalent results for the respective idealised case; exactly zero redshift bias ($\nztrue$), and no applied nuisance model (`do nothing'). We also make estimates of the global MAP parameters for each forecast, starting the Nelder-Mead \citep{NelderMead1965} algorithm from the fiducial parameter set in order to avoid local minima in the parameter space. We thus verify that the MAP remains at the fiducial parameter set for idealised cases, and observe varying degrees of bias in the best-fit model for all other cases.

We present our findings in the following section, focusing for Stage-III on the accuracy and precision of recovery of the $S_8$ parameter by cosmic shear alone, and the $\seom$ plane for other probe combinations. This is done to reflect the fact that Stage-III cosmic shear constraints on $\Omega_{\rm m}$ are typically prior-dominated (e.g.~\citealt{Joudaki:2016mvz, Joudaki:2019pmv}). For $S_8$ alone, we quote the mean parameter, and the interval $\sigma_{68}$; the difference between the weighted 0.16 and 0.84 quantiles of the posterior distribution. 1-dimensional biases on $S_8$ (relative to the idealised case) are then given in ratio to $2\sigma_{68}$. For $\seom$, we quote the Figure of Merit (FoM), and the 2-dimensional parameter bias is calculated according to
\begin{equation}
    \chi^{2}_{S_8-\Omega_{\rm{m}}} =
    \begin{pmatrix}
    \widehat{S_8}-S_8\\
    \widehat{\Omega_{\rm{m}}}-\Omega_{\rm{m}}
    \end{pmatrix}^{\rm{T}}\,
    \tens{C}^{-1}_{S_8,\,\Omega_{\rm{m}}}\,
    \begin{pmatrix}
    \widehat{S_8}-S_8\\
    \widehat{\Omega_{\rm{m}}}-\Omega_{\rm{m}}
    \end{pmatrix}\,,
    \label{eq:cosmological_parameter_bias}
\end{equation}
where hats indicate the estimated statistic; unmarked parameters represent the target parameter values (those found by the idealised case); $\tens{C}_{S_8,\,\Omega_{\rm m}}$ denotes the covariance matrix of $S_8$ and $\Omega_{\rm m}$ as computed from the weighted chains; and we convert the deviance criterion $\chi^{2}_{S_8-\Omega_{\rm{m}}}$ first into a $p$-value, assuming a $\chi^2$ distribution with two degrees of freedom, and then into a significance given in units of $\sigma$. The FoM is computed as $\sqrt{\det \tens{C}^{-1}_{S_8,\,\Omega_{\rm m}}}$.

For our Stage-IV-like forecasts, we found that the typical non-linear `banana' degeneracy on the $\siom$ plane is no longer present, even for cosmic shear alone. We therefore switch to $\siom$ statistics when discussing all Stage-IV results, which are computed simply by swapping $\sigma_8$ in for $S_8$ in the above parameter bias (Eq. \ref{eq:cosmological_parameter_bias}) and FoM formulae.

\section{Results}
\label{sec:results}

\begin{figure*}
    \centering
    \includegraphics[width=0.8\textwidth]{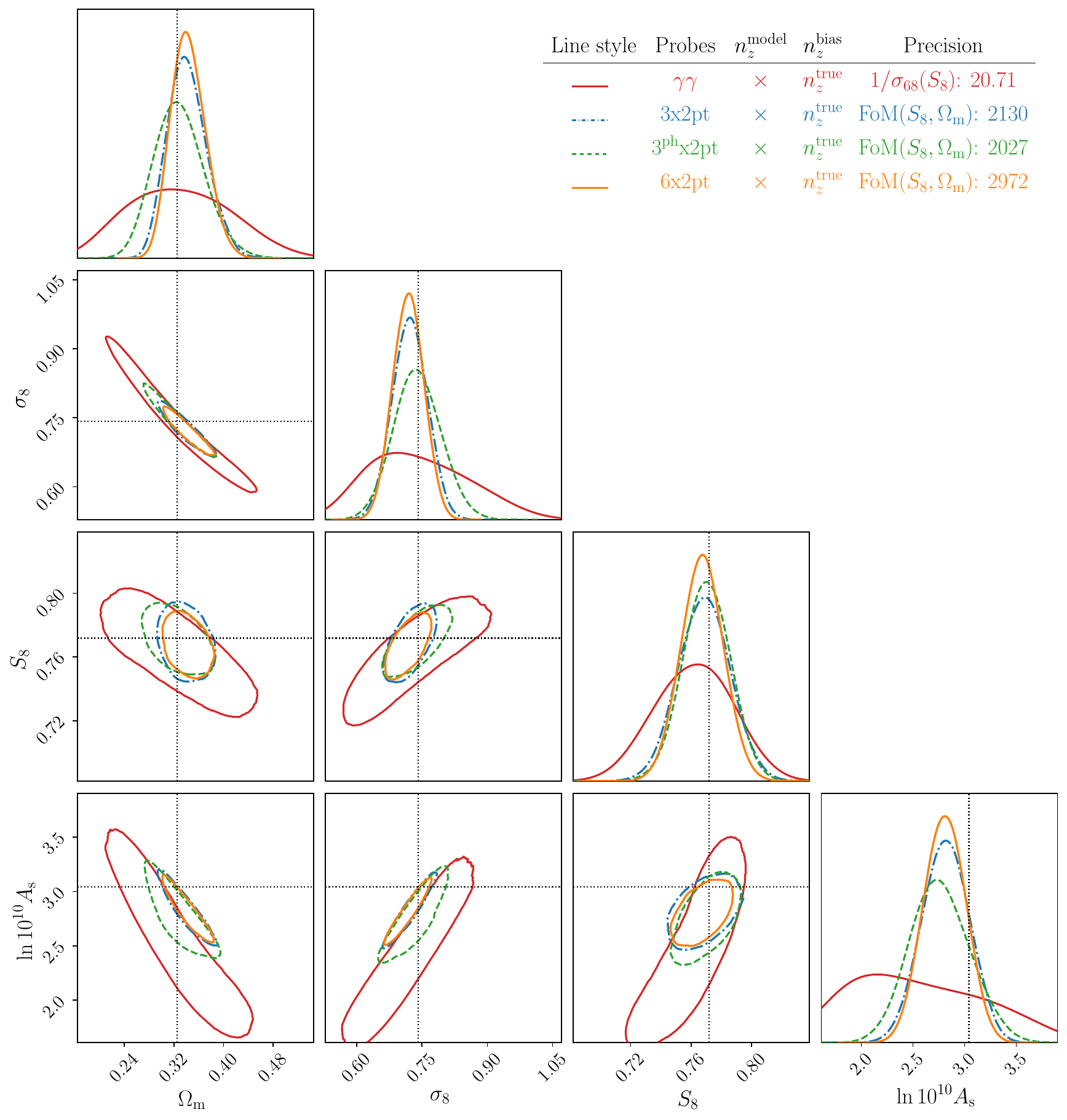}
    \caption{$1\sigma$ confidence contours for cosmological structure parameters $\Om,\sigma_8,S_8$, and $\lnas$, constrained by the forecast configurations having no redshift bias (Table \ref{tab:configurations}; `exactly zero') and applying no nuisance model to the redshift distributions (Table \ref{tab:configurations}; `do nothing'). Red contours and curves give the shear-only constraints ($\GG$); blue dot-dashed the spectroscopic $\thr$; green dashed the photometric $\pp$; and orange the full $\six$. Black dashed lines indicate the fiducial parameter centres (Table \ref{tab:priors}). Marginalised posterior probability distributions are seen to extend away from the fiducial centres due to projection effects (discussed in Sect. \ref{sec:results}), particularly in the case of $\lnas$.     }
    \label{fig:exact_nz_contours}
\end{figure*}

We summarise the results of our Stage-III forecasts in Table~\ref{tab:forecast_summary}, which gives for each $\{{\rm probes}:n(z)\,{\rm bias}:n(z)\,{\rm model}\}$ configuration; the signal-to-noise ratio S/N of the data-vector, calculated as a $\chi^2$ against a null signal hypothesis; the best-fit $\chi^2$ at the global MAP (Eq. \ref{eq:data_chi2}); and either the $S_8$ parameter bias and the inverse uncertainty $1/\sigma_{68}$ on $S_8$ (for $\gamma\gamma$), or the parameter bias (Eq \ref{eq:cosmological_parameter_bias}) and the FoM on the $\seom$ plane (for all other probe combinations).

We begin by considering the baseline Stage-III cosmological parameter constraints under idealised conditions, where the $n(z)$ bias is exactly zero, and no redshift nuisance model is marginalised over (bold rows in Table \ref{tab:forecast_summary}). The resulting $1\sigma$ confidence contours are shown in Fig. \ref{fig:exact_nz_contours} for the primary cosmological parameters constrained by weak lensing: $S_8,\,\Omega_{\rm m},\,\sigma_8$, where $S_8\equiv\sigma_8\sqrt{\Omega_{\rm m}/0.3}$ and $\sigma_8$ is derived from the $z=0$ linear power spectrum, emulated by {\tt CosmoPower} using the primordial amplitude $\lnas$ (also shown).

We note that whilst the fiducial centres (Table \ref{tab:priors}) are recovered by all configurations, the majority of the marginalised posterior volume for $\lnas$ extends away from the MAP towards lower values. Meanwhile, the posterior PDFs of late-time structure growth parameters $\sigma_8$ and $S_8$ are more centred upon the fiducial values. These are projection effects, where marginalisation over the constrained variables results in a preference for low $A_{\rm s}$, which is not constrained by late-time LSS probes (see \citealt{Troster2021}, who showed that fixing $A_{\rm s}$ during sampling of $S_8$ does not result in significantly different inferences for $S_8$ in the case of KiDS-1000, and also \citealt{Joudaki:2016mvz, Joudaki:2019pmv, Joachimi21, Longley22} for a discussion of $A_{\rm s}$ priors). We move forward considering only the constrained cosmological parameters.

The figure legend reports the inverse error on $S_8$ (for $\GG$ only; red solid) or the FoM on the $\seom$ plane for all other sets of Stage-III probe combinations: spectroscopic $\thr$ (blue dot-dashed); photometric $\thr$ ($\pp$; green dashed); and $\six$ (orange solid). 
In this first setup, our shear-only FoM is $\sim30\%$ larger than that found by \cite{Asgari21} owing to the lack of marginalisation over redshift nuisance parameters. We separately verify that our pipeline is able to reproduce the best-fit parameters and posterior distribution of \cite{Asgari21} to sub-percent accuracy when analysis choices (including the data-vector and covariance) are homogenised. 

In ratio to the idealised $\thr$ configuration, the FoM for the $\pp$ configuration is $\sim 5\%$ smaller. While the photometric density tracer is deeper and has fewer galaxy biases to constrain than the spectroscopic sample, it covers only $\sim1/12$ of the area. In the case of the $\six$ configuration, the FoM is  $\sim 1.4$ times larger (see Table \ref{tab:forecast_summary}). The $\six$ analysis might therefore offer up to a $\sim 40\%$ gain in constraining power, relative to a spectroscopic $\thr$ analysis under idealised conditions. 

The results are similar for the unbiased redshift distribution $\nzunb$ without any nuisance model, though we begin to see slight biases on the $\seom$ plane relative to the idealised cases. These increase in size for the $\pp$ and $\six$ configurations -- those which use the photometric sample as a density tracer -- by up to $\sim0.5\sigma$ depending on the probes/nuisance model (Table \ref{tab:forecast_summary}). The FoMs are almost unchanged with respect to the idealised scenarios given in Fig. \ref{fig:exact_nz_contours}. For both the $\pp$ and $\six$ configurations, we find weak correlations between the amplitude of intrinsic alignments $A_1$ and the primary cosmological parameters $\Om,S_8,\sigma_8$.
 
Not shown are the dimensionless Hubble parameter, $h$, primordial spectral tilt, $n_{\rm s}$, and {\tt HMCode2016} halo mass-concentration relation amplitude, $A_{\rm bary}$, as these are only weakly constrained by most of the probe configurations, and principally included for marginalisation purposes. The $\six$ configuration is the exception, where $h$ approaches a $1\sigma$ error of $\sim0.02$ and $n_{\rm s}$ begins to avoid the boundaries of the prior volume, the latter revealing correlations (of given sign) with $\Om\,(-),\,\lnas\,(+),\,\sigma_8\,(+),$ and the galaxy biases $(-)$. For next-generation experiments, extended LSS probe configurations such as the $\six$ could offer more competitive constraints on secondary cosmological variables.

\begin{table*}[!h]
    \centering
    \renewcommand{\arraystretch}{1.1}
    \tiny
    \begin{tabular}{c|c|c|r|r|r|r}
    \hline
    \hline
    Probes & $n(z)$ bias & $n(z)$ model & $\delta{}S_{8}\,[\sigma]$ & $1/\sigma_{68}(S_{8})$ & $\chi^{2}_{\rm MAP}\,$ & S/N\\
\hline
\textbf{\multirow{13}{*}{$\gamma\gamma$}} & \textbf{${\rm true}$} & \textbf{\multirow{4}{*}{${\rm do\,nothing}$}} & \textbf{0.00} & \textbf{20.71} & \textbf{0.00} & \textbf{592}\\
 & ${\rm unbiased}$ &  & -0.01 & 20.59 & 0.00 & 590\\
 & ${\rm incoherent}$ &  & -0.29 & 20.73 & 0.40 & 625\\
 & ${\rm coherent}$ &  & 0.32 & 20.00 & 0.59 & 676\\
\cline{2-7}
 & ${\rm unbiased}$ & \multirow{3}{*}{$\delta{z}_{i}$} & -0.09 & 20.50 & 0.00 & 590\\
 & ${\rm incoherent}$ &  & -0.10 & 20.35 & 0.61 & 625\\
 & ${\rm coherent}$ &  & -0.10 & 22.03 & 0.55 & 676\\
\cline{2-7}
 & ${\rm unbiased}$ & \multirow{3}{*}{$\mathcal{N}_{\rm comb}+\delta{z}_{i}$} & -0.09 & 20.25 & 0.00 & 590\\
 & ${\rm incoherent}$ &  & -0.05 & 20.40 & 0.00 & 625\\
 & ${\rm coherent}$ &  & -0.03 & 21.14 & 0.00 & 676\\
\cline{2-7}
 & ${\rm unbiased}$ & \multirow{3}{*}{$\mathcal{N}_{\rm comb}$} & 0.00 & 20.20 & 0.00 & 590\\
 & ${\rm incoherent}$ &  & -0.01 & 20.44 & 0.00 & 625\\
 & ${\rm coherent}$ &  & 0.04 & 21.47 & 0.00 & 676\\
\hline
Probes & $n(z)$ bias & $n(z)$ model & $\delta{}S_{8},\Omega_{\rm m}\,[\sigma]$ & FoM & $\chi^{2}_{\rm MAP}\,$ & S/N\\
\cline{2-7}
\hline
\textbf{\multirow{13}{*}{${\rm 3x2pt}$}} & \textbf{${\rm true}$} & \textbf{\multirow{4}{*}{${\rm do\,nothing}$}} & \textbf{0.00} & \textbf{2130} & \textbf{0.00} & \textbf{9854}\\
 & ${\rm unbiased}$ &  & 0.00 & 2136 & 0.03 & 9851\\
 & ${\rm incoherent}$ &  & 0.07 & 2044 & 1.27 & 9885\\
 & ${\rm coherent}$ &  & 0.35 & 2170 & 1.32 & 9933\\
\cline{2-7}
 & ${\rm unbiased}$ & \multirow{3}{*}{$\delta{z}_{i}$} & 0.01 & 2041 & 0.02 & 9851\\
 & ${\rm incoherent}$ &  & 0.04 & 2091 & 1.43 & 9885\\
 & ${\rm coherent}$ &  & 0.15 & 2189 & 1.97 & 9933\\
\cline{2-7}
 & ${\rm unbiased}$ & \multirow{3}{*}{$\mathcal{N}_{\rm comb}+\delta{z}_{i}$} & 0.00 & 2098 & 0.05 & 9851\\
 & ${\rm incoherent}$ &  & 0.00 & 2144 & 0.09 & 9885\\
 & ${\rm coherent}$ &  & 0.00 & 2196 & 0.14 & 9933\\
\cline{2-7}
 & ${\rm unbiased}$ & \multirow{3}{*}{$\mathcal{N}_{\rm comb}$} & 0.00 & 2131 & 0.05 & 9851\\
 & ${\rm incoherent}$ &  & 0.01 & 2121 & 0.09 & 9885\\
 & ${\rm coherent}$ &  & 0.00 & 2181 & 0.14 & 9933\\
\cline{2-7}
\hline
\textbf{\multirow{13}{*}{${\rm 3^{ph}x2pt}$}} & \textbf{${\rm true}$} & \textbf{\multirow{4}{*}{${\rm do\,nothing}$}} & \textbf{0.00} & \textbf{2027} & \textbf{0.00} & \textbf{7146}\\
 & ${\rm unbiased}$ &  & 0.23 & 2055 & 4.47 & 7094\\
 & ${\rm incoherent}$ &  & 2.36 & 1480 & 46.97 & 7099\\
 & ${\rm coherent}$ &  & 2.21 & 1783 & 76.52 & 7888\\
\cline{2-7}
 & ${\rm unbiased}$ & \multirow{3}{*}{$\delta{z}_{i}$} & 0.00 & 1809 & 2.06 & 7094\\
 & ${\rm incoherent}$ &  & 0.55 & 1721 & 19.41 & 7099\\
 & ${\rm coherent}$ &  & 0.56 & 2325 & 29.96 & 7888\\
\cline{2-7}
 & ${\rm unbiased}$ & \multirow{3}{*}{$\mathcal{N}_{\rm comb}+\delta{z}_{i}$} & 0.09 & 1825 & 0.41 & 7094\\
 & ${\rm incoherent}$ &  & 0.23 & 1956 & 0.83 & 7099\\
 & ${\rm coherent}$ &  & 0.35 & 2097 & 0.87 & 7888\\
\cline{2-7}
 & ${\rm unbiased}$ & \multirow{3}{*}{$\mathcal{N}_{\rm comb}$} & 0.22 & 2041 & 0.50 & 7094\\
 & ${\rm incoherent}$ &  & 0.02 & 1980 & 0.98 & 7099\\
 & ${\rm coherent}$ &  & 0.17 & 2220 & 1.10 & 7888\\
\cline{2-7}
\hline
\textbf{\multirow{13}{*}{${\rm 6x2pt}$}} & \textbf{${\rm true}$} & \textbf{\multirow{4}{*}{${\rm do\,nothing}$}} & \textbf{0.00} & \textbf{2972} & \textbf{0.00} & \textbf{16134}\\
 & ${\rm unbiased}$ &  & 0.21 & 3046 & 8.60 & 16087\\
 & ${\rm incoherent}$ &  & 1.01 & 2942 & 79.42 & 16107\\
 & ${\rm coherent}$ &  & 2.05 & 3080 & 104.36 & 16102\\
\cline{2-7}
 & ${\rm unbiased}$ & \multirow{3}{*}{$\delta{z}_{i}$} & 0.01 & 2787 & 6.34 & 16087\\
 & ${\rm incoherent}$ &  & 0.33 & 2598 & 43.12 & 16107\\
 & ${\rm coherent}$ &  & 0.33 & 2658 & 67.90 & 16102\\
\cline{2-7}
 & ${\rm unbiased}$ & \multirow{3}{*}{$\mathcal{N}_{\rm comb}+\delta{z}_{i}$} & 0.37 & 2997 & 14.21 & 16087\\
 & ${\rm incoherent}$ &  & 0.01 & 2851 & 24.38 & 16107\\
 & ${\rm coherent}$ &  & 0.02 & 2658 & 39.18 & 16102\\
\cline{2-7}
 & ${\rm unbiased}$ & \multirow{3}{*}{$\mathcal{N}_{\rm comb}$} & 0.49 & 3300 & 14.31 & 16087\\
 & ${\rm incoherent}$ &  & 0.02 & 3008 & 24.53 & 16107\\
 & ${\rm coherent}$ &  & 0.14 & 2899 & 39.54 & 16102\\

    \hline
    \hline
    \end{tabular}
    \caption{Summary statistics for our Stage-III forecasts, conducted with various sets of cosmological probes, redshift biases, and redshift recalibration nuisance models (left-most columns; see also Table \ref{tab:configurations}). The remaining columns give: the posterior mean parameter bias (in units of $\sigma$, see the end of Sect. \ref{sec:forecast}) on $S_8$, or on the $\seom$ plane (Eq. \ref{eq:cosmological_parameter_bias}), relative to the idealised case (${\rm probes}:\nztrue:{\rm do\,nothing}$; boldface rows); the inverse error $1/\sigma_{68}$ on $S_8$, or the $\seom$ Figure of Merit (FoM); the best-fit $\chi_{\vec{d}}^2$ at the MAP parameter set (Eq. \ref{eq:data_chi2}); and the S/N of the data-vector (hence the numbers for different nuisance models are the same when sharing the same data-vector). Numbers for forecasts conducted with exactly zero redshift bias ($\nztrue$) are given in boldface. As expected, we see a perfect recovery of the fiducial parameters by the MAP when the bias is exactly zero, signified by zero $\chi^2$ (some biased shear-only cases show zero $\chi^2$ due to rounding).
    }
    \label{tab:forecast_summary}
\end{table*}

We now turn to the impacts of our simulated redshift calibration failures, described in Sect. \ref{sec:stage_3_photometric}. 

\subsection{No nuisance model}
\label{sec:do_nothing_model}

\begin{figure*}
    \centering
    \includegraphics[width=0.8\textwidth]{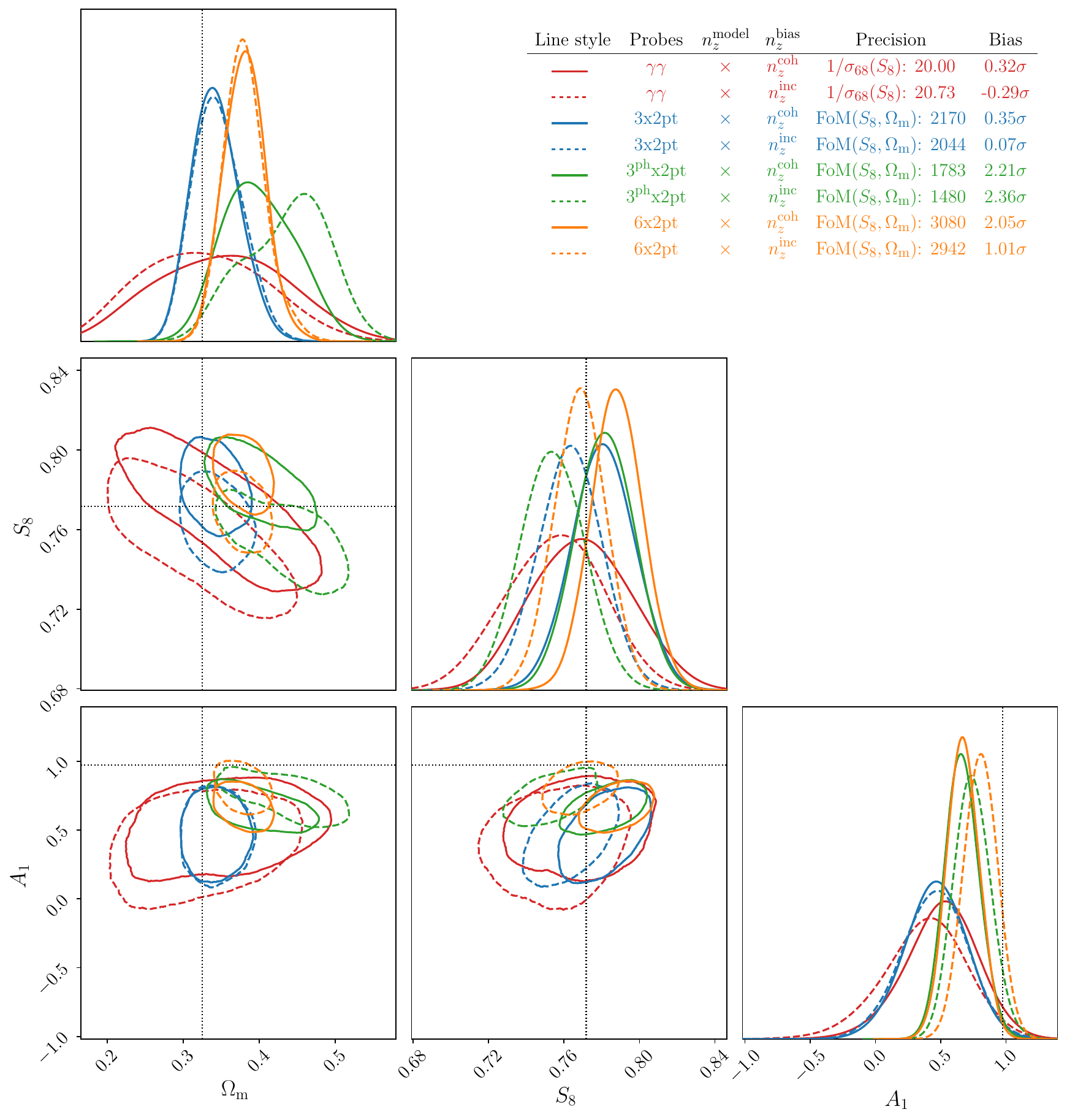}
    \caption{$1\sigma$ confidence contours for cosmological structure parameters $\Om,\sigma_8,S_8$, and the intrinsic alignment amplitude $A_1$, constrained by the $\GG, \, \thr, \, \pp$, and $\six$ forecast configurations with `incoherently' (dashed contours/curves) and `coherently' (solid contours/curves) biased redshift distributions, and no nuisance model (`${\rm do\,nothing}$') for redshift recalibration (see Table \ref{tab:configurations} for a summary of forecast configurations). Colours here denote probe configurations as in Fig. \ref{fig:exact_nz_contours} (also given by the legend), whilst line styles denote the type of redshift bias (as described in Sect. \ref{sec:stage_3_photometric}). Dashed black cross-hairs display the fiducial, true parameter values. The legend gives for each contour set: the probe combination; the true redshift distribution (i.e. the type of redshift bias); the $n(z)$ nuisance model (given as `$\times$' here, for the `${\rm do\,nothing}$' model); the FoM on the $\seom$ plane  (Sect.~\ref{sec:forecast}); and the parameter bias in the same plane (comparing posterior means to their equivalents for the $\nztrue:{\rm do\,nothing}$ cases, Sect. \ref{sec:forecast}; Eq. \ref{eq:cosmological_parameter_bias}). Significant mis-estimations are seen in almost all cases, with variable errors according to the type of redshift bias.
    }
    \label{fig:coh_incoh_donothing_contours}
\end{figure*}

\begin{figure*}
    \centering
    \includegraphics[width=0.9\textwidth]{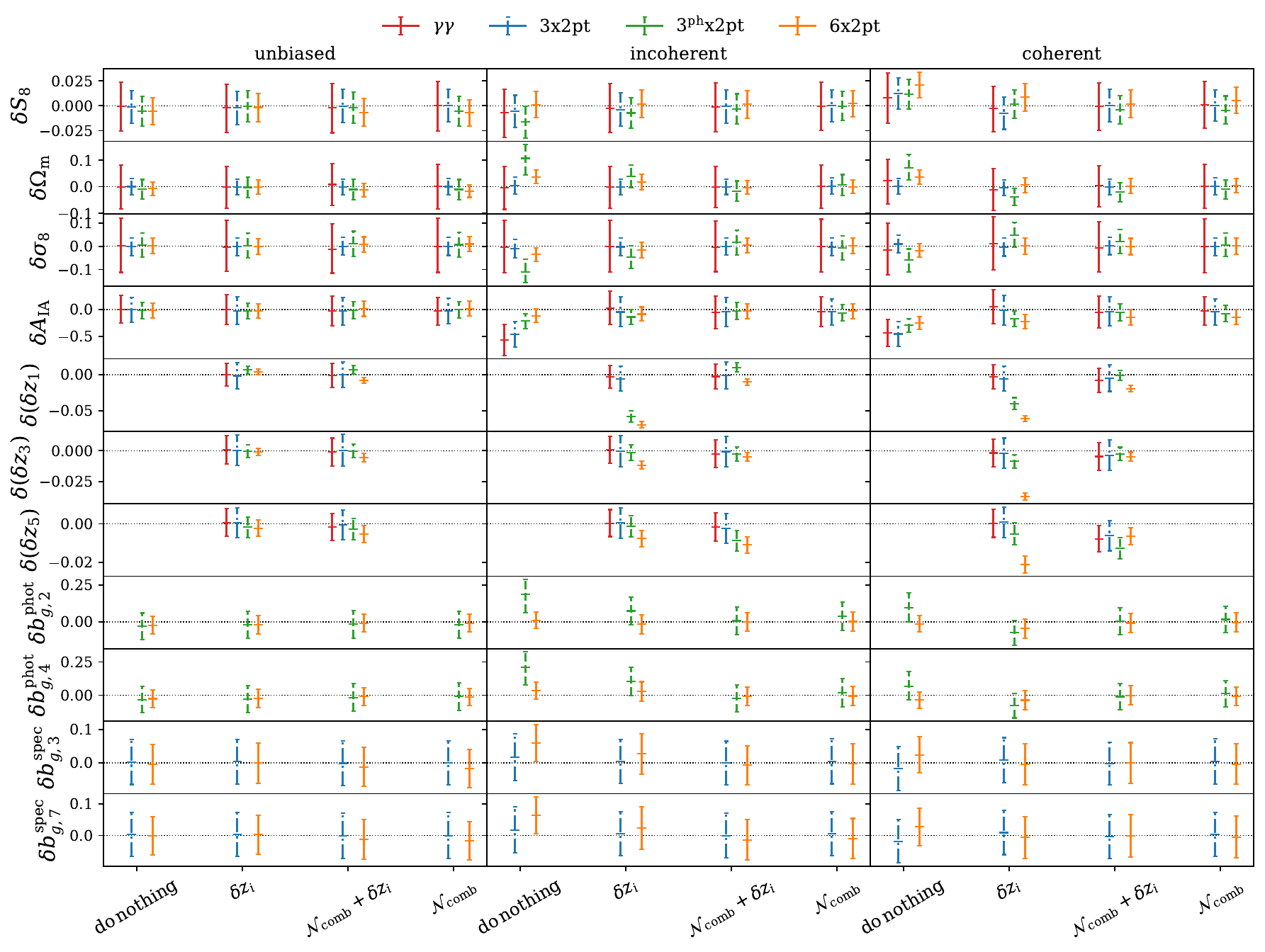}
    \caption{Marginalised one-dimensional constraints for cosmological and (a selection of) nuisance parameters, found by each of our Stage-III forecast configurations: $\GG, \, \thr, \, \pp$, and $\six$, with colours and line-styles as in Fig. \ref{fig:exact_nz_contours} (also given in the legend). The three columns give the unbiased, incoherently biased, and coherently biased forecasts, respectively, and points are grouped on the $x$-axis according to the nuisance model employed by each forecast (see Table \ref{tab:configurations} for a summary of forecast configurations). Error bars illustrate the $\sigma_{68}$ interval, within which the posterior means are given as horizontal markers. $y$-axes then give the difference  $\delta{p}=p-p_{\rm ideal}$ for parameter $p$ relative to the idealised ($\nztrue:{\rm do\,nothing}$) case for each probe set (except for shift model parameters $\deltazi$, where the difference is given relative to the fiducial truth, i.e. the actual differences in the tomographic means, as we did not conduct shift model forecasts for $\nztrue$). 
    }
    \label{fig:1d_constraints}
\end{figure*}

Let us first consider forecasts where we attempt no recalibration of the biased photometric sample tomographic redshift distributions (the `do nothing' model; Table \ref{tab:configurations}). In Fig. \ref{fig:coh_incoh_donothing_contours}, we display $1\sigma$ confidence contours (swapping $\lnas$ for the NLA amplitude $A_1$ in the figure) for both the incoherently biased case ($\nzinc$, dashed contours) and the coherently biased case ($\nzcoh$, solid contours). Colours denote the same probe sets as in Fig. \ref{fig:exact_nz_contours}. 
Recall that our biased redshift distributions (Fig. \ref{fig:redshift_distributions}; coloured curves) modify the target data-vector, and that the starting distribution informing the theory-vector is $N(z)$ (black curves in Fig. \ref{fig:redshift_distributions}). In Fig. \ref{fig:coh_incoh_donothing_contours} (and subsequent contour figures) we isolate errors in cosmological inference arising from redshift biases, as opposed to projection effects, by quoting parameter biases relative to those observed for the corresponding idealised case; the $\nztrue:{\rm do\,nothing}$ case given in Fig. \ref{fig:exact_nz_contours}, for which the target distribution $\nztrue$ is identical to the starting distribution $N(z)$.

For several of these forecasts, we see significant biases on the $\seom$ plane when failing to recalibrate the biased redshift distributions. Each of the probe configurations reveals the coherent redshift bias to manifest more strongly on the $\seom$ plane than the incoherent bias -- with the exception of $\pp$, where the biases are similar in magnitude. This is due to an increased dependence upon the full shape of the redshift distribution for $\pp$ and $\six$, where the former derives a large proportion of its constraining power from photometric auto-clustering. The incoherently biased distribution $\nzinc$ features the largest overall deviation in the full shape (as reckoned by $\Delta_{\rm full}$; Eq.~\ref{eq:delta_full}), and results in a more significant bias on the $\seom$ plane for the $\pp$ configuration.

The amplitude of intrinsic alignments, $A_1$, is significantly under-estimated by all probe configurations, for each of the two redshift biases. In this controlled scenario of known and realistic redshift calibration failures, with all else held constant, we thus demonstrate that the IA model compensates for the redshift errors \citep{vanUitert18,Li2021,Fischbacher2022}. The under-estimation of $A_1$ has the net effect of limiting the GI suppression of the shear signal, more significantly for the lower-redshift and off-diagonal correlations (e.g. between bins 1 and 5) where the GI contribution is stronger\footnote{The II contribution is generally subdominant to GI, particularly at higher redshifts from which the bulk of the constraining power is derived.}. The inferred shear signal is therefore asymmetrically boosted, mimicking the shift of lower-$z$ distributions to higher redshifts seen in the upper panels of Fig. \ref{fig:redshift_distributions}. For consistency, we have separately verified that in a shear-only case where the redshift distributions are coherently biased towards lower redshifts, the $A_1$ amplitude is correspondingly biased high.

The shear-only (Fig. \ref{fig:coh_incoh_donothing_contours}; red contours) and spectroscopic $\thr$ (blue contours) configurations are the most robust to redshift errors, as might be expected given their reliance upon the photometric redshift distribution only at the level of the lensing efficiency $q(\chi)$. Through a combination of this broad, less sensitive kernel, and a biased IA amplitude $A_1$ (bottom panels) to improve the fit to the shear correlations, the shear-only configuration remains biased at just \{$\rm incoherent:-0.29\sigma$, $\rm coherent:0.32\sigma$\} in $S_8$. Extending to $\thr$, the unaffected spectroscopic clustering anchors the $\sigma_8$ (and thus $\Om$) constraints (left panels), yielding biases at \{$\rm incoherent:0.07\sigma$, $\rm coherent:0.35\sigma$\} on the $\seom$ plane.

Both the $\pp$ (green contours) and $\six$ (orange contours) configurations show substantial biases on the $\seom$ plane. These derive from over-estimations of both $S_8$ and $\Om$ in the coherent cases ($\pp:2.21\sigma$, $\six:2.05\sigma$). In the incoherent cases, they derive from (over-) under-estimation of ($\Om$) $S_8$ for the $\pp$ ($2.36\sigma$) configuration, and over-estimation of $\Om$ for the $\six$ configuration ($1.01\sigma$). These results demonstrate that the sensitivity of photometric density probes to unmitigated full-shape errors in the redshift distribution has a significant and persistent impact upon cosmological parameter inference.

The inability of the $\pp$ to constrain $\Om$, and hints of bimodality in the posterior (Fig. \ref{fig:coh_incoh_donothing_contours}; top-left panel, green curves), conspire to degrade the FoM by $\sim 20-30\%$ relative to the $\thr$ configuration, whilst the $\six$ configurations gain more than $40\%$ in each of the in/coherent bias cases. 

Fig.~\ref{fig:1d_constraints} displays the one-dimensional posterior means and $\sigma_{68}$ intervals for various cosmological and nuisance parameters (rows), relative to the idealised ($\nztrue:{\rm do\,nothing}$) case for each probe set. Columns give the three redshift bias scenarios (titles), with the four groups of points in each panel corresponding to the applied redshift nuisance models ($x$-axis labels), and colours/line-styles denoting the same probe configurations as in Fig. \ref{fig:exact_nz_contours} (also given in legend).

Under the `do nothing' model (left-most points in the column), the incoherently biased redshift distribution (middle column), without recalibration, causes the $\pp$ and $\six$ configurations to under-estimate $\sigma_8$ and over-estimate $\Om$. This behaviour stems from the forms of the biased correlations, a subset of which are shown in Fig. \ref{fig:dvec_figure} (note that the impacts of redshift biases upon correlations vary across the data-vector). For the lower redshift bins, where $\nzinc$ is most deviant from $N(z)$ (Fig. \ref{fig:redshift_distributions}), photometric clustering statistics typically demand a lower amplitude of correlation to match $\nzinc$ (orange curves in Fig.~\ref{fig:dvec_figure}), such that either $\sigma_8$ or the galaxy bias $\bgphot$ must decrease (increasing $\Om$ changes the scale dependence of the clustering power spectrum). 

Meanwhile, the spectroscopic and low-$z$ photometric\footnote{Increases in the shear sample redshifts can only boost the GGL signal amplitude when the lenses are spectroscopic, i.e. fixed in redshift. However, simultaneous increases to the lens redshifts -- as in the case of photometric GGL where the shear sample is also the density sample -- source reductions in the GGL signal amplitude due to decreased galaxy bias at higher redshift. The net effect is for the photometric GGL signal to be boosted at low-to-intermediate $z$ (particularly in the coherent bias case) where the gain in lensing efficiency is dramatic, and slightly reduced at high-$z$, where it is more modest compared with the loss of galaxy bias. We note that this manifestation of balance between lensing efficiency and lens bias may be particular to our sample definitions.} GGL statistics see an increase in signal as the shear sample is pushed to higher redshift, thus placing upward pressure on $\bgphot$; $\sigma_8$ must decrease to satisfy the clustering, whilst $\bgphot$ must increase to satisfy the GGL, and $\Om$ is forced high (and $A_1$ low) in order to maintain the shear correlation amplitude. This results in a bias on the $\seom$ plane, and over-estimations of the galaxy bias, both photometric and spectroscopic, to compensate for the reduced $\sigma_8$. 

In the case of the $\pp$ configuration, the photometric galaxy bias is not anchored by cross-correlations with spectroscopic samples; larger increases in $\bgphot$ force $\sigma_8$ even lower, and result in an under-estimation of $S_8$. 
For the $\six$ configuration, the cross-correlations forbid such large values for the photometric galaxy bias; decreases in $\sigma_8$ are limited, and $S_8$ remains unbiased.

In the coherent bias scenario, the redshift distributions are similar to those of the incoherent bias for the first two tomographic bins but are then shifted more significantly for the last three bins (Fig.~\ref{fig:redshift_distributions}). As a result, the coherently biased data-vector requires higher shear correlation amplitudes at intermediate-to-high redshifts, due to the significant offsets in the tomographic means. In this scenario, a severely under-estimated alignment amplitude $A_1$ is unable to provide a sufficient boost to the higher-$z$ correlations (which dominate the cosmic shear signal-to-noise). While $S_8$ remained unbiased for the incoherent case (at least for the $\six$ configuration), here it must assume a higher value in order to describe the high-$z$ shear correlations. This is achieved through a slight increase in the inferred value of $\sigma_8$ (relative to the incoherent case), with clustering reduction demands compensated by reductions in $\bgphot$, whilst $\Om$ is held at a similar value to that of the incoherent case; hence $S_8$ increases and the $\seom$ bias is increased relative to the incoherent case. Incidentally, the $\bgphot$ are better recovered in the coherent case than the incoherent, even though we might expect $\nzinc$ to more strongly affect clustering statistics and thus $\bgphot$ inferences, given its larger full-shape deviation from the starting distribution ($\Delta_{\rm full}$; Eq. \ref{eq:delta_full}). 

These forecasts demonstrate that, {\it when left untreated}, different types of errors in redshift calibration can cause the inference of cosmological parameters to be variably biased according to different combinations of weak lensing and density probes. Parameter constraints from configurations using the biased $n(z)$ only for shear statistics (cosmic shear and spectroscopic $\thr$) are less sensitive to incoherent shifting of redshift distributions, whilst those making use of the same $n(z)$ for density statistics can suffer large errors in the inference of $\sigma_8,\Om,$ and $b_{\rm g}$. 

In the shear-only, $\thr$, and $\six$ incoherently-biased cases, we find that $S_8$ remains relatively unbiased, but that this results from the IA model compensating for the redshift uncertainties. While this may be specific to our choice of a biased redshift distribution, \citet{Leonard24} have found similar results for the $\thr$ in their analysis. If the mean redshift offsets (Fig. \ref{fig:redshift_distributions}; vertical orange lines) were not decreasing with redshift (i.e. were not primarily manifested in the first two tomographic bins, in contrast to the coherent bias case, which shifts the mean in every bin), then the resulting errors in shear two-point functions would possibly be less amenable to correction by a modified IA contribution. This is part of the challenge for the coherently biased forecasts, which show larger parameter biases for most configurations, partially due to the inability of the IA model to absorb the redshift bias. Moreover, the form of contamination of $A_1$ here is dictated by the predominant direction of the redshift errors, which place the true distributions at higher redshifts; as previously mentioned, an opposite redshift bias -- where the photometric redshifts are overestimated -- demands lower shear correlation amplitudes, and with all else held constant, forces $A_1$ high so as to suppress the shear correlations.

We now turn to the application of nuisance models that attempt to correct for mis-estimation of galaxy redshift distributions.

\subsection{Shift model}
\label{sec:shift_model}

\begin{figure*}
    \centering
    \includegraphics[width=0.8\textwidth]{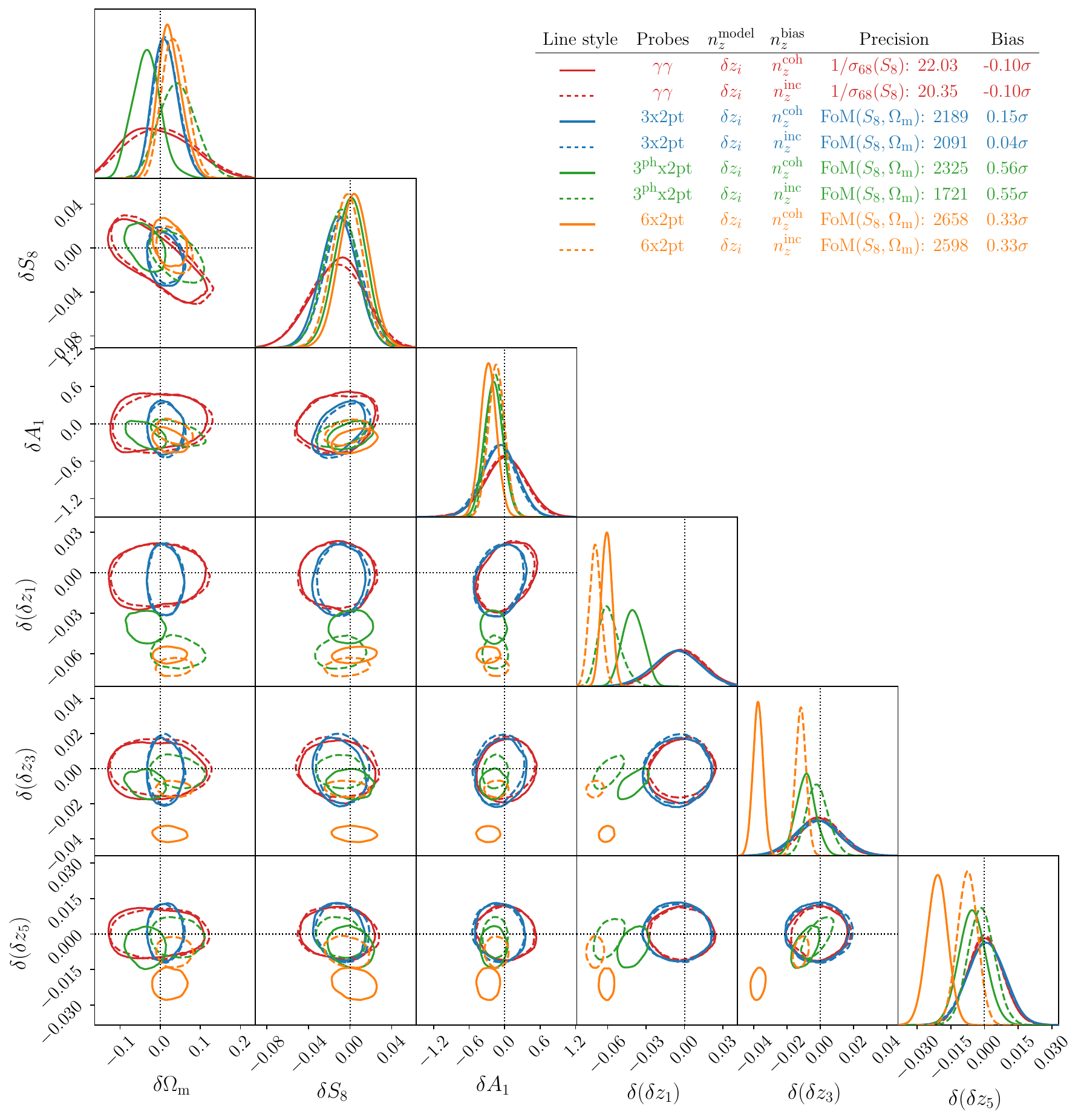}
    \caption{$1\sigma$ confidence contours for cosmological and nuisance parameters, as in Fig. \ref{fig:coh_incoh_donothing_contours}, now for forecasts employing the shift model for internal redshift recalibration (Sect. \ref{sec:generic_spectra}) -- we thus include contours for the first, third, and fifth scalar shifts, $\deltazi$, and exclude the power spectrum normalisation, $\sigma_8$, in the figure for increased visibility. We show here the difference $\delta{p}={}p-p_{\rm ideal}$ for parameter $p$ relative to the idealised ($\nztrue:{\rm do\,nothing}$) case for each probe set (or the fiducial truth, i.e. the actual difference in the tomographic means, in the case of $\deltazi$ parameters), since the correct shifts $\deltazi$ depend upon the in/coherent form of the redshift bias. Most $\deltazi$ forecasts see a reduction in bias and a loss of Figure of Merit on the $\seom$ plane (given by the legend) relative to the `do nothing' model (Fig. \ref{fig:coh_incoh_donothing_contours}).
    }
    \label{fig:coh_incoh_shiftmodel_contours}
\end{figure*}

\begin{figure*}
    \centering
    \includegraphics[width=\textwidth]{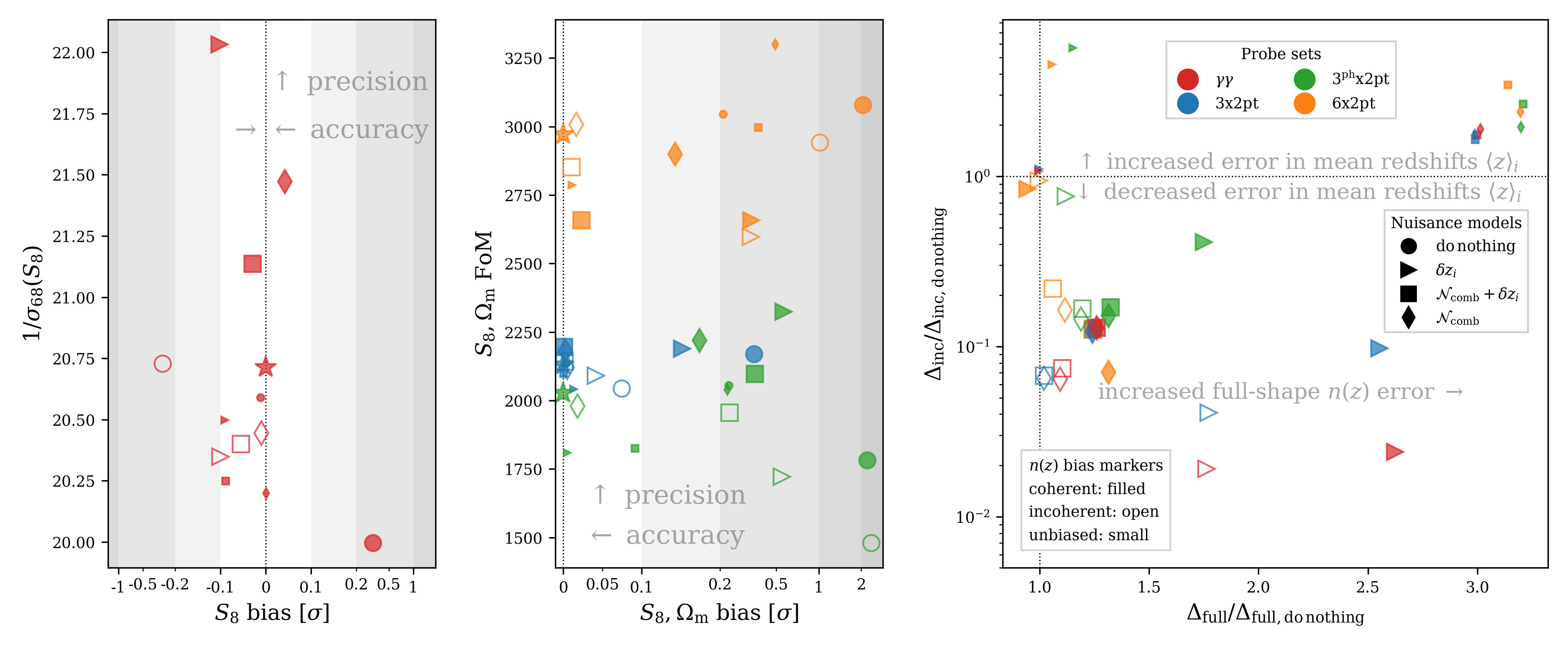}
    \caption{Summary statistics describing the accuracy and precision of the cosmological parameter inference, and corresponding recovery of the true redshift distribution via internal recalibration, for each of our Stage-III forecast configurations. Colours denote different probe sets, whilst marker-styles give the form of the redshift bias and the applied redshift nuisance model (legends in right-panel; star markers denote the $\nztrue:{\rm do\,nothing}$ case for each probe set). 
    \emph{Left/middle:} The accuracy and precision of the cosmological constraints are given in terms of the $S_8$ bias and inverse $\sigma_{68}$ for shear-only (\emph{left}), and the two-dimensional parameter bias and the Figure of Merit (Sect. \ref{sec:forecast}) on the $\seom$ plane for other probe sets (\emph{middle}). Biases are given relative to the idealised case for each respective probe set (see Eq. \ref{eq:cosmological_parameter_bias} for the 2-d bias). The $x$-axes ($S_8$ or $\seom$ bias) scale linearly between $0$ and $|0.2\sigma|$, and logarithmically thereafter. \emph{Right:} The $n(z)$ recovery is characterised by the redshift distribution bias metrics $\Delta_{\rm full}$ and $\Delta_{\rm incoherent}$ defined in Sect. \ref{sec:stage_3_photometric} (Eqs. \ref{eq:delta_incoh_coh} \& \ref{eq:delta_full}). Each describes the deviance of the recovered $n(z)$ with respect to the true distribution ($\equiv\nzinc$ for open points, $\nzcoh$ for filled points, or $\nzunb$ for small points) via a Euclidean distance, with $\Delta_{\rm full}$ computed over the full $n(z)$ shapes, and $\Delta_{\rm incoherent}$ computed between the 5-vectors of tomographic mean redshifts. These are given after the respective normalisation to $\Delta_{\rm incoherent}$ and $\Delta_{\rm full}$ found for the `do nothing' model (Table~\ref{tab:configurations}), corresponding to circles in the left/middle-panels. Circles/stars are not shown in the right-panel for the `do nothing' model, where each would sit at the centre of the dotted cross-hair. The spread of points shows that maximal parameter accuracy (towards zero in the left-panel, and leftward in the middle panel) and precision (upward in the left/middle-panels) most often come from a balance between reductions in $\Delta_{\rm incoherent}$ and minimal increases in $\Delta_{\rm full}$ (towards the bottom-left in the right-panel), and that this is most achievable through use of the comb model (squares and diamonds), particularly for configurations that probe photometric densities (green and orange). 
    }
    \label{fig:nz_err_vs_constraints}
\end{figure*}

\begin{figure*}
    \centering
    \includegraphics[width=\textwidth]{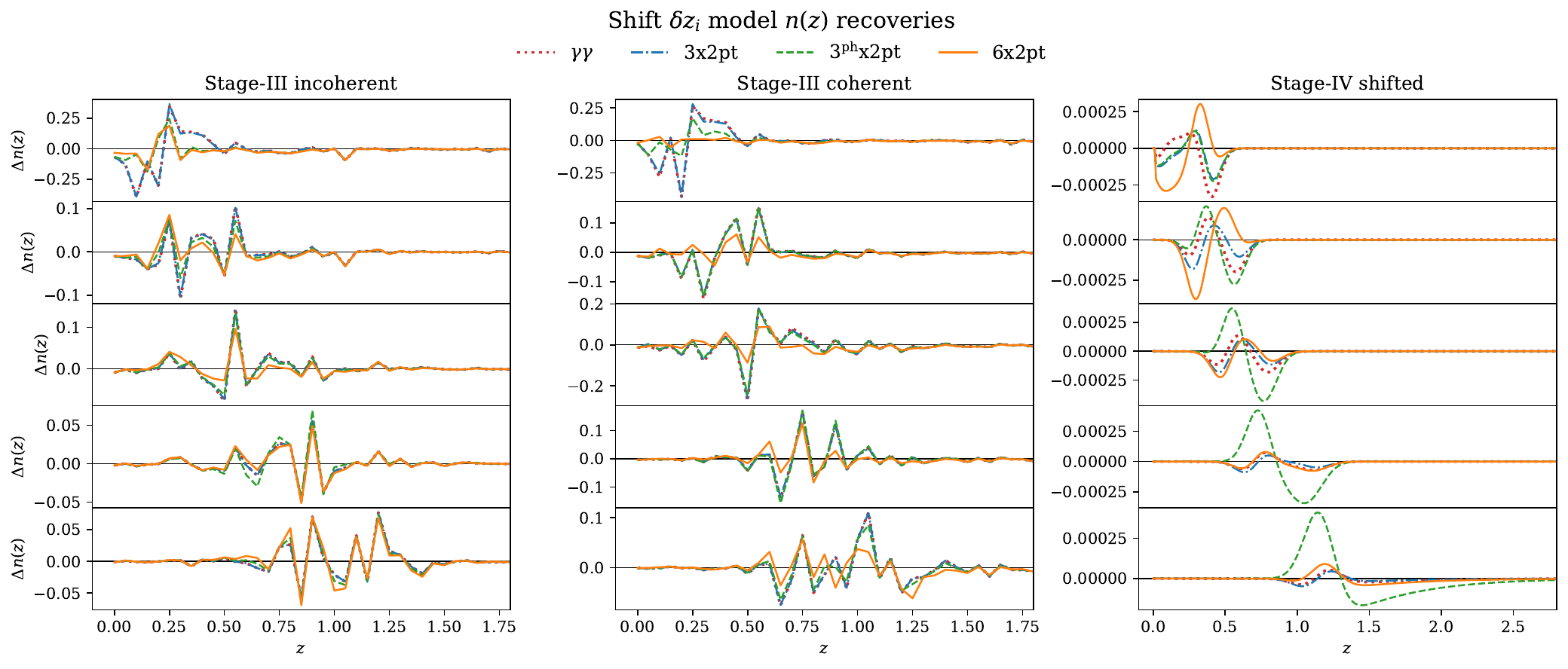}
    \caption{The residual five-bin (\emph{rows}, 1-5 descending) tomographic redshift distribution errors, $\Delta{}n(z)$, found for the best-fit (MAP) shift models (Sect. \ref{sec:modelling_choices}) constrained by each of our Stage-III and Stage-IV redshift-biased forecasts (see Table \ref{tab:configurations} for a summary of configurations). Colours and line-styles give the probe configurations as in previous figures (except $\GG$, indicated here by dotted curves for clarity; see legend). For the Stage-III configuration, columns show the attempted recoveries of the incoherently (\emph{left}) and coherently (\emph{middle}) biased redshift distributions (Sect. \ref{sec:stage_3_photometric}), which feature full-shape $n(z)$ deviations. For the Stage-IV configurations (\emph{right}), the shifted distribution features pure shifts in the tomographic means (Sect. \ref{sec:stage_4_photometric}). The shift model enjoys generally superior recoveries of the true $n(z)$ upon inclusion of spectroscopic-photometric cross-clustering correlations within the $\six$ configurations. We note that the pure-shift biases examined for the Stage-IV forecasts are well-compensated (residual errors are at the sub-percent level), and that the shift model offers a comparatively more accurate treatment of these Stage-IV biases than the full-shape $n(z)$ errors in the Stage-III forecasts, where the largest residuals exceed $100\%$.
    }
    \label{fig:nz_recovery}
\end{figure*}

\begin{table}
    \tiny
    \centering
    \renewcommand{\arraystretch}{1.1}
    \begin{tabular}{lccccc}
    \hline
    \hline
    \multirow{2}{*}{Configuration} & $\sigma_{68}(\delta z_1)$ & $\sigma_{68}(\delta z_2)$ & $\sigma_{68}(\delta z_3)$ & $\sigma_{68}(\delta z_4)$ & $\sigma_{68}(\delta z_5)$ \\
 & $\times{}10^2$ & $\times{}10^2$ & $\times{}10^2$ & $\times{}10^2$ & $\times{}10^2$ \\
\hline
Stage-III &  &  &  &  &  \\
$\gamma\gamma:\rm coh.$ & 3.28 & 1.98 & 2.25 & 1.71 & 1.45 \\
$\gamma\gamma:\rm incoh.$ & 3.18 & 1.94 & 2.20 & 1.71 & 1.39 \\
${\rm 3x2pt}:\rm coh.$ & 3.44 & 2.13 & 2.47 & 1.88 & 1.58 \\
${\rm 3x2pt}:\rm incoh.$ & 3.42 & 2.17 & 2.53 & 1.86 & 1.57 \\
${\rm 3^{ph}x2pt}:\rm coh.$ & 1.59 & 1.16 & 1.10 & 1.00 & 1.14 \\
${\rm 3^{ph}x2pt}:\rm incoh.$ & 1.58 & 1.28 & 1.23 & 1.14 & 1.11 \\
${\rm 6x2pt}:\rm coh.$ & 0.78 & 0.60 & 0.58 & 0.63 & 0.87 \\
${\rm 6x2pt}:\rm incoh.$ & 0.88 & 0.60 & 0.61 & 0.63 & 0.85 \\
\hline
Stage-IV &  &  &  &  &  \\
$\gamma\gamma:\rm shift.$ & 1.21 & 1.67 & 2.34 & 3.08 & 5.14 \\
${\rm 3x2pt}:\rm shift.$ & 0.91 & 0.83 & 0.89 & 1.01 & 1.67 \\
${\rm 3^{ph}x2pt}:\rm shift.$ & 0.47 & 0.56 & 0.62 & 0.65 & 0.99 \\
${\rm 6x2pt}:\rm shift.$ & 0.25 & 0.19 & 0.16 & 0.16 & 0.40 \\

    \hline
    \hline
    \end{tabular}
    \caption{The sizes of intervals $\sigma_{68}(\deltazi)$ for shift parameters $\deltazi$, as constrained by each of our redshift-biased forecast configurations (left-most column), each multiplied by $10^2$ for clarity. One sees dramatic gains in precision coming purely from including spectroscopic-photometric cross-clustering in the $\six$.
    }
    \label{tab:shift_precision}
\end{table}

Let us assess whether or not the shift $\deltazi$ model -- the currently-preferred method of redshift distribution recalibration in weak lensing analyses \citep{Hikage19,Hamana:2019etx, Asgari21,Amon2022,Secco2022} -- is sufficient to correct redshift biases in our Stage-III forecasts (Stage-IV $\deltazi$ forecasts are discussed in Sect. \ref{sec:stage_4_forecasts}). 
The parameter contours are displayed in Fig.~\ref{fig:coh_incoh_shiftmodel_contours} for both biased redshift scenarios. This includes $\Om$, $S_8$, and $A_1$, as previously, along with a subset of the shift parameters $\deltazi$.

We first note that there is a reduction in bias on the $\seom$ plane (and in $S_8$ alone) in all of the $\deltazi$ forecasts relative to the `do nothing' model. Despite the concerns highlighted in Sect. \ref{sec:forecast} \& Fig. \ref{fig:comb_vs_shifts}, the shift model is unlikely to significantly worsen the accuracy of the $S_8$ and $\seom$ inference -- though we also note that our applied Gaussian priors for $\deltazi$ parameters are optimistic in their construction, each reflecting the actual difference in the mean tomographic redshifts (see Sect. \ref{sec:stage_3_photometric}).

As in Sect.~\ref{sec:do_nothing_model}, the coherent redshift bias tends to manifest more strongly in the parameter inference than the incoherent bias. We note with interest that several non-$\six$ configurations are able to \emph{gain} in FoM with respect to their `do nothing' model counterparts (Fig.~\ref{fig:coh_incoh_donothing_contours} and Table~\ref{tab:forecast_summary}), despite marginalising over five additional parameters on application of the shift model.
This is due to an interplay between (i) degradation of constraints due to the addition of parameters, and (ii) increased signal-to-noise in the shear-sensitive parts of the theory-vector, owing to the application of positive shift parameters (aided by Gaussian priors; Sect \ref{sec:stage_3_photometric}) that increase the mean tomographic redshifts of the photometric sample. For some forecasts (notably the coherently-biased $\GG$ and $\pp$ configurations), the latter effect greatly outweighs the former, and results in significantly increased constraining power. This is reflected in the data-vector S/N (Table \ref{tab:forecast_summary}), which reveals the $\nzcoh$ data-vector to be far more constraining than corresponding $\nztrue$ and $\nzunb$ data-vectors for $\GG$ and $\pp$. Conversely, for $\nzunb$ configurations, the shifts $\deltazi$ are close to zero; the application of the shift model only results in reduced constraining power relative to the $\rm do\,nothing$ cases.

In the case of the shift model, with optimistically constructed 
priors for parameters $\deltazi$ (Table \ref{tab:priors}; Sect. \ref{sec:stage_3_photometric}), the resulting theory-vectors are translated to higher redshifts where possible. This releases much of the pressure on the intrinsic alignment amplitude $A_1$ to boost the shear signal amplitude by itself (as the photometric redshift distributions of the data and theory have effectively been forced to overlap in terms of the mean), and Figs. \ref{fig:1d_constraints} \& \ref{fig:coh_incoh_shiftmodel_contours} consequently reveal much-improved recoveries of the true alignment amplitude by the $\GG$ and $\thr$ shift model configurations.

However, $\pp$ (green) and $\six$ (orange) contours in Fig.~\ref{fig:coh_incoh_shiftmodel_contours} begin to reveal the deficiencies of the shift model for configurations depending on photometric density tracers, between which correlations are far more sensitive to the shape of the $n(z)$ than its mean. Where shift parameters $\deltazi$ are able to effectively mitigate the bias in the tomographic mean redshifts for $\GG$ and $\thr$ configurations (red/blue contours, lower-right panels), resulting in $S_8$ and $\seom$ biases at $\leq0.15\sigma$, they are less able to do so for the $\pp$ and $\six$ configurations, yielding biases at $\sim0.5\sigma$ and $\sim0.3\sigma$, respectively.

Figs. \ref{fig:1d_constraints} \& \ref{fig:coh_incoh_shiftmodel_contours} show that, whilst the `true shifts' -- the actual differences in tomographic mean redshifts between $N(z)$ and $\nzcoh$ or $\nzinc$ -- are accurately recovered by the shift model for the $\GG$ and $\thr$ configurations, they are significantly under-estimated in many cases by the $\pp$ and in all cases by the $\six$ configurations, where the latter also constrains the shifts to as much as $4\times$ greater precision (Table \ref{tab:shift_precision}). As highlighted in Sect.~\ref{sec:forecast} and Fig.~\ref{fig:comb_vs_shifts}, the correct shifts help the $\GG$ and $\thr$ configurations by reducing the error in the lensing efficiency $q(z)$, but they \emph{amplify} errors in the full-shape redshift distribution $n(z)$. The shift model therefore has a disproportionately negative impact upon density statistics -- principally the photometric clustering -- that are sensitive not only to the mean but also to the shape of the redshift distributions.

The $\pp$ configuration is able to use the shifts to fit higher amplitudes from shear correlations (particularly at at high-$z$), yielding minor improvements to the recovery of $A_1$ (relative to the `do nothing' case), and a sharp increase in FoM. However, these shifts cause a strong suppression of the density statistics, resulting in a compensatory increase in $\sigma_8$ (coupled to a decrease in $b_{\rm g}$; Fig.~\ref{fig:1d_constraints}), and (to maintain the shear amplitude) a decrease in $\Om$. These changes are so large for the $\nzcoh$ case, which requires larger $\deltazi$ at intermediate-to-high-$z$, that the signs of biases in $\sigma_8, \, \Om$, and $b_{\rm g}$ are flipped relative to the `do nothing' case (Fig. \ref{fig:1d_constraints}). 

Meanwhile, the $\six$ configuration cannot use the shift model to improve the fit to the shear correlations because the spectroscopic-photometric cross-clustering would then be poorly described; the $\deltazi$ parameters break out of the Gaussian prior, and are constrained to be close to zero. As a result, the $\six$ configuration continues to underestimate $A_1$, thereby boosting the shear amplitude, and remains largely unbiased on the $\seom$ plane for both $\nzinc$ and $\nzcoh$ -- but at the cost of some $11-13\%$ of its FoM, thanks to the addition of shift parameters that are unaccompanied by any increases in S/N.

We further elucidate these trends by considering the redshift distribution bias metrics defined in Sect. \ref{sec:stage_3_photometric}, $\Delta_{\rm incoherent}$ and $\Delta_{\rm full}$, now computed for the final redshift distributions with best-fit nuisance models applied (the trends derived for $\Delta_{\rm incoherent}, \, \Delta_{\rm coherent}$ are quite similar, hence we focus only on $\Delta_{\rm incoherent}$ here). The left/middle panels in Fig. \ref{fig:nz_err_vs_constraints} show for our Stage-III forecasts the inverse $\sigma_{68}(S_8)$ and FoM on the $\seom$ plane vs. the respective parameter biases (relative to the idealised case for each respective probe set) according to Eq. \ref{eq:cosmological_parameter_bias} (for the 2-d bias). The righthand panel then shows the corresponding final deviation in tomographic mean redshifts $\Delta_{\rm incoherent}$ vs. the final full-shape deviation $\Delta_{\rm full}$, each normalised to the relevant untreated (`do nothing') case.

Marker styles give the nuisance model and redshift bias scenarios as outlined in the middle-right and bottom-left legends, respectively, whilst colours denote the probe configurations as in previous figures (also given in the top legend). Circular points denote the `do nothing' model, and these are not shown in the right-hand panel, where each would sit at the centre of the cross-hair. Parameter bias thresholds of $0.1\sigma, \, 0.2\sigma, \, 1\sigma, \, 2\sigma$ are illustrated in the first two panels with grey shading, such that the unshaded regions indicate higher accuracy of cosmological parameter inference. The precision of parameter inference increases towards the top of these panels. In the right-hand panel, mean redshift $\meanz_i$ errors decrease towards the bottom, whilst full-shape $n(z)$ errors decrease towards the left of the panel.

Each non-circular point in Fig. \ref{fig:nz_err_vs_constraints} can be thought of as an attempt to improve upon the corresponding circle (or the cross-hair) via some redshift nuisance model, calibrated by the various two-point data-vectors (denoted by different colours). The figure shows that, whilst the best-fit shift model is able to reduce the cosmological parameter bias in all cases (triangles compared to circles in left/middle-panels), it yields a much poorer recovery of the full-shape $n(z)$ (as reckoned by $\Delta_{\rm full}$) when not constrained by the cross-clustering (orange as compared to red, blue, and green triangles; right-panel). Under the $\six$ configurations, which do include the cross-clustering, each shift $\deltazi$ is constrained to a significantly smaller magnitude at much higher precision. 
The shift model is thus only able to produce small reductions in $\Delta_{\rm incoherent}$ and $\Delta_{\rm full}$ -- though the latter being the only such reductions seen for any nuisance model (only the orange triangles are leftward of the vertical dotted line in the righthand panel).

The impact of the best-fit shifts upon the $n(z)$ recoveries can be seen in Fig. \ref{fig:nz_recovery}, where the first two columns show the final redshift distribution error $\Delta{}n(z)$ for the Stage-III in/coherently biased forecast configurations.
In almost every panel, the $\six$ best-fit shifts yield the closest recovery of the true redshift distribution, followed by the photometric density-probing $\pp$, and then the photometric shear-only configurations $\GG$ and $\thr$. 
The relative precision of shift parameter constraints is given in Table \ref{tab:shift_precision}, which quotes $\sigma_{68}$ for shifts $\deltazi$, for each of the shift model configurations. The $\six$ probe combination is seen to yield constraints on the shift $\deltazi$ parameters that are up to $\sim4$ times more precise (depending on the tomographic bin) than the spectroscopic $\thr$ constraints.

The right panel of Fig. \ref{fig:nz_err_vs_constraints} demonstrates that, whilst the shift model (triangles) can be effective in mitigating errors in tomographic mean redshifts, it does so at the cost of increased errors in the full shape of the $n(z)$. This is not a problem for probe combinations with weak sensitivity to the $n(z)$ shape, like $\GG$ (red) and $\thr$ (blue), but for configurations probing the photometric density field -- $\pp$ (green) and $\six$ (orange) -- such increases cannot be tolerated. The $\six$ in particular, anchored by the cross-clustering, does not allow the shift parameters $\deltazi$ to assume significantly non-zero values, thus yielding little change in $\Delta_{\rm inc}$ and $\Delta_{\rm full}$ relative to the $\rm do\,nothing$ model (cross-hair). For these configurations, more accurate cosmological inference is achieved through the use of a more flexible nuisance model: the Gaussian `comb' model (squares \& diamonds; Sect. \ref{sec:comb_spectra}), to which we now turn.

\subsection{Comb model}
\label{sec:comb_model}

\begin{figure*}
    \centering
    \includegraphics[width=0.8\textwidth]{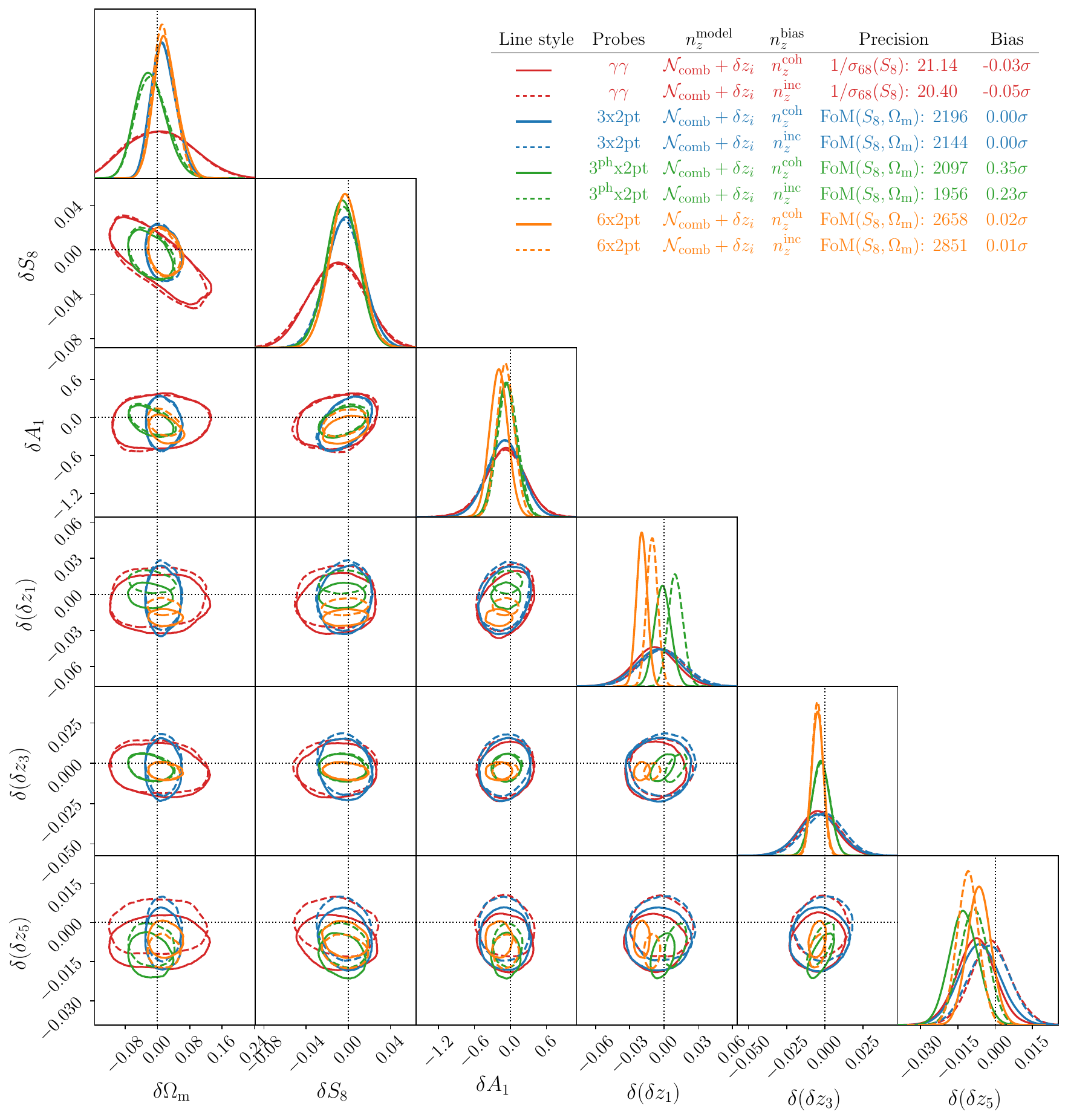}
    \caption{$1\sigma$ confidence contours for cosmological nuisance parameters, as in Fig. \ref{fig:coh_incoh_shiftmodel_contours}, now for forecasts employing the comb+shift model $\hcomb$ for internal redshift calibration (Sect. \ref{sec:comb_spectra}). The comb+shift recalibration reduces biases on the $\seom$ plane with respect to the `do nothing' (Fig. \ref{fig:coh_incoh_donothing_contours}) and shift (Fig. \ref{fig:coh_incoh_shiftmodel_contours}) models, and recovers some of the FoM lost to the shift model even whilst marginalising over scalar shifts.
    }
    \label{fig:coh_incoh_combmodel_contours}
\end{figure*}

Fig. \ref{fig:coh_incoh_combmodel_contours} displays confidence contours for $\Om, \, S_8, \, A_1, \, \deltazi$, as in Fig. \ref{fig:coh_incoh_shiftmodel_contours}, now for the comb+shift ($\hcomb$) model -- this model finds an optimised redshift distribution $n_{\rm comb,opt}(z)$ via Eq. (\ref{eq:data_chi2}) before likelihood sampling (see Sects.~\ref{sec:comb_spectra}~\&~\ref{sec:forecast}), and then applies the shift model to the optimised $n(z)$ during sampling.

First, considering the $\seom$ plane (top-left), the in/coherently biased forecast constraints are almost identical for a given probe configuration, and almost completely unbiased, at $\lesssim0.1\sigma$ in most cases. The exception is the $\pp$ case, where the maximum bias is still relatively small at $0.35\sigma$. This is a significant result, as the data-vectors differ fundamentally between $\nzinc$ and $\nzcoh$, in such ways as to source the various biases found in previous sections, and yet the ${\hcomb}$ model is able to homogenise the accuracy and precision of the constraints (as is the $\fcomb$ model, not shown).

This mitigation of redshift bias is effective regardless of the probe configuration. As seen in the right panel of Fig. \ref{fig:nz_err_vs_constraints} for every $\nzcoh$ and $\nzinc$ configuration (large squares and diamonds), the comb models are able to reduce the error in $\Delta_{\rm inc}$, whilst causing minimal damage to $\Delta_{\rm full}$ relative to the corresponding shift models (triangles; some of which are unable to significantly improve $\Delta_{\rm inc}$, as discussed in the previous section). We reiterate that the shifts are still in place for the ${\hcomb}$ model (squares), and yet the $n(z)$ recovery is similar to that seen for the $\fcomb$ model (diamonds); the gain, therefore, comes almost exclusively from the flexible optimisation of the $n(z)$ against the data-vector, and is not predicated upon the inclusion of cross-clustering correlations (though they help to further improve the recovery). This is also reflected in the final redshift distributions; our analysis of the best-fit $\hcomb$ and pure $\fcomb$ models shows smaller $\Delta{}n(z)$ performance differentials between the $\six$ and less complex configurations.

We find some differences in the constraints according to the form of the redshift bias. First, $A_1$ is still slightly under-estimated by the ${\six}:\nzcoh:{\hcomb}$ configuration (at $\sim1\sigma$; see the third group of points in the right panel of Fig. \ref{fig:1d_constraints}), though to a lesser extent than for the `do nothing' and $\deltazi$ models (biased at $\lesssim1.5-2\sigma$). This would suggest that the recovery of $\nzcoh$ by the optimised comb model -- whilst superior to the shift model (see Fig.~\ref{fig:nz_err_vs_constraints}) -- is not quite sufficient to yield large enough shear amplitudes, such that the alignment contribution must continue to compensate. In future work, we intend to increase the flexibility of the comb model -- first by increasing the number and decreasing the width, of Gaussian components in the mixture model -- in order to alleviate the contamination of the alignment model.

We further find that the best-fit shifts, $\deltazi$, found by the $\hcomb$ configurations are still not equal to the `correct' shifts (recalculated after optimisation of the comb to reflect the actual difference in the tomographic mean redshifts), now also for the coherently biased $\thr$ and $\GG$ forecasts.
Instead, they are uniformly consistent with zero, for all tomographic bins across all forecasts. This is partially due to the application of zero-mean Gaussian priors for the shifts $\deltazi$ (with the same widths $\sigma_i$ carried over from the ensemble $\{n(z)\}_X$; Sect. \ref{sec:stage_3_photometric}). As discussed in Sect. \ref{sec:forecast}, this was done to preserve a fair comparison between the $\hcomb$ model and the shift model (for which priors limit the freedom of $\deltazi$) but also because the original centres for the shifts would be inappropriate after optimisation of the comb.\footnote{One could compute new non-zero centres for the shift priors after comb optimisation, but this would involve returning to the hypothetical external redshift calibration. Instead, we assume that the mean correction has been applied by the comb optimisation, and proceed with zero-mean priors for the shifts.}

To investigate the sensitivity of $\hcomb$ \emph{maxima a posteriori} models to the recentring (to zero-mean) of the $\deltazi$ priors, we experiment with priors of $2\times$ and $3\times$ the width; flat priors between $\pm3\sigma_i$; and with no priors (i.e. a pure maximum-likelihood search). We find that the application of zero-mean priors results in stronger reductions of $\Delta_{\rm incoherent}$, and smaller biases on the $\seom$ plane, in all cases. When the priors are widened/removed, some configurations start to prefer non-zero shifts $\deltazi$, but this always results in a larger $\seom$ bias, and $\six$ configurations always prefer post-comb shifts that are consistent with zero. 
This suggests that widening the priors for shifts in the $\hcomb$ model does not necessarily yield the correct mean redshifts in the end, and that the primary correction occurs during the comb optimisation, as previously indicated.

We test this interpretation by also considering the comb model without shifts, simply fixing the optimised $n(z)$ for use by the sampler. Under most $\nzcoh$ and $\nzinc$ configurations, both the $\hcomb$ and $\fcomb$ models (squares \& diamonds in Fig. \ref{fig:nz_err_vs_constraints}) see reduced parameter bias as well as matched or increased precision in comparison with the corresponding shift model (triangles). In addition, the $\fcomb$ tends to make small gains in precision relative to the corresponding $\hcomb$ model (squares), whilst typically not appreciably losing accuracy. The only exceptions are $\GG : \nzcoh$ and ${\pp}:\nzcoh$, where the shift model reports higher precision, though at the cost of increased parameter bias. Hence, the strong performance of either of the two comb models relative to the shift model reinforces our expectation that a more flexible recalibration of biased redshift distributions is generally preferred over the application of scalar shifts to the tomographic mean redshifts.

That said, we identify shortcomings in our implementation of comb models when applied to the \emph{unbiased} data-vector (corresponding to $\nzunb$; Sect. \ref{sec:stage_3_photometric}), particularly for the $\six$ configuration. In this case, the resulting redshift distribution poorly recovers $\nzunb$ compared to the shift or `do nothing' models (top-right of right-panel in Fig.~\ref{fig:nz_err_vs_constraints}), resulting in best-fit parameters that are biased up to $\sim0.5\sigma$ on the $\seom$ plane. 
As Fig. \ref{fig:comb_vs_shifts} shows, the optimised comb (solid green curves) outperforms the shift model (dotted red curves) in recovering both the lensing efficiency $q(z)$ and the redshift distribution $n(z)$, but it cannot avoid slightly amplifying the error in $n(z)$ relative to that seen for the unmodified distribution $N(z)$ (purple dashed curves), as the fine structure is effectively smoothed-over by the Gaussian components. When the true distribution is unbiased relative to the starting distribution, this causes both the initial and optimised comb models to significantly worsen the already-small redshift distribution errors $\Delta{}n(z)$ -- as seen for the small points in top-right corner of the right-panel in Fig. \ref{fig:nz_err_vs_constraints}, where $\Delta_{\rm full}$ is effectively a sum over the square of $\Delta{}n(z)$. 
Put differently, the redshift-space width of comb components is not sufficiently narrow to capture the intricate structure of the redshift distribution, and the `do nothing' case consequently offers a closer recovery. We expect that a smoother redshift distribution, or a finer resolution for comb components, could alleviate this concern.

A higher-resolution comb model might be more successful in handling very small redshift distribution errors $\Delta{}n(z)$, though we note that there may be associated trade-offs to consider; the optimisation procedure would probably be more time-consuming, and a higher density of comb components could increase the space of degenerate $n(z)$ solutions for a given data-vector. Some tuning of the comb resolution \citep[as in][]{Stolzner2020} is thus likely to be necessary for the description of redshift distributions with varying degrees of bias, under various combinations of cosmological probes.

We note that the comb model is not alone in failing to improve the $n(z)$ recovery for an initially unbiased estimate of the distribution -- all small symbols in Fig.~\ref{fig:nz_err_vs_constraints} are located upward (and many rightward) of the cross-hair, signifying that the `do nothing' model, i.e. $N(z)$ unmodified, offers the best recovery of $\nzunb$ in terms of both the mean redshifts and the full shape (as reckoned by our metrics $\Delta_{\rm incoherent}$ and $\Delta_{\rm full}$).

The requirement for tuning of a flexible redshift distribution model, and the possibility of \emph{introducing} biases into pristinely-calibrated prior distributions, further motivate the need for more flexible $n(z)$ nuisance modelling, and for marginalisation over uncertainty in the $n(z)$. As previously discussed (Sect. \ref{sec:comb_spectra}), we do not implement an analytic marginalisation over the comb model uncertainty \citep[see][]{Stolzner2020} in this work, owing to development needed for extended analyses, and questions concerning appropriate priors for comb amplitudes -- though analytic marginalisation is certainly a primary avenue for improvement.

Another possibility is the {\tt Hyperrank} approach of \cite{Cordero2022} (see similar approaches in \citealt{Hildebrandt2016} and \citealt{Zhang2022}), which consistently and efficiently marginalises over a space of possible redshift distributions \citep[generated through the external calibration procedure;][]{Myles2021} by rank-ordering them via hyper-parameters (typically defined as functions of the tomographic mean redshifts $\meanz_i$) which are then sampled-over in the chain. Inspired by this approach, we suggest a complementary strategy making use of the comb, or other flexible $n(z)$ models. Since the data-vector $\vec{d}$ and corresponding covariance are typically given ultimate authority to arbitrate between possible recalibrations of the $n(z)$, we propose that the data-vector itself be used to generate the $n(z)$ ensemble. Re-sampling $\vec{d}$ many times, each sample could be used to constrain a flexible $n(z)$ model -- e.g. the optimised comb model -- assuming the best-fit cosmological/nuisance model for the fiducial data-vector. The resulting ensemble of $n_{\rm comb,opt}(z)$ realisations would then sample the space of redshift distributions preferred by the observational data, which could be efficiently marginalised over during sampling of the cosmological/nuisance model with the {\tt Hyperrank} method.

Such an approach could be implemented for any flexible $n(z)$ model, and a comparison of constraints derived with marginalisation over the externally-calibrated $n(z)$ ensemble \citep[as done by][]{Cordero2022} against a data-vector-driven $n(z)$ ensemble would provide a useful internal consistency check for a given survey. The former rests upon redshift calibration methods, whilst the latter would be sensitive to aspects of data-vector and covariance estimation, e.g. shear measurements and beyond-Gaussian covariance contributions. Ultimately, a hybrid of the two approaches could be considered, where the former allows for a prior to be imposed on the latter in the absence of cross-covariances.

\subsection{Stage-IV forecasts}
\label{sec:stage_4_forecasts}

\begin{table*}
\tiny
    \centering
    \renewcommand{\arraystretch}{1.5}
    \begin{tabular}{c|c|c|r|r|r|r}
\hline
\hline
Probes & $n(z)$ bias & $n(z)$ model & $\delta{}\sigma_{8},\Omega_{\rm m}\,[\sigma]$ & FoM & $\chi^{2}_{\rm MAP}\,$ & S/N\\
\hline
\textbf{\multirow{3}{*}{$\gamma\gamma$}} & \textbf{${\rm true}$} & \textbf{\multirow{2}{*}{${\rm do\,nothing}$}} & \textbf{0.00} & \textbf{31197} & \textbf{0.00} & \textbf{42620}\\
 & ${\rm shifted}$ &  & 12.57 & 35055 & 118.20 & 41856\\
\cline{2-7}
 & ${\rm shifted}$ & \multirow{1}{*}{$\delta{z}_{i}$} & 0.32 & 6084 & 0.00 & 41856\\
\cline{2-7}
\hline
\textbf{\multirow{3}{*}{${\rm 3x2pt}$}} & \textbf{${\rm true}$} & \textbf{\multirow{2}{*}{${\rm do\,nothing}$}} & \textbf{0.00} & \textbf{82854} & \textbf{0.00} & \textbf{116120}\\
 & ${\rm shifted}$ &  & 11.40 & 63358 & 275.58 & 115112\\
\cline{2-7}
 & ${\rm shifted}$ & \multirow{1}{*}{$\delta{z}_{i}$} & 0.00 & 48111 & 0.00 & 115112\\
\cline{2-7}
\hline
\textbf{\multirow{3}{*}{${\rm 3^{ph}x2pt}$}} & \textbf{${\rm true}$} & \textbf{\multirow{2}{*}{${\rm do\,nothing}$}} & \textbf{0.00} & \textbf{80312} & \textbf{0.00} & \textbf{248263}\\
 & ${\rm shifted}$ &  & 9.64 & 88409 & 2108.51 & 222605\\
\cline{2-7}
 & ${\rm shifted}$ & \multirow{1}{*}{$\delta{z}_{i}$} & 0.01 & 44556 & 0.02 & 222605\\
\cline{2-7}
\hline
\textbf{\multirow{3}{*}{${\rm 6x2pt}$}} & \textbf{${\rm true}$} & \textbf{\multirow{2}{*}{${\rm do\,nothing}$}} & \textbf{0.00} & \textbf{96626} & \textbf{0.00} & \textbf{315421}\\
 & ${\rm shifted}$ &  & 6.02* & - & 4932.95 & 316056\\
\cline{2-7}
 & ${\rm shifted}$ & \multirow{1}{*}{$\delta{z}_{i}$} & 0.00 & 88546 & 0.08 & 316056\\

\hline
\hline
    \end{tabular}
    \caption{The same as Table \ref{tab:forecast_summary}, but now for our Stage-IV forecast configurations (left-most column; see also Table \ref{tab:configurations}). Note that, as discussed at the end of Sect.~\ref{sec:forecast}, we now quote statistics for the $\siom$ plane for all configurations including $\GG$, given the increase in $\Om$ constraining power, and reduction in non-linearity of the $\siom$ plane, for Stage-IV-like statistics. 
    We also note that the ${\six}:\nzshift:{\rm do\,nothing}$ forecast was unable to achieve convergence when sampling the posterior distribution; the parameter bias is quoted in this case for the MAP model (marked by an asterisk), and the FoM is accordingly omitted.
    }
    \label{tab:forecast_summary_s4}
\end{table*}

\begin{figure*}
    \centering
    \includegraphics[width=0.8\textwidth]{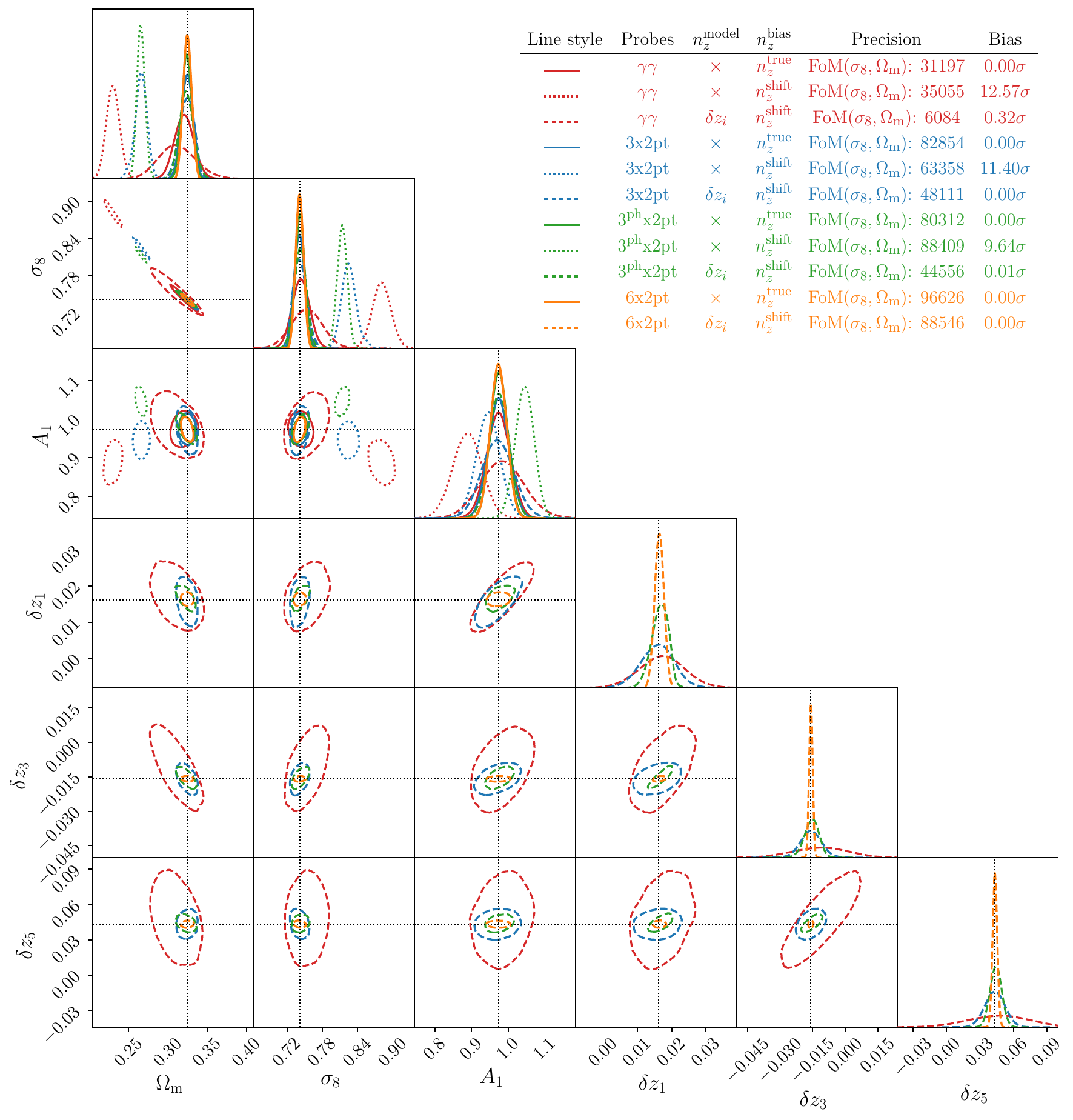}
    \caption{$1\sigma$ confidence contours for cosmological and nuisance parameters, as in Fig. \ref{fig:coh_incoh_shiftmodel_contours}, now for all forecasts conducted using synthetic Stage-IV-like galaxy samples (Sects. \ref{sec:stage_4_photometric} \& \ref{sec:stage_4_spectrosopic}), including: the true redshift distribution $\nztrue$ with the `do nothing' model (i.e. the idealised case; solid contours); the shifted distribution $\nzshift$ with the `do nothing' model (the uncorrected case; dotted contours); and the shifted distributions with the shift $\deltazi$ model (the corrected case; dashed contours); each performed for the $\GG$ (red), $\thr$ (blue), $\pp$ (green), and $\six$ (orange) configurations. Contours for the ${\six}:{\nzshift}:{\rm do\,nothing}$ configuration are excluded from this figure owing to difficulties in reaching stable convergence within the inference chains, though the MAP estimate reveals a strong bias, at $>6\sigma$ on the $\seom$ plane (Table \ref{tab:forecast_summary_s4})
    The addition of clustering cross-correlations within the $\six$ configuration increases the precision of constraints upon the shift parameters, $\deltazi$, thereby preserving $>90\%$ of the idealised FoM whilst erasing the bias on the $\sigma_8-\Omega_{\rm m}$ plane (given in the legend). Conversely, the ${\{\GG,\thr,\pp\}}:\deltazi$ forecasts compensate the parameter bias, but see reductions in the FoM by factors of $\sim1.8-5$.
    }
    \label{fig:stage_4_contours}
\end{figure*}

As a final demonstration of the redshift calibration power of spectroscopic-photometric multi-probe analyses, we turn to our Stage-IV analysis forecasts, the results of which are summarised in Table \ref{tab:forecast_summary_s4}. 

Our Stage-IV synthetic samples (Table \ref{tab:sample_details}) take the form of five-bin Rubin LSST Y1-like tomography for the photometric sample (Sect. \ref{sec:stage_4_photometric}), and DESI-like bright galaxy, emission line galaxy, and luminous red galaxy spectroscopic samples (Sect. \ref{sec:stage_4_spectrosopic}). Besides overall increases to survey depths, areas, and number densities, these mock data also feature greater area/redshift overlaps between spectroscopic and photometric samples, in comparison with our Stage-III-like samples. We note that the final results are sensitive to these survey characteristics, which ultimately determine the signal-to-noise ratios of the different probes explored in this analysis.

To simulate redshift biases, we simply draw random scalar shifts from five zero-mean Gaussian distributions, with widths increasing as a function of redshift\footnote{The resulting shifts were $[0.016, -0.012, -0.016, -0.043,  0.043]$, respectively.}, and apply these to the fiducial photometric distributions. This simpler model for redshift bias was adopted owing to a lack of $n(z)$ covariance estimates for our Stage-IV-like samples, which would have enabled us to follow a similar procedure to that outlined in Sect. \ref{sec:stage_3_photometric}. Given that these are not full-shape $n(z)$ biases, which must be expected for real analyses, we note that these forecasts are more optimistic about the performance of the shift model than those performed for Stage-III surveys.

We sub-divide the spectroscopic sample into $11$ partially-overlapping tomographic bins of width $\Delta{}z\geq0.2$, as described in Sect. \ref{sec:stage_4_spectrosopic}, and conduct forecasts for the subset of configurations given in Table \ref{tab:configurations}. These include the exactly known $\nztrue$ and shifted $\nzshift$ redshift distributions; the `do nothing' and scalar shift $\deltazi$ (with optimistic priors; Sect. \ref{sec:stage_3_photometric}) nuisance models; and the four probe configurations: $\GG$, $\thr$, $\pp$, and $\six$. Similarly to the Stage-III survey setup, we neglected to run Stage-IV shift model forecasts where the $n(z)$ was exactly zero ($\deltazi:\nztrue$).

Fig. \ref{fig:stage_4_contours} displays the collated results of our Stage-IV forecasts, with red contours corresponding to $\GG$, blue to $\thr$, green to $\pp$, and orange to $\six$. Solid contours display $n^{\rm true}(z)$ forecasts without any nuisance model; the idealised scenarios. Note that the blue, green, and orange contours are often overlapping. We find that the non-linear $\siom$ degeneracy seen for Stage-III analyses is linearised for our Stage-IV setup, and quote statistics for the $\siom$ plane accordingly. Relative to $\GG$, the FoM$(\siom)$ is seen to increase by factors of $\sim2.7$ ($\thr$), $\sim2.6$ ($\pp$), and $\sim3.1$ ($\six$).

Dotted contours give the $\nzshift$ forecasts\footnote{In this case, we refrain from showing the $\six$ configuration due to difficulty in finding convergence.}, also without any nuisance model for the $n(z)$. The induced biases on the $\siom$ plane are seen to be severe. Similar to our Stage-III forecasts, the alignment amplitude $A_1$ is significantly mis-estimated as a consequence of unmitigated redshift biases. In contrast to the Stage-III $\pp$ configurations, $A_1$ is shifted high here due to the different form of redshift bias, with bins 2-4 shifted low, and bins 1 \& 5 shifted high. Whilst the net requirement for the $\GG$ and $\thr$ configurations is for increased shear amplitudes, resulting in underestimations of $A_1$ (similarly to Stage-III), the $\pp$ must also satisfy a demand for reduction of photometric GGL signals, and thus yields an overestimation of $A_1$.

The dashed contours in Fig. \ref{fig:stage_4_contours} show the $\nzshift$ forecasts with the application of the shift model for redshift recalibration. The scalar shifts are found to effectively mitigate the $\siom$ bias for each of the configurations. However, weak $\deltazi$ constraints and inter-parameter correlations (between $\deltazi$, and with $A_1$ and cosmological parameters) cause the $\GG$, $\thr$, and $\pp$ configurations to lose significant proportions of their FoM in exchange for the correction. In contrast, the $\six$ configuration is able to constrain the shift parameters at up to $\sim19\times$, $6\times$, and $4\times$ the precision seen for $\GG$, $\thr$, and $\pp$, respectively (Table \ref{tab:shift_precision}), thereby retaining $>90\%$ of its idealised FoM. As a result, the FoM($\siom$) gain factor, relative to $\GG$, is $\sim7.9$ ($\thr$), $\sim7.3$ ($\pp$), and $\sim14.6$ ($\six$) -- the $\six$ analysis allows for far superior retention of constraining power throughout $n(z)$ recalibration.

We reiterate that the artificial redshift distribution bias implemented for these forecasts is particularly generous to the shift model (see Fig. \ref{fig:nz_recovery}, where the $n(z)$ error is reduced to sub-percent level by each configuration). One should expect the recalibration to suffer in all cases under more realistic redshift calibration errors, such as those implemented in our Stage-III forecasts (Sect. \ref{sec:stage_3_photometric}). However, improved recalibration upon inclusion of cross-clustering correlations should be expected for any form of redshift bias, as demonstrated by the dramatic increases in precision for nuisance parameters constrained by the Stage-III and Stage-IV $\six$ forecasts, independent of the suitability of the model for correction of the bias.

We note here that, as discussed in Sect. \ref{sec:modelling_choices}, Stage-IV forecasts are far more capable of meaningfully constraining the halo model amplitude $A_{\rm bary}$, when compared with their Stage-III counterparts -- particularly the $\six$ and $\pp$ configurations, for which the upper edge of the prior (indicating a zero-feedback universe) sits beyond $4\sigma$ in the marginalised 1-d posterior distribution (not shown). Our work therefore indicates that, modulo the approximations made here, the next generation of weak lensing experiments is indeed promising for the future of baryonic feedback modelling.

\subsection{Speed of convergence}
\label{sec:cross_clustering_slow_chains}

\begin{figure}
    \centering
    \includegraphics[width=\columnwidth]{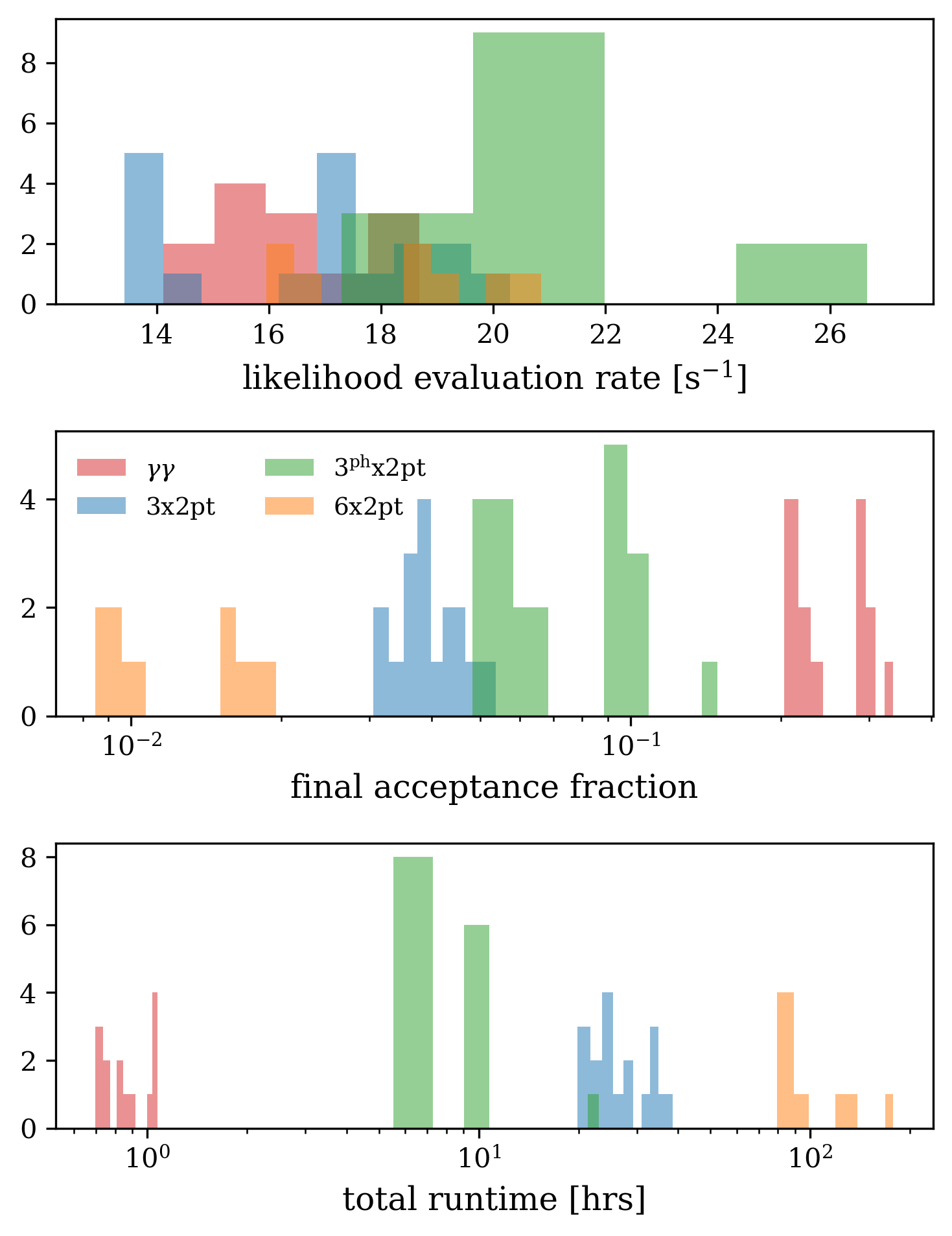}
    \caption{Histograms showing the number of likelihood evaluations per second (\emph{top}), the final accepted fraction of evaluations (\emph{middle}), and the total forecast run-time in hours (\emph{bottom}) of our Stage-III forecasts, coloured according to the probe configurations (as in previous figures, and also given in the legend; see Table \ref{tab:configurations}). We note that these are indicative estimates for runtimes, excluding some forecasts that were rerun at later stages, and including some variability stemming from the number of parallel threads utilised for each specific forecast. As a result, some forecast timings were unrepresentative ($<10$ evaluations ${\rm s}^{-1}$), and these are not shown.
    } 
    \label{fig:convergence_stats}
\end{figure}

We remark here upon the requirements and consequences posed by the inclusion of spectroscopic-photometric cross-clustering correlations within the joint LSS analysis. We emphasise that our conclusions here might be specific to the choice of a nested sampler and the priors adopted in our analyses and they might change if either of these choices are modified.

One of the major lessons learned during this study is that the cross-clustering works to improve nuisance recalibration of redshift distributions by dramatically decreasing acceptance fractions during the sampling of the posterior probability distribution. This reflects the much smaller space of parameters that are able to simultaneously satisfy the cross-clustering -- pinned to the spectroscopic $n(z)$ -- and the demands from shear correlations for modified redshift distributions. 

The time required for convergence of the chains is found to increase accordingly. In order to enable the running of many forecast configurations including the cross-clustering, this motivated us to make the series of approximations described in Sect.~\ref{sec:approximations}. We illustrate this point in Fig.~\ref{fig:convergence_stats}, which shows the likelihood evaluation rates (top), final acceptance fractions (middle), and total run-times\footnote{We note that minor timing overheads are incurred by the optimisation process for each of our Stage-III comb model configurations. The magnitude of the overhead is dependent upon the choice of algorithm, its hyperparameters (e.g.~tolerance, maximum function evaluations, etc.), and the probes--redshift bias combination. Regardless, the time to optimise is subdominant to that required by likelihood sampling, and the optimisation timing is not included in Fig. \ref{fig:convergence_stats}.} (bottom) of our Stage-III forecasts (colours denote probe configurations; see Table \ref{tab:configurations}). 
Our approximations enabled us to evaluate the likelihoods of extended data-vectors $\sim20$ times per second, on-average, and yet the low acceptance rate (middle panel; leftward) of the $\six$ configuration resulted in total run-times of between a few and several days, with the slowest approaching 10 days (40-48 cores). If we are to utilise the redshift calibration power from cross-clustering in future analyses, further advances in the realms of power spectrum emulation and fast computation of other cosmological functions would be a great boon.

\section{Conclusions}
\label{sec:conclusions}

We have conducted simulated likelihood forecasts for the fully combined analysis of weak lensing and galaxy clustering from overlapping Stage-III- and Stage-IV-like photometric and spectroscopic galaxy surveys. In particular, we assessed the potential for positional cross-correlations between the spectroscopic and photometric galaxies to improve the precision and accuracy of the calibration of shear sample redshift distributions, thereby reducing biases in the cosmological parameter inference and increasing the statistical constraining power of the surveys.

Synthetic Stage-III samples were defined to have similar photometric statistics to those reported for the 4th Data Release of the Kilo Degree Survey \citep{Kuijken2019}, supplemented by spectroscopic galaxies from the completed Baryon Oscillation Spectroscopic Survey \citep{Eisenstein2011} and 2-degree Field Lensing Survey \citep{Blake2016a}, assuming $661\sqdeg$ of overlap \citep{Asgari21}. For our Stage-IV configuration, we mimicked the expected statistics for the first year of Rubin Observatory: Legacy Survey of Space and Time observations \citep{TheLSSTDarkEnergyScienceCollaboration2018},
supplemented by spectroscopic observations resembling the Dark Energy Spectroscopic Instrument bright galaxy, emission line galaxy, and luminous red galaxy samples \citep{DESICollaboration2016}, assuming $4000\sqdeg$ of overlap.

We simulated redshift calibration failures for Stage-III surveys by sampling many thousands of 5-bin redshift distributions with variations over the full $n(z)$ shape and selecting extreme outliers in terms of the differences in tomographic mean redshifts. We distinguished between samples according to the `coherence' of the mean redshift errors; whether the bulk distribution is shifted in one direction, or features more internal scattering and full-shape $n(z)$ errors. For our Stage-IV configuration, we simply drew random, uncorrelated shifts to apply to the means of tomographic distributions, thus more closely resembling a Stage-III `incoherent' bias, but lacking full-shape errors.

We explored the capabilities of the commonly-used tomographic bin-wise shift parameters in recalibrating redshift distributions to accommodate such biases, in contrast to a flexible Gaussian mixture model for the $n(z)$, as well as a `do nothing' case where the biases were left unaddressed. Our central proposal was that any redshift recalibration model should be more precisely and accurately constrained by analyses that include clustering cross-correlations with spectroscopic samples, thereby `pinning' the nuisance variables to the well-known spectroscopic $n(z)$. 
To this end, we explored joint analyses of various combinations of up to six cosmological observables, probing the cosmic shear field traced by photometric galaxies and the cosmic density field traced by both photometric and spectroscopic galaxies.

We modelled the matter power spectrum $P(k)$ with the {\tt CosmoPower} emulator \citep{Mancini2021}, trained against a library of power spectra featuring baryonic contributions according to the 1-parameter {\tt HMCode2016} model \citep{Mead2015,Mead2021}. We used the {\tt CCL} library \citep{Chisari2018} in ``calculator'' mode to compute Limber-approximated angular power spectra for cosmic shear $C_{\gamma\gamma}(\ell)$, galaxy-galaxy lensing $C_{n\gamma}(\ell)$, and galaxy clustering $C_{nn}(\ell)$, including all contributions from lensing magnification and intrinsic alignments. We considered the multipole range $\ell\in[100,1500]$ for a Stage-III (and $\ell\in[100,3000]$ for a Stage-IV) survey setup, modelling all scales for cosmic shear correlations and limiting positional probes to an $\ell_{\rm max}$ corresponding to $k_{\rm max}=0.3\,h\,{\rm Mpc}^{-1}$ at a given redshift.

We performed simulations for a flat-$\Lambda$CDM cosmological model, sampling the posterior probability distribution for five cosmological parameters $\Omega_{\rm c}h^2,\Omega_{\rm b}h^2,\lnas,h,n_{\rm s}$ (then deriving $\sigma_8$ and $S_8\equiv\sigma_8\sqrt{\Om/0.3}$), one amplitude each for intrinsic alignments $A_1$ and baryonic contributions $A_{\rm bary}$, (if applicable) one linear galaxy bias $b_{\rm g}$ per tomographic sample, and (if applicable) one scalar shift $\deltazi$ in the mean redshift per photometric tomographic sample. We used an analytical Gaussian covariance in the process. Our primary findings are summarised as follows:
\begin{itemize}
    \item Under idealised conditions, without redshift biases or nuisance models, a Stage-III $\six$ analysis offers a $\sim40\%$ gain in Figure of Merit (FoM) on the $\seom$ plane, relative to the spectroscopic $\thr$ analysis, and a $\sim45\%$ gain relative to the photometric $\pp$ analysis (noting these figures may be specific to our sample definitions).

    \item The relative coherence of biases in the redshift distribution is important in determining the manifestation of the bias at the level of cosmological parameters. Coherent biases -- uni-directional shifts in the bulk distribution -- are more likely to affect the determination of $S_8$ by demanding large changes to the shear correlation amplitudes across all redshifts. Conversely, incoherent biases -- multi-directional shifts, and full-shape errors -- are more harmful to inferences of $\sigma_8,\Om$ and the photometric galaxy bias $\bgphot$, as they manifest more strongly in density correlations that depend differently upon $\sigma_8$ and $\Om$.
    
    \item If we `do nothing' to the (either coherently or incoherently) biased redshift distributions, the intrinsic alignment amplitude $A_1$ is misestimated in order to compensate for the redshift errors. Modifications to $A_1$ can suppress or amplify the shear signal (primarily via GI contributions to off-diagonal correlations), and thereby mimic the shifting of lower-$z$ bins. This is less effective at higher redshifts, where intrinsic alignments are weaker, and where most of the S/N for cosmic shear lies. We confirm that the direction of the error in $A_1$ corresponds to the direction of the bias in tomographic mean redshifts; under-estimation of mean redshifts results in an under-estimation of $A_1$, though this picture is complicated when considering multiple probes beyond cosmic shear, or when tomographic mean errors are variable, i.e. multi-directional.
    
    \item Even when the IA amplitude is misestimated in this way, there remain significant biases on the $\seom$ plane, especially for the photometric $\pp$ ($\gtrsim2\sigma$) and $\six$ ($\gtrsim1\sigma$) configurations. Among all configurations, the spectroscopic $\thr$ is the most robust to redshift biases, owing to its weaker sensitivity to the fine structure of the photometric $n(z)$.

    \item The `shift model' is often able to reduce the overall error $\Dinc$ in the tomographic mean redshifts, but does so whilst typically increasing the full-shape $n(z)$ error $\Dfull$. Cosmic shear $\GG$ and the spectroscopic $\thr$ configuration (with spectroscopic lens galaxies) are sensitive to the photometric redshift distribution only via the lensing efficiency, which is largely determined by the mean redshift; increases to $\Dfull$ are relatively unimportant here, and the shift model is effective in mitigating cosmological parameter biases.
    
    \item In contrast, the photometric $\pp$ and $\six$ configurations are sensitive to the full shape of the $n(z)$ via photometric density probes. In these cases, significant increases to $\Dfull$ cannot be tolerated by the data-vector. The shift model is thus less able to reduce $\Dinc$, such that the $n(z)$ recovery is relatively poor, and small cosmological parameter biases remain, at $\sim0.5\sigma$ ($\pp$) and $\sim0.3\sigma$ ($\six$).

    \item A more flexible model for $n(z)$ recalibration -- the `comb', a Gaussian mixture model -- is able to effectively mitigate tomographic mean errors $\Dinc$, without also expanding the full-shape errors $\Dfull$. With or without the additional application of scalar shifts, this model yields the most accurate recoveries of in/coherently biased redshift distributions, and lowest cosmological parameter biases, per configuration of probes. However, more work is needed to ensure that the comb does not yield adverse impacts for near-perfectly calibrated redshift distributions, i.e. where $n(z)$ biases are not significant.

    \item For the Stage-III configuration, the inclusion of spectroscopic-photometric cross-clustering correlations within the full $\six$ analysis yields up to $\sim4\times$ and $\sim2\times$ tighter constraints upon shift model parameters $\deltazi$, relative to the $\thr$ and $\pp$ analyses, respectively. For our Stage-IV setup, these figures rise to $\sim6\times$ and $\sim4\times$, respectively. This gain in nuisance calibration power translates into double the $\siom$ Figure of Merit compared to $\thr/\pp$, and only an $\sim8\%$ FoM loss compared to the idealised, zero-bias $\six$ scenario. Large-scale structure analyses seeking to maximise constraining power whilst effectively mitigating redshift calibration errors should therefore consider making use of cross-clustering correlations for their enhanced recalibration potential.
    
\end{itemize}

\subsection{Outlook}

One of the challenges we expect for the application of this method is the computational cost of the likelihood evaluations. This will be particularly important for Stage-IV surveys where the modelling will have to be more accurate (and costly) to match the precision of the data. Most of the cost derives from the computation of cross-clustering correlations between spectroscopic and photometric data sets, which enlarge the data-vector but also diminish the fraction of accepted parameter samples that make up the chains. The former could in principle be improved by pre-computation and emulation of the angular power spectra (as opposed to the matter power spectrum), but the latter is likely to pose a consistent challenge. We emphasise that this challenge might be specifically linked to our choice of a nested sampler.

Another aspect to be explored is the configuration of the comb model. In our work, we followed closely the set-up of \cite{Stolzner2020}, which applied the methodology to cosmic shear alone. In principle, the addition of clustering and GGL statistics could enable a finer comb to more accurately recover the $n(z)$, which we have found to be desirable for the handling of residual $A_1$ biases or initially unbiased redshift distributions, as discussed in Section \ref{sec:comb_model}. 
However, a finer comb incurs additional computational costs in the context of analytic marginalisation, where the number of required operations per likelihood evaluation is dramatically increased. For the $\hcomb$ and $\fcomb$ approaches explored here, 
the additional cost would be minimal. The impact of a more motivated marginalisation over uncertainty in the comb model remains to be explored for $\rm (3\mbox{-}6)\times2pt$ analysis configurations. This could be achieved following the methodology in \citep{Stolzner2020} or by incorporating a strategy more similar to {\tt Hyperrank} \citep[see][and the discussion at end of Sect. \ref{sec:comb_model}]{Cordero2022}. Ideally, we would also explore more variably biased redshift distributions; the coherent and incoherent cases could be more dissimilar, with a different net direction of bias, and a distribution maximising the full-shape error (as opposed to the error in the means) would be of interest.

The need for fast theory-vector evaluations in order to run our full likelihood analyses of the $\six$ correlations has led us to make some simplified model choices. Although we acknowledge that the accuracy needs of next-generation data require more complex large-scale structure modelling, we emphasise that the choices were homogeneous throughout the different probe combinations, $\rm (3\mbox{-}6)\times2pt$. This allowed us to make meaningful comparisons between them. However, in the future, it will be important to enable extensions of our pipeline to include more accurate models of baryons \citep{VanDaalen2011,Chisari19,vanDaalen20,Mead2021,Salcido23}, intrinsic alignments \citep{Blazek19,Vlah20,Fortuna21,Bakx23,Maion24}, redshift space distortions \citep{Kaiser87, TH96, Ross11}, non-Limber angular power spectra \citep{Campagne17,Fang2020,Leonard23}, non-linear bias \citep{Baldauf2011, Pandey2020, Mead2021, Nicola2024} and fingers-of-God \citep{Jackson72}, and of their redshift evolution. 
The increased constraining power of the $\six$ configuration also makes it interesting to open up the cosmological parameter space, and to determine the constraining power of this probe combination in terms of curvature, neutrino mass, evolving dark energy, and other extensions \citep{Joudaki:2016kym, Troster2021,DESExtensions,desi2024}.

In this work, we have fixed the number of spectroscopic sample redshift bins (for each Stage). It would be useful for future work to explore whether or not fewer bins could achieve the same performance in recalibrating the photometric redshifts of the shear sample, whilst reducing the cost of each likelihood evaluation. 

As the main motivation of this work has been to understand the gains from including spectroscopic-photometric cross-correlations, we decided to work with projected angular power spectra of the clustering of these galaxies instead of their three-dimensional power spectra. In the future, this alternative could also be explored. This would increase the cosmological constraining power in the spectroscopic sample alone \citep[e.g.][]{Joudaki2017,heymans21},  specifically by breaking degeneracies with $\Omega_{\rm m}$ and the spectroscopic bias parameters (through the redshift space distortions; \citealt{Kaiser87}). The challenge lies in computing and validating the covariance matrix when the estimators used for the different tracers are not the same, which has thus far been achieved by creating a large suite of numerical simulations \citep{hdvw15, Joachimi21, Joudaki2017, heymans21}. Recent work moreover suggests a possible avenue for its analytic computation \citep{Taylor:2022rgy}. 

We have considered the same, linear galaxy bias function for the spectroscopic and photometric samples. It is likely that the galaxy bias will differ in the real data due to the differences in the redshift coverage and depth of each sample. This can have advantages in terms of cosmological constraints, as differently biased tracers that sample the same underlying field can mitigate the impact of cosmic variance on the cosmological constraints \citep{McDonald09b}.

The increased constraining power from the $\six$ probe combination results in higher sensitivity to modelling choices. Each element of the model should be tested in the future, such as whether the $\rm 6\times2pt$ analysis is sensitive to models of intrinsic alignments with higher complexity than NLA, or to fundamental parameters like neutrino mass \citep{FontRibera14}. In this work, except for the case of photometric redshift distributions, we have always matched the model in the synthetic data vector with that in the theoretical prediction, but these can be deliberately mismatched to assess potential biases in the cosmological parameters due to assuming overly simplistic models given the precision of the data. 
The same applies to models of galaxy bias, baryonic feedback, and other astrophysical or instrumental effects. This should further be coupled to an exploration of the range of scales being used in the analysis, i.e. $\ell_{\rm max}$ (e.g. \citealt{Krause2016}).

\begin{acknowledgements}
This paper went through a KiDS internal review process.

We are grateful to Markus Michael Rau and Elena Sellentin for helpful discussions, and we also thank Alessio Spurio Mancini for his development of and continued improvements to the {\tt CosmoPower} emulator, without which this work would not have been possible. 

We thank members of the KiDS consortium for stimulating discussions and support throughout this work. 

HJ, NEC acknowledge support from the Delta ITP consortium, a program of the Netherlands Organisation for Scientific Research (NWO) that is funded by the Dutch Ministry of Education, Culture and Science (OCW), project number 24.001.027. This publication is also part of the project ``A rising tide: Galaxy intrinsic alignments as a new probe of cosmology and galaxy evolution'' (with project number VI.Vidi.203.011) of the Talent programme Vidi which is (partly) financed by the Dutch Research Council (NWO). 

SJ acknowledges the Dennis Sciama Fellowship at the University of Portsmouth and the Ram\'{o}n y Cajal Fellowship from the Spanish Ministry of Science.

AL acknowledges support from the research project grant `Understanding the Dynamic Universe' funded by the Knut and Alice Wallenberg Foundation under Dnr KAW 2018.0067.

BJ acknowledges support by the ERC-selected UKRI Frontier Research Grant EP/Y03015X/1 and by STFC Consolidated Grant ST/V000780/1.

RR, JLvdB and AD are supported by an ERC Consolidator Grant (No. 770935). 

KK acknowledges support from the Royal Society and Imperial College.

BS, ZY and SU acknowledge support from the Max Planck Society and the Alexander von Humboldt Foundation in the framework of the Max Planck-Humboldt Research Award endowed by the Federal Ministry of Education and Research (Germany).

TT acknowledges funding from the Swiss National Science Foundation under the Ambizione project PZ00P2$\_$193352.

AHW is supported by the Deutsches Zentrum für Luft- und Raumfahrt (DLR), made possible by the Bundesministerium für Wirtschaft und Klimaschutz, and acknowledges funding from the German Science Foundation DFG, via the Collaborative Research Center SFB1491 ``Cosmic Interacting Matters - From Source to Signal''.

HH is supported by a DFG Heisenberg grant (Hi 1495/5-1), the DFG Collaborative Research Center SFB1491, an ERC Consolidator Grant (No. 770935), and the DLR project 50QE2305.

MA acknowledges support from the UK Science and Technology Facilities Council (STFC) under grant number ST/Y002652/1 and the Royal Society under grant numbers RGSR2222268 and ICAR1231094.

MB is supported by the Polish National Science Center through grants no. 2020/38/E/ST9/00395, 2018/30/E/ST9/00698, 2018/31/G/ST9/03388 and 2020/39/B/ST9/03494.

BG acknowledges support from a UK STFC Consolidated Grant and from the Royal Society of Edinburgh through the Saltire Early Career Fellowship (ref. number 1914).

CH acknowledges support from the European Research Council under grant number 647112, from the Max Planck Society and the Alexander von Humboldt Foundation in the framework of the Max Planck-Humboldt Research Award endowed by the Federal Ministry of Education and Research, and the UK Science and Technology Facilities Council (STFC) under grant ST/V000594/1.

LL is supported by the Austrian Science Fund (FWF) [ESP 357-N].

HYS acknowledges the support from CMS-CSST-2021-A01 and CMS-CSST-2021-A04, NSFC of China under grant 11973070, and Key Research Program of Frontier Sciences, CAS, Grant No. ZDBS-LY-7013 and Program of Shanghai Academic/Technology Research Leader.

MWK acknowledges support from the Science and Technology Facilities Council.

CM acknowledges support under the grant number PID2021-128338NB-I00 from the Spanish Ministry of Science, and from the European Research Council under grant agreement No. 770935.

Part of this work is based upon public data products produced by the KiDS consortium, derived from observations made with ESO Telescopes at the La Silla Paranal Observatory under programme IDs 100.A-0613, 102.A-0047, 179.A-2004, 177.A-3016, 177.A-3017, 177.A-3018, 298.A-5015. 

Figures in this work were produced using {\tt matplotlib} \citep{Hunter2007} and {\tt ChainConsumer} \citep{Hinton2016}. 

Author contributions: all authors contributed to the development and writing of this paper. The author list is given in three groups: the lead authors (HJ, NEC, SJ, RR, BS), followed by two alphabetical groups. The first alphabetical group includes those who are key contributors to both the scientific analysis and the data products. The second group covers those who have either made a significant contribution to the data products or the scientific analysis. 

For the purpose of open access, a CC BY public copyright license is applied to any Author Accepted Manuscript version arising from this submission.

\end{acknowledgements}

\bibliographystyle{aa}
\bibliography{main}

\end{document}